\documentclass[aps,amsmath,amssymb,nofootinbib]{revtex4}
\usepackage{graphicx,color}
\usepackage{dcolumn}% Align table columns on decimal point
\usepackage{bm}% bold math
\definecolor{labelkey}{cmyk}{.4,.2,0,0}

\usepackage{hyperref}

\newcommand{\be}{\begin{equation}}
\newcommand{\ee}{\end{equation}}
\newcommand{\bea}{\begin{eqnarray}}
\newcommand{\eea}{\end{eqnarray}}

\newcommand{\nn}{\nonumber }

\newcommand{\JN}{{\mathbb N}}

\newcommand{\JZ}{{\mathbb Z}}

\newcommand{\JR}{{\mathbb R}}

\newcommand{\sE}{{\sf E}}
\newcommand{\sU}{{\sf U}}
\newcommand{\sV}{{\sf V}}
\newcommand{\su}{{\sf u}}
\newcommand{\sv}{{\sf v}}
\newcommand{\sff}{{\sf f}}

\usepackage{amsthm}

\theoremstyle{plain}
  \newtheorem{theorem}{Theorem}[section]
  
  \newtheorem{proposition}[theorem]{Proposition}

\theoremstyle{definition}
  \newtheorem{definition}{Definition}[section]

\theoremstyle{remark}

\numberwithin{equation}{section}

\begin{document}

\title{Stationary measures for two dual families of finite and zero temperature models of directed polymers on the square lattice}

\author{Thimoth\'ee Thiery } \affiliation{CNRS-Laboratoire
de Physique Th{\'e}orique de l'Ecole Normale Sup{\'e}rieure, PSL Research University, 24 rue
Lhomond,75231 Cedex 05, Paris, France}

\begin{abstract}
We study the recently introduced Inverse-Beta polymer, an exactly solvable, anisotropic finite temperature model of directed polymer on the square lattice, and obtain its stationary measure. In parallel we introduce an anisotropic zero temperature model of directed polymer on the square lattice, the Bernoulli-Geometric polymer, and obtain its stationary measure. This new exactly solvable model is dual to the Inverse-Beta polymer and interpolates between models of first and last passage percolation on the square lattice. Both stationary measures are shown to satisfy detailed balance. We also obtain the asymptotic mean value of (i) the free-energy of the Inverse-Beta polymer; (ii) the optimal energy of the Bernoulli-Geometric polymer. We discuss the convergence of both models to their stationary state. We perform simulations of the Bernoulli-Geometric polymer that confirm our results.
\end{abstract}

\maketitle

\section{Introduction}

The directed polymer (DP) problem, i.e. the statistical mechanics problem of directed paths in a random environment, has been the subject of intense studies both from the physics and mathematics community (see e.g. \cite{HuseHenley1985,Kardar1987} for early physics work). The DP is a classical example of equilibrium statistical mechanics of disordered systems, but its importance goes well beyond this field, notably because of its connection with the Kardar-Parisi-Zhang (KPZ) universality class \cite{KPZ} (for recent reviews see \cite{SpohnReview, HalpinReview,CorwinReview}). In the field of DPs in dimension $1+1$, important progresses have been possible thanks to the existence of models with {\it exact solvability properties}, that is models for which, for one or several reasons, exact computations are possible. Examples of such properties include notably Bethe ansatz (BA) integrability, existence of combinatorial mappings (Robinson-Schensted-Knuth (RSK) correspondence and geometric RSK (gRSK) correspondence) and the exact solvability property (ESP) which is the focus of this work, {\it an exactly known stationary measure} (SM). A given model can have one or several of those properties. The continuum directed polymer is BA solvable \cite{kardareplica} and its SM is also known: starting from an initial condition such that the free-energy of the DP performs a Brownian motion, it remains so at all time \cite{HuseHenleyFisher,ForsterNelsonStephen}. Geometric and exponential last passage percolation are exactly solvable using the RSK correspondence \cite{Johansson2000}, can also be mapped (see e.g. \cite{KriecherbauerKrug}) onto the totally asymmetric exclusion process (TASEP), which is exactly solvable by BA, and its SM is also exactly known. The SM of the O'Connell-Yor semi-discrete DP is also known \cite{OConnellYor,SeppalainenValko} and it is solvable using the gRSK correspondence \cite{OConnellToda}. The first discovered exactly solvable model of DP on the square lattice at finite temperature, the Log-Gamma polymer, was introduced because of the possibility of writing down exactly its SM \cite{Seppalainen2012}. It was later shown that the model is exactly solvable using the gRSK correspondence \cite{logsep2} and BA \cite{usLogGamma}. The shortly after introduced Strict-Weak polymer also enjoy all three properties \cite{StrictWeak1,StrictWeak2} while for the recently discovered Beta \cite{BarraquandCorwinBeta,ThieryLeDoussal2016b} and Inverse-Beta polymer \cite{usIBeta} only BA solvability has been shown (although a work on the SM of the Beta polymer is currently in preparation \cite{SeppalainenInprep}).

\smallskip

The links between these different types of ESPs are not yet understood. As such, the discovery of an ESP in a model is of great interest, even when the model already has one known ESP. This is true from a mathematical point of view since these properties are signs of a rich underlying mathematical structure, but it is also important from the perspective of calculating relevant physical observables since each ESP has interesting applications. In particular, although most of the recent focus on exactly solvable models of DP has been on the derivation of the exact distribution of the fluctuations of the free-energy at large scale, an information which is not contained in the SM and for which RSK/gRSK correspondence \cite{Johansson2000,logboro,StrictWeak2} and BA solvability \cite{we,dotsenko,ThieryLeDoussal2016b,usIBeta,BarraquandCorwinBeta,usLogGamma} are more adapted, the exact knowledge of the SM is of great interest. The SM indeed contains information on the multi-point correlations of the DP free-energy at large scale. These are notoriously hard to obtain using other analytical techniques. More generally the SM allows to study different questions in a complementary fashion to other ESPs. An important historical example of application of the knowledge of the SM of the continuum DP can be found in \cite{HuseHenleyFisher,ForsterNelsonStephen}: together with the Galilean invariance, it provided the first (and probably still the simplest) derivation of the critical exponents of the KPZ universality class. More recently in the Log-Gamma case, the SM was e.g. used to obtain a rigorous derivation of the critical exponents of the DP \cite{Seppalainen2012}, or also to derive a precise characterization of the localization properties of the DP \cite{CometsNguyen2015}.

\smallskip

\begin{figure}
\centerline{\includegraphics[width=7cm]{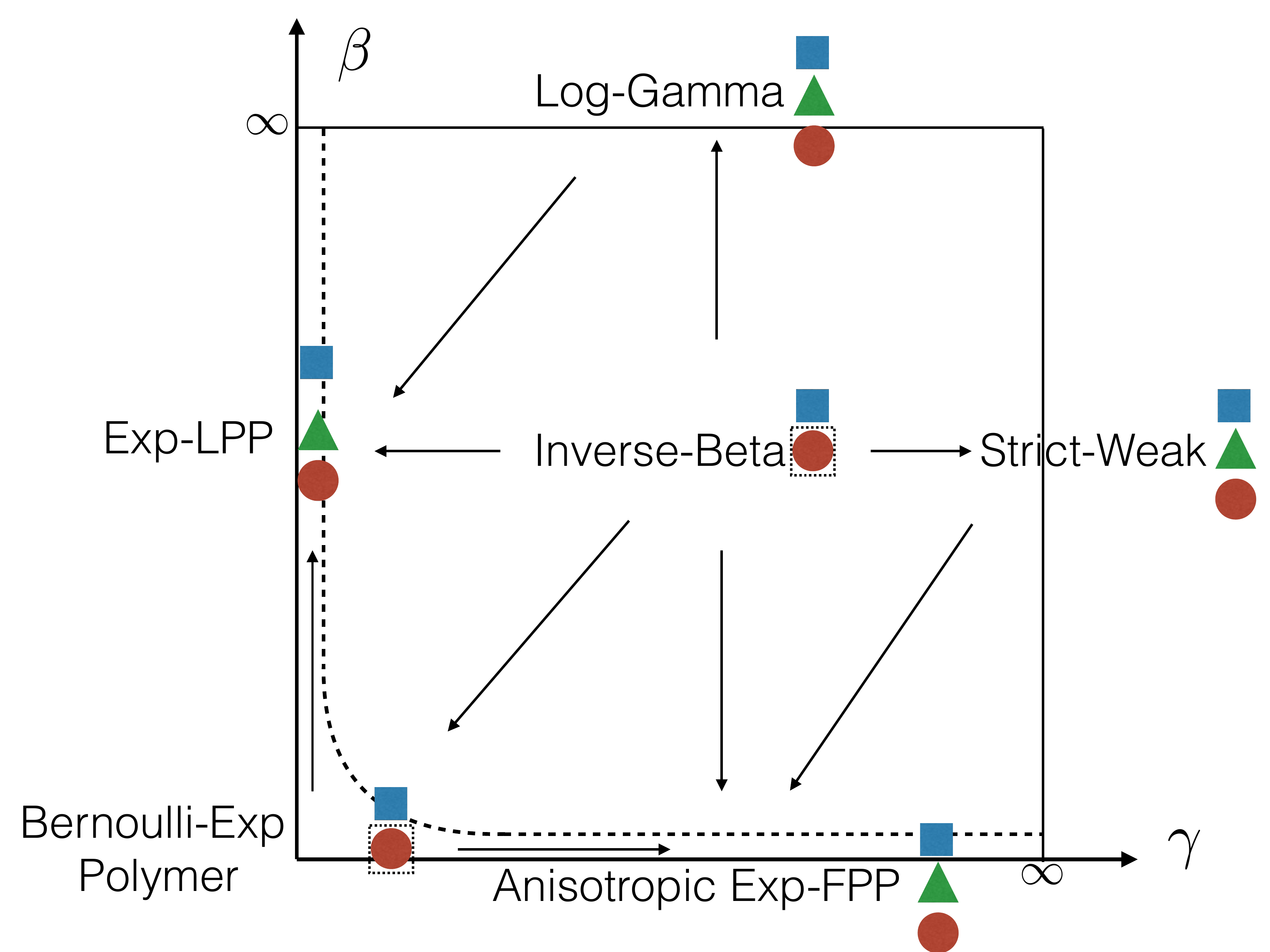}  \includegraphics[width=7cm]{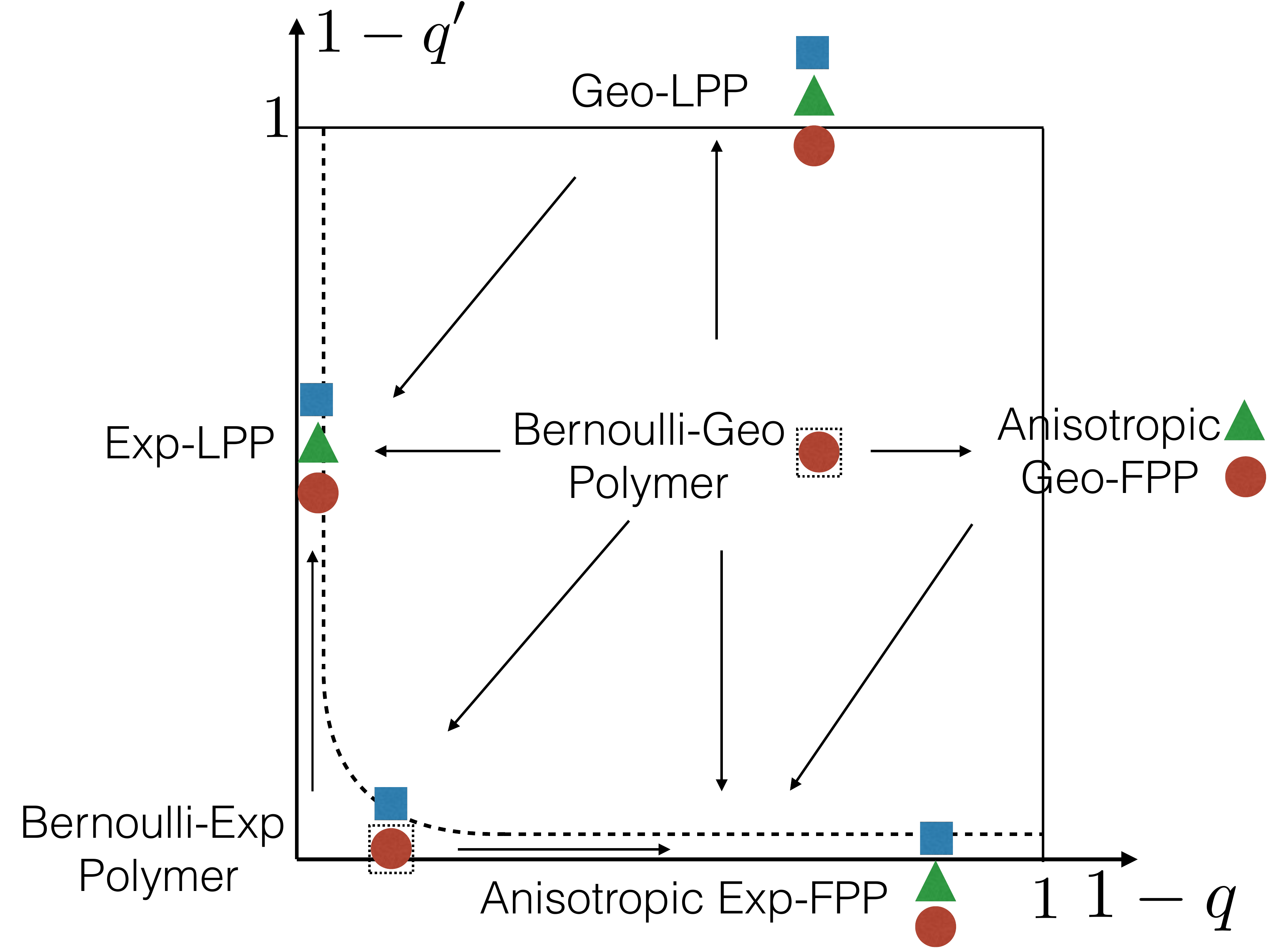} } 
\caption{The Inverse-Beta polymer is at the center of a large class of finite temperature exactly solvable models of DP on the square lattice with continuous random energies (left), but also admits zero-temperature limiting models (`below the dashed-line', in the limit ${\rm min(\gamma , \beta) } \to 0$). Conversely, the Bernoulli-Geometric polymer introduced in this paper is at the center of a large class of zero-temperature models of DP on the square lattice with discrete random energies (left), but it also admits limiting models with continuous energies (`below the dashed-line', in the limit ${\rm max(q , q') } \to 1$), which coincide with the limiting zero-temperature models of the Inverse-Beta polymer. Arrows indicate the possibility of taking a limit from one model to another. The models shown in this picture are defined in Sec.~\ref{Sec:Overview}, Sec.~\ref{subsec:relation} and Sec.~\ref{subsec:T0relation}. Different known exact solvability properties of the various models are here indicated by blue squares for BA solvability, green triangles for RSK or gRSK solvability and by red dots for exactly known stationary measures. The red dots enclosed by a dashed-line (as well as the definition of the Bernoulli-Geometric polymer) are some of the results of this work.}
\label{fig:spaceofmodels}
\end{figure}

The goal of this paper is twofold. First we obtain the stationary measure of the recently discovered Inverse-Beta polymer. In a few words in the stationary state the free-energy of the DP performs a random walk with Inverse-Beta distributed increments, thus generalizing in a discrete setting the stationary measure of the continuum DP. The existence of this stationary measure is rather natural since the Inverse-Beta polymer is an anisotropic finite temperature model (with two parameters $\gamma, \beta >0$) of DP on the square lattice which in different limits converges in law to the Log-Gamma ($\beta \to \infty$, the isotropic limit) and Strict-Weak polymer ($\gamma \to \infty$, the strongly anisotropic limit). These two models possess an exactly known SM that we will generalize to the Inverse-Beta polymer, and our approach will have a strong methodological and conceptual overlap with the one used by Sepp\"{a}l\"{a}inen in \cite{Seppalainen2012}. Secondly we introduce a new anisotropic $0$ temperature model of DP on the square lattice (with two parameters $q,q'\in [0,1[$), which we call the Bernoulli-Geometric polymer, and obtain exactly its stationary measure. This model interpolates between the exactly solvable geometric {\it first} passage percolation problem studied in \cite{OConnell2005} (in the $q \to 0$ limit) and the geometric {\it last} passage percolation problem e.g. studied in \cite{Johansson2000} (in the $q' \to 0$ limit). The existence of this model was already suggested in \cite{usIBeta} following the fact that a $0$ temperature limit ($\gamma = \epsilon \gamma'$ $\beta = \epsilon \beta'$ and $\epsilon \to 0$) of the Inverse-Beta polymer gave an anisotropic generalization of {\it exponential} last passage percolation. Since (isotropic) last passage percolation is exactly solvable both for geometric and exponential distribution of random waiting times (the exponential case being the limit $q = 1 - \gamma' \epsilon$ with $\epsilon \to 0$ of the geometric case), it was rather natural to conjecture that an exactly solvable anisotropic generalization of {\it geometric} last passage percolation should exist. This motivated the search for such a model. The Bernoulli-Geometric polymer introduced in this paper appears as this missing model, and we thus complement the rich universe of exactly solvable models of DP on the square lattice (see Fig.~\ref{fig:spaceofmodels}). For the finite temperature case in particular, the only known model not present in this framework is the Beta polymer, which somehow lives in a different class since it has the peculiarity of also being a model of random walk in a random environment \cite{BarraquandCorwinBeta,ThieryLeDoussal2016b}. 

Before we give the main results of the paper and define the Inverse-Beta and Bernoulli-Geometric polymers in Sec.~\ref{Sec:Overview}, let us start by explaining more precisely the general question that is tackled in this article on a simpler model. 

\section{Recall: stationary measure of the Log-Gamma polymer} \label{sec:Recall}

In this section for pedagogical purposes we recall the stationary measure of the Log-Gamma polymer. The results that we obtain on the stationary measure of the Inverse-Beta polymer can be seen as a generalization of the known results presented in this section to a richer model, and we believe it can be useful for non-specialists to first recall here those simpler results. Specialists on the other hand are encouraged to jump directly to Sec.~\ref{Sec:Overview}. 

\smallskip

\begin{figure}
\centerline{  \includegraphics[width=7cm]{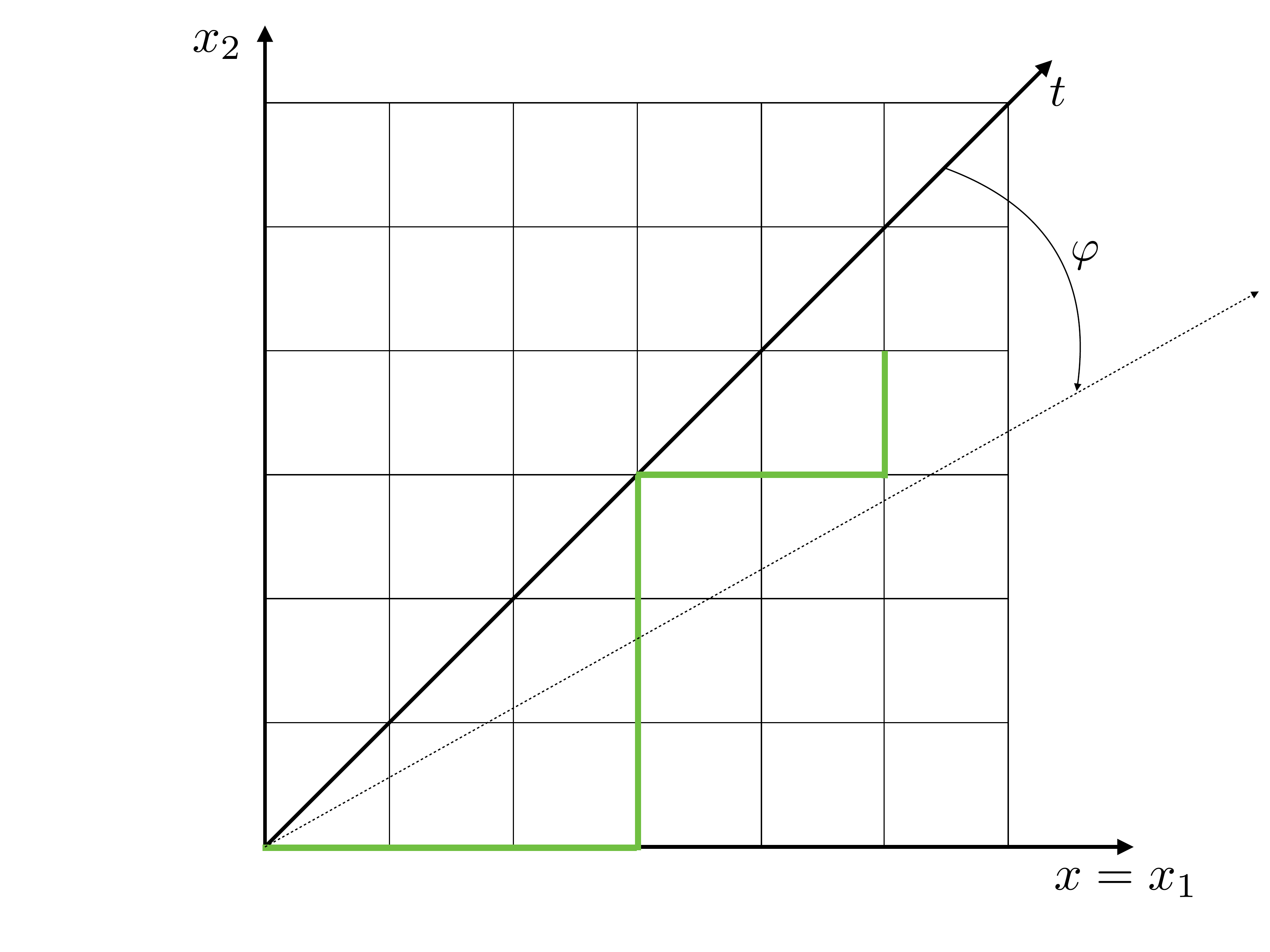}} 
\caption{A model of directed polymer on $\JZ^2$. Green: an admissible (i.e. up/right) polymer path of length $t=9$ with starting point $(0,0)$ and endpoint $(x_1,x_2) = (5,4)$. We are interested in the stationarity properties that are reached in the limit of long polymers in a given direction $\varphi$.}
\label{fig:introLogGamma}
\end{figure}

Let us first consider the case of an abstract, homogeneous model of directed polymer on the square lattice with on-site disorder: the random environment is defined by drawing random Boltzmann weights $W_{x_1,x_2} > 0$ at each point $(x_1,x_2)$ of $\mathbb{N}^2$. Boltzmann weights on different lattice sites are supposed to be independent and homogeneously distributed as a positive random variable (RV) $W$ with a probability distribution function $P_W(W)$. The partition sum of DP with starting point $(x_1,x_2) = (0,0)$ and endpoint $(x_1,x_2) \in \mathbb{N}^2$ is defined by
\bea \label{firstsec1}
Z_{x_1,x_2} := \sum_{\pi :(0,0) \to (x_1 , x_2) } \prod_{(x_1',x_2') \in \pi} W_{x_1',x_2'} \ ,
\eea
where the sum $\sum_{\pi :(0,0) \to (x_1 , x_2) }$ is over all directed paths, also called up-right paths, from $(0,0)$ to $(x_1,x_2)$. Those are the paths such that the only jumps allowed are to the right, i.e. as $(x_1 ,x_2) \to (x_1 + 1 , x_2) $ or upward, i.e. as $(x_1 ,x_2) \to (x_1  , x_2+1) $ (see Fig.~\ref{fig:introLogGamma}). For a given model of DP, one would like e.g. to characterize the asymptotic properties of $Z_{x_1,x_2}$ in the limit of long polymers $t=x_1+x_2 \to \infty$. In this paper we focus on the horizontal and vertical ratios of partition sums defined as
\bea
U_{x_1,x_2} := \frac{Z_{x_1,x_2}}{Z_{x_1-1,x_2}} \quad , \quad V_{x_1,x_2} := \frac{Z_{x_1,x_2}}{Z_{x_1,x_2-1}}   \  . 
\eea
Introducing the variables $t=x_1+x_2$ (the length of the polymers) and $x=x_1$ and the notations $U_t(x):= U_{x , t-x}$ and $V_t(x) := V_{x,t-x}$, we are interested in obtaining the distribution of these RVs in the limit of long polymers in a given direction. That is, for a given $\varphi \in ]-1/2 , 1/2[$ (see Fig.~\ref{fig:introLogGamma}) and $\forall T \in \JN^*$, $X \in \JN^*$, we are interested in the set of RVs
\bea
\left( \tilde{U}_{t'}(x')  , \tilde{V}_{t'}(x') \right)_{t' \in [0,T], x' \in [-X,X]} := \lim_{t \to \infty} \left(U_{t +t'}((1/2 +\varphi)t + x') , V_{t +t'}((1/2 +\varphi)t + x')  \right)_{t' \in [0,T], x' \in [-X,X]} \ .
\eea
The only known finite temperature model of DP on $\JZ^2$ with on-site disorder (i.e. defined as above) for which characterizing exactly the properties of the asymptotic process $\left( \tilde{U}_{t'}(x')  , \tilde{V}_{t'}(x') \right)_{t' \in [0,T], x' \in [-X,X]} $ is possible is the Log-Gamma polymer. In this case, the random Boltzmann weights are distributed as the inverse of a gamma random variable: $W \sim Gamma(\gamma)^{-1}$. Here $\sim$ means `distributed as' and we recall that a RV $x$ is gamma distributed with parameter $\alpha>0$ if its PDF is $p(x) = \frac{1}{\Gamma(\alpha)} x^{-1 + \alpha} e^{-x} \theta(x)$ ($\Gamma$ is the Euler's gamma function and $\theta$ is the Heaviside theta function). For this special choice of distribution, although it is not mathematically fully proven, the (mathematically rigorous) results of \cite{Seppalainen2012,Seppalainen2015} lead to the conjecture that in this case

\medskip

At fixed $t' \in [0,T]$ the variables $(\tilde{U}_{t'}(x'))_{x' \in [-X ,  X]}$ and $(\tilde{V}_{t'}(x'))_{x' \in [-X,X]}$ are all independent and distributed as $\tilde{U}_{t'}(x') \sim Gamma(\gamma-\lambda)^{-1}  $ and $\tilde{V}_{t'}(x') \sim Gamma(\lambda)^{-1}$. The additional parameter $\lambda \in ]0 , \gamma[$ depends on $\varphi$ and is the solution of the equation $0= -(1/2 +\varphi) \psi'(\gamma -\lambda)  + (1/2 - \varphi)\psi'(\lambda) $, where $\psi(x) = \Gamma'(x)/\Gamma(x)$ is the digamma function.

\medskip

An additional property of reversibility of the process is known from \cite{Seppalainen2012}. These properties rely on a non trivial property of gamma distributions (see Lemma 3.2 in \cite{Seppalainen2012}). Moreover, Lemma 3.2 of \cite{Seppalainen2012} also suggests that the Log-Gamma polymer is the only model with on-site disorder for which it is possible to obtain exactly the stationary measure. One of the purposes of this paper is to show that it is also possible to obtain exactly the stationary measure in the Inverse-Beta polymer, an anisotropic finite temperature model of DP on $\JZ^2$ with on-edge disorder that generalizes both the Log-Gamma and Strict-Weak models. We also obtain similar results for the Bernoulli-Geometric polymer, a related zero temperature model that we introduce in this paper.

\section{Overview: definitions, main results and outline} \label{Sec:Overview}

\subsection{Definitions of the models of directed polymers} \label{SecOverDef}

\subsubsection{{\bf General notations}}

All the models of DPs considered in this paper live on the square lattice $\JZ^2$. We will consider two coordinate systems on $\JZ^2$, the usual Euclidean coordinates $(x_1,x_2) \in \JZ^2$ and the $(t,x)$ coordinates $t=x_1+x_2$ and $x=x_1$ (see Fig.\ref{fig:IBeta}). The variable $t$ will often corresponds to the length of the polymers. To avoid confusion, an arbitrary real function on the lattice, $f: (x_1,x_2) \in  \JZ^2 \to  f(x_1,x_2) \in \JR$ will be denoted either as $f_{x_1,x_2} := f(x_1,x_2)$, or as $f_t(x) = f(x_1=x , x_2 = t-x)$. The random environment will live on the {\it edges} of $\JZ^2$ and we will generally note by $e$ an edge of $\JZ^2$.

\subsubsection{{\bf Finite temperature models: The Inverse-Beta polymer(s)}} \label{SubSecOverDefIB}

We now define three versions of the Inverse-Beta (IB) polymer. The first is the usual point to point IB polymer introduced in \cite{usIBeta}. Its partition sum will be noted $Z_{x_1,x_2}$. The second is the IB polymer with boundaries, a model which possesses a stationarity property and whose definition is original to this work. Its partition sum will be noted $\hat Z_{x_1,x_2}$. The third model is the IB polymer with a stationary initial condition, with partition sum $\check Z_{x_1,x_2}$, which also possesses a stationary property and whose definition is original to this work. It is intimately linked with the IB polymer with boundaries and is closer in spirit to the stationary models considered for the continuum DP. The first model will be defined by choosing {\it two} parameters $(\gamma,\beta) \in \JR_+^2$ (henceforth referred to as the bulk parameters). The others have one additional parameter $\lambda\in]0,\gamma[$, which will specify one stationary measure among a family of stationary measures at fixed $(\gamma,\beta)$ (henceforth referred to as the stationarity or boundary parameter). Throughout this work the use of the hat and check notations will permit to distinguish between quantities associated to each model. 

\smallskip

\begin{definition}
{\bf The point to point IB polymer}\label{Def:ptopIB}
We recall here the definition of the point to point IB polymer partition sum as studied using Bethe ansatz in \cite{usIBeta}. To each vertex $(x_1,x_2) \in \JZ^2$ of the square lattice is associated a random variable $W_{x_1,x_2}\in \mathbb{R}_+$. The set of RVs $\{W_{x_1 ,x_2} , (x_1,x_2) \in \mathbb{Z}^2 \}$ consists of independent, identically distributed (iid) RVs distributed as $W \sim \frac{1}{B} -1$ where $B\in [0,1]$ is a Beta RV of parameters $\gamma$ and $\beta>0$. The PDF $P(B)$ of a Beta random variable is
\bea \label{DistBeta}
B \sim Beta(\gamma , \beta )  \Longleftrightarrow P(B) = \frac{\Gamma(\gamma+\beta)}{\Gamma(\gamma)\Gamma(\beta)} B^{\gamma-1} (1-B)^{\beta-1} \theta(B) \theta(1-B)\ ,
\eea
where here and throughout the rest of the paper $\sim$ means `distributed as', $\Gamma$ is the Euler's gamma function and $\theta$ is the Heaviside theta function. Given a random environment specified by a drawing of $W_{x_1,x_2}$ at each vertex $(x_1,x_2) \in \mathbb{Z}^2$, we associate to each edge of the square lattice $e$ a random Boltzmann weight (BW) $w(e)$ as follows. The random BWs $w(e)$ on horizontal (resp. vertical) edges will be denoted by the letter $u$ (resp. $v$) and indexed by the vertex to which they lead (see left of Fig.~\ref{fig:IBeta} and beware that we use here the opposite convention compared to \cite{usIBeta}), and given in terms of $W_{x_1,x_2}$ by
\bea \label{Distuv}
&& w((x_1-1, x_2) \to (x_1,x_2)) = u_{x_1,x_2} = W_{x_1,x_2}   > 0  \ , \nn \\
&& w((x_1-1, x_2) \to (x_1,x_2)) = v_{x_1,x_2} = W_{x_1,x_2} +1  > 1  \ .
\eea
\begin{figure}
\centerline{  \includegraphics[width=9cm]{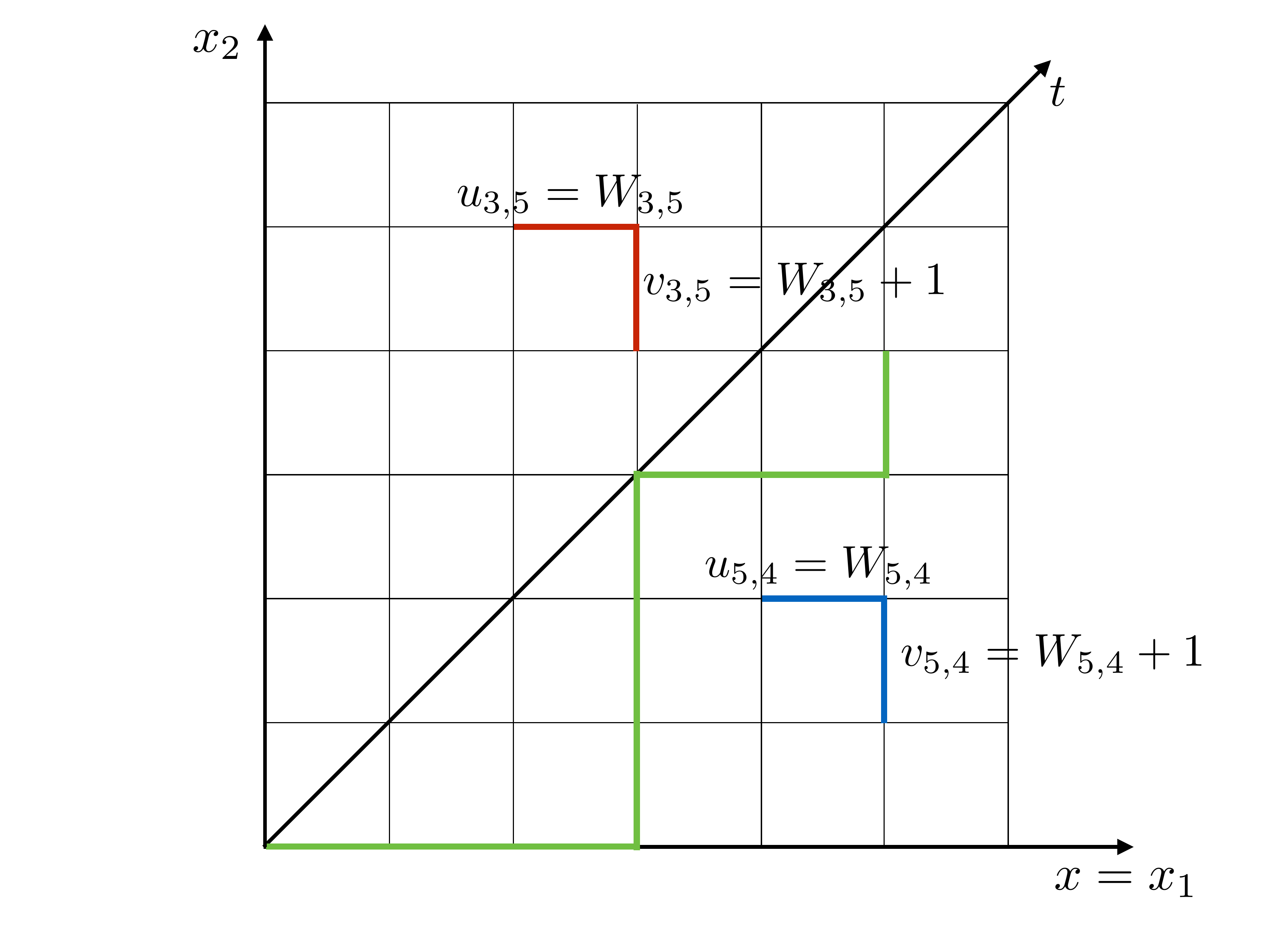} , \includegraphics[width=9cm]{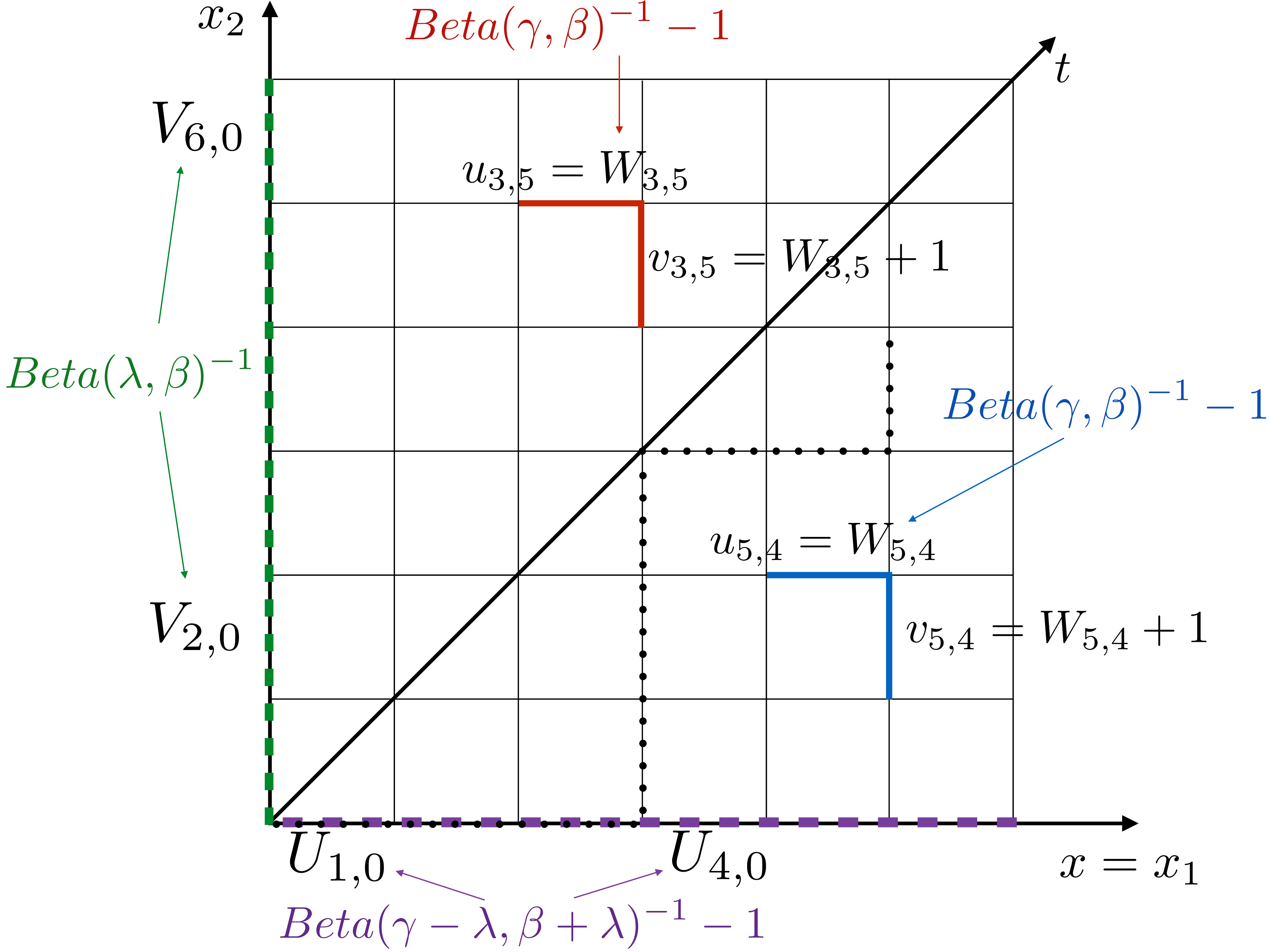} } 
\caption{Left: The point to point Inverse-Beta polymer. Blue (resp. Red) : couple of correlated Boltzmann weights on edges arriving at $(x_1 ,x_2) =(5,4)$ (resp. $(x_1 ,x_2) =(3,4)$). Green: an admissible (i.e. up/right) polymer path of length $t=9$ with starting point $(0,0)$ and endpoint $(x_1,x_2) = (5,4)$. Right: The Inverse-Beta polymer with boundaries. The Boltzmann weights in the bulk (blue and red) are the same as in the model without boundaries and are distributed as in (\ref{Statio-Inv-Beta3}). The random Boltzmann weights on the vertical (dashed-green) and horizontal (dashed-purple) boundaries are distributed as in (\ref{Statio-Inv-Beta2}). The dotted line represents a possible polymer path from $(x_1,x_2) = (0,0)$ to $(x_1,x_2) = (5,4)$.  }
\label{fig:IBeta}
\end{figure}

Hence in this model the BWs on different edges are correlated if and only if they lead to the same vertex, since in this case $v = u+1$, and the vertical direction is always favored compared to the horizontal one. The model is thus anisotropic. It interpolates between two other known exactly solvable models of DP on $\JZ^2$: the Log-Gamma polymer (isotropic $\beta \to \infty$ limit) and the Strict-Weak polymer ($\gamma \to \infty$ limit, see Sec.~\ref{subsec:relation}). Let us write here for clarity the PDF of $W$, noted $P_W(W)$:
\bea \label{Statio-Inv-Beta3}
W \sim \frac{1}{Beta(\gamma , \beta )} -1 >0 \  \quad , \quad P_W(W) = \frac{\Gamma(\gamma+\beta)}{\Gamma(\gamma)\Gamma(\beta)}  \left(1-\frac{1}{W+1}\right)^{\beta -1} \left(\frac{1}{W+1}\right)^{\gamma +1} \theta(W).
\eea
% The isotropic limit of the model is the Log-Gamma polymer model. It is obtained by sending $\beta \to \infty$ (see \cite{usIBeta} and Sec.~\ref{subsec:relation}). For this reason $\beta$ will be called the anisotropy parameter.
 Given a random environment, the partition sum of the point-to-point IB polymer with starting point $(0,0)$ and endpoint $(x_1 \geq 0,x_2 \geq 0)$ is defined as
\bea
Z_{x_1,x_2} = \sum_{\pi :(0,0) \to (x_1 , x_2) } \prod_{e \in \pi} w(e) \ ,
\eea
where here and throughout the rest of this work the sum $\sum_{\pi :(0,0) \to (x_1 , x_2) }$ is over all directed paths, also called up-right paths, from $(0,0)$ to $(x_1,x_2)$. Those are the paths such that the only jumps allowed are either to the right, i.e. as $(x_1 ,x_2) \to (x_1 + 1 , x_2) $, or upward, i.e. as $(x_1 ,x_2) \to (x_1  , x_2+1) $ (see Fig.~\ref{fig:IBeta}). Equivalently, using the $(t,x)$ coordinate system the partition sum $Z_t(x) = Z_{x , t-x}$ is defined recursively as, for $t \geq 0$,
\bea \label{ZrecTime}
&& Z_t(x) = u_{t}(x) Z_{t-1} (x-1)  + v_t(x) Z_{t-1}(x)  \quad \text{ for } t \geq 1 \nn \\
&& Z_{t=0}(x) = \delta_{x, 0}  \ ,
\eea
where $\delta_{i,j}$ is the Kronecker delta symbol. Following (\ref{ZrecTime}), the length of the polymers $t$ will also be thought of as a time-like variable, (\ref{ZrecTime}) being then thought of as a Markov process. The latter is a discrete version of the stochastic-heat-equation satisfied by the partition sum of the continuum DP. 
\end{definition}

\begin{definition}\label{Def:IBbound}
{\bf The IB polymer with boundaries} We define a second version of the IB polymer by changing the BWs on the boundaries of $\mathbb{N}^2$. The random BWs are now denoted by $\hat w(e)$ and given by
\bea \label{DistuvBoundary}
&& \hat w((x_1-1, x_2) \to (x_1,x_2) ) = u_{x_1,x_2} = W_{x_1,x_2}   > 0 \quad , \quad \text{ if }  x_2 \geq 1 \ , \nn \\
&& \hat w((x_1, x_2-1) \to (x_1,x_2) ) = v_{x_1,x_2} = W_{x_1,x_2} +1  > 1 \quad , \quad \text{ if }  x_1 \geq 1    \   , \nn \\
&& \hat w((x_1-1, 0) \to (x_1,0) ) = U_{x_1,0}   \   , \nn \\
&& \hat w((0, x_2-1) \to (0,x_2) ) = V_{0,x_2}   \   . 
\eea
Here the random BWs in the bulk $u_{x_1,x_2}= W_{x_1,x_2} $ and $v_{x_1,x_2} = W_{x_1, x_2}+1$ for $x_1,x_2 \geq 1$ are distributed as before (see (\ref{Distuv})), and the BWs on the boundaries are all independent and distributed as $U_{x_1 ,0} \sim U$ and $V_{0,x_2} \sim V$ where
\bea \label{Statio-Inv-Beta2}
&& U \sim  \frac{1}{Beta(\gamma- \lambda , \beta + \lambda)} -1  >0    \quad  , \quad P_U(U) = \frac{\Gamma(\gamma+\beta)}{\Gamma(\gamma-\lambda)\Gamma(\beta + \lambda)}  \left(1-\frac{1}{U+1}\right)^{\beta +\lambda -1} \left(\frac{1}{U+1}\right)^{\gamma -\lambda +1}  \theta(U) ,   \nn \\
&& V \sim \frac{1}{Beta(\lambda , \beta)} >1 \quad,  \quad  P_V(V)= \frac{\Gamma(\lambda+\beta)}{\Gamma(\lambda)\Gamma(\beta)}   \left(1-\frac{1}{V}\right)^{\beta -1} \left(\frac{1}{V}\right)^{\lambda +1}  \theta(V-1)\ .
\eea 
Here $0 < \lambda < \gamma$ is an additional parameter and we have written explicitly the PDF $P_U(U)$ and $P_V(V)$ of $U$ and $V$ that easily follow from (\ref{DistBeta}). In the following we will refer to $\lambda$ as the boundary or stationarity parameter. We consider again the partition sum for polymers with starting point $(0,0)$ and endpoint $(x_1 \geq 0,x_2 \geq 0)$, defined as
\bea
\hat Z_{x_1,x_2} = \sum_{\pi :(0,0) \to (x_1 , x_2) } \prod_{e \in \pi} \hat w(e) \ .
\eea
\end{definition}

\begin{definition} \label{def:IBstat}
{\bf The IB polymer with stationary initial condition} We define a third version of the IB polymer. Following the recursion equation (\ref{ZrecTime}), we define the DP partition sum $\check Z_t(x)$ for $t \geq -1$ and $x\in \JZ$ as 
\bea \label{ZcheckrecTime}
&& \check Z_t(x) = u_{t}(x) \check Z_{t-1} (x-1)  + v_t(x) \check Z_{t-1}(x)  \quad \text{ for } t \geq 1 
\eea
and with the initial condition
\bea \label{IBStat:CI}
\check Z_{0}(0) = 1 \quad , \quad \frac{\check Z_0(x)}{\check Z_{-1}(x-1)} =  U(x) \quad , \quad \frac{\check Z_0(x)}{\check Z_{-1}(x)} =  V(x) \quad \text{ for } x  \in \JZ \ .
\eea
Where $(U(x))_{x\in \JZ}$ and $(V(x))_{x\in \JZ}$ are two sets of iid RVs distributed as $U(x) \sim U$ and $V(x) \sim V$ with $U,V$ distributed as (\ref{Statio-Inv-Beta2}), while the RVs $(u_t(x),v_t(x))$ are distributed as before (\ref{Distuv}). The definition of $Z_t(x)$ for $t=-1$ is for future notational convenience and for what concerns $Z_t(x)$ for $t \geq 0$ it is equivalent to set the initial condition as $\check Z_0(x+1)/\check Z_0(x)=U(x+1)/V(x)$. This model is analogous to the point to Brownian continuum DP.

\end{definition}

\subsubsection{{\bf Zero temperature models: The Bernoulli-Geometric polymer(s)}} \label{SubSecOverDefBG}

We now define as previously for the Inverse-Beta polymer three versions of the Bernoulli-Geometric (BG) polymer: the point to point BG polymer, the BG polymer with boundaries and the BG polymer with stationary initial condition. The first model will be defined by choosing {\it two} (`bulk') parameters $(q,q') \in [0,1[^2$. The others have one additional (`boundary' or `stationarity') parameter $q_b\in]q,1[$, which will specify one stationary measure among a family of stationary measures at fixed $(q,q')$. All definitions of this section are to our knowledge original to this work. Here and throughout the paper the similarities between these models and the IB polymers will be highlighted using similar notations, with the convention that we reserve sans-serif letters for the BG polymers. The connection between the IB and BG polymers, which was the main motivation for introducing the BG polymer, was already mentioned in the introduction. It will be made more precise in Sec.~\ref{subsec:T0relation}.

\begin{definition} \label{Def:ptopBG}
{\bf The point to point Bernoulli-Geometric polymer} We now define the Bernoulli-Geometric polymer. We assign to each edge $e$ of $\JZ^2$ a discrete random energy ${\cal E}(e) \in \JZ $. Depending on whether the edge is horizontal or vertical, the random energies are drawn from different probability distributions. Let us introduce the notation
\bea
&& {\cal E} \left((x_1,x_2) \to (x_1+1 ,x_2) \right) = \su_{x_1+1 , x_2}  \ , \nn \\
&& {\cal E} \left((x_1,x_2) \to (x_1 ,x_2+1) \right)  = \sv_{x_1 , x_2 +1}  \ ,
\eea
hence $\su$ (resp. $\sv$) denotes a random energy on an horizontal (resp. vertical) edge. We suppose that {\it couples} of random variables indexed by the endpoint of the edges $(\su_{x_1, x_2},\sv_{x_1, x_2})$ are iid RVs distributed as $(\su_{x_1, x_2},\sv_{x_1, x_2}) \sim (\su , \sv)$ where the couple $(\su , \sv)$ is distributed as
\bea \label{T0DistEnergyuv1}
&& \su \sim  (1- \zeta_{\su \sv}) (1+ G_{q'})  -\zeta_{\su \sv} G_{q} \in \JZ    \   ,   \nn \\
&& \sv \sim -\zeta_{\su \sv} G_{q} \in \JZ_-    \  , 
\eea
where $0<q<1$ and $0<q'<1$ are the two parameters of the models and $G_q$, $G_{q'}$ and $\zeta_{\su \sv}$ are independent RVs distributed as follows. $G_{q} \in \JN$ and $G_{q'}\in \JN$ are geometric RVs with parameters $q$ and $q'$ with the convention
\bea \label{T0probaGeo}
Proba(G_{q} = k \in \JN) = (1- q) q^k  \ ,
\eea
and similarly for $G_{q'}$ with the exchange $q \to q'$. $\zeta_{\su \sv} \in \{0,1\}$ is a Bernoulli RV with parameter $p_{\su \sv}$ given by
\bea \label{T0puv}
p_{\su \sv} = \frac{1-q'}{1-q q'} \in ] 0, 1[  \   , 
\eea
and thus
\bea \label{T0probaBernou}
Proba(\zeta_{\su \sv} = 1) =  p_{\su \sv}  \quad , \quad Proba(\zeta_{\su \sv} = 0) =  1- p_{\su \sv} \ .
\eea
As such, $\su \geq \sv$ (equality occurring whenever $\zeta_{\su \sv}=1$) and note that $\su \in \JZ$ can be positive or negative while $\sv \in  \JZ_{-}$ is always negative (or zero). $\su$ and $\sv$ are correlated RVs since they are both functions of the same Bernoulli RV $\zeta_{\su \sv}$. Note that one can also add correlations between $G_q$ and $G_{q'}$: since $\zeta_{\su \sv} \in \{0 ,1\}$ one easily shows that correlations between $G_q$ and $G_{q'}$ do not affect the PDF of $(\su , \sv)$. The latter can be written as
\bea \label{T0DistEnergyuv2}
Proba(\su = k_{\su} \in \JZ , \sv =  k_{\sv} \in \JZ_{-}) =  && p_{\su \sv} \delta(k_{\su} =k_{\sv}) \delta(k_{\sv} \leq 0 ) (1-q) q^{-k_{\sv} } \nn \\
&& +  (1- p_{\su \sv} )  \delta(k_{\sv} = 0 )  \delta(k_{\su} \geq 1 ) (1-q') (q')^{k_{\su}-1} \ ,
\eea
 where here and throughout the rest of the paper the symbol $\delta$ is used to denote the indicator function of the set specified inside the $\delta$. Finally, given a random environment specified by a drawing of the random energies $(\su_{x_1, x_2},\sv_{x_1, x_2})$, we are interested in the optimal energy to go from the origin $(0,0)$ to the point $(x_1,x_2)$
\bea
 \sE_{x_1,x_2} = {\rm \min} \left\{ {\cal E}(\pi) = \sum_{e \in \pi} {\cal E}(e) , \pi : (0,0) \to (x_1,x_2)  \right\}  \ .
\eea
Where as before the minimization is over up-right paths. Assigning for convention the value $\sE_{x_1,x_2} = + \infty$ for vertices $(x_1,x_2)$ with either $x_1<0$ or $x_2<0$, the model can also be recursively defined as, using the $(t,x)$ coordinates
\bea \label{T0rec}
&& \sE_{t}(x) = {\rm \min} \left( \sE_{t-1}(x-1) + \su_{t}(x) ,  \sE_{t-1}(x) + \sv_{t}(x) \right)  \text{ for } t \geq 1  \ ,  \nn \\
&& \sE_{t=0}(0) = 0 \quad \text{and} \quad \sE_{t=0}(x) = + \infty \quad \text{for} \quad x \neq 0 \ .
\eea
The definition of this model is, to our knowledge, original to this work. The model can be defined for any value of the parameter $p_{\su \sv} \in [0, 1]$, but it is only for the value given by (\ref{T0puv}) that we can write down exactly its stationary measure. This precise value thus makes the model special, in the sense that it possesses an ESP. In this model the parameters $q$ and $q'$ do not play symmetric roles: $q'$ can be thought of as an anisotropy parameter which favors the vertical edges by sometimes (with probability $1- p_{\su \sv}=q' \frac{1-q}{1-q q'}$) putting a penalty on horizontal edges. Two important limits are an isotropic limit of the model which is obtained by setting $q' \to 0$, and an anisotropic limit which is obtained for $q \to 0$. In the isotropic limit $p_{\su \sv} =1$ and the model corresponds to a problem of last passage percolation, while in the anisotropic limit $p_{\su \sv} =0$ and the model corresponds to a problem of first passage percolation (see Sec.~\ref{subsec:T0relation}). More generally the Bernoulli-Geometric polymer thus mixes an optimization problem of the first-passage type with an optimization problem of the last-passage type. In this interpretation $p_{\su \sv}$ is a mixing parameter which must have the precise value (\ref{T0puv}) for the model to be exactly solvable. Interesting continuous limits are also obtained by letting $q , q' \to 1$. There the model converges to the zero temperature limit of the Inverse-Beta polymer. This will be further discussed in Sec.~\ref{subsec:T0relation}.
\end{definition}

\begin{figure}
\centerline{\includegraphics[width=9cm]{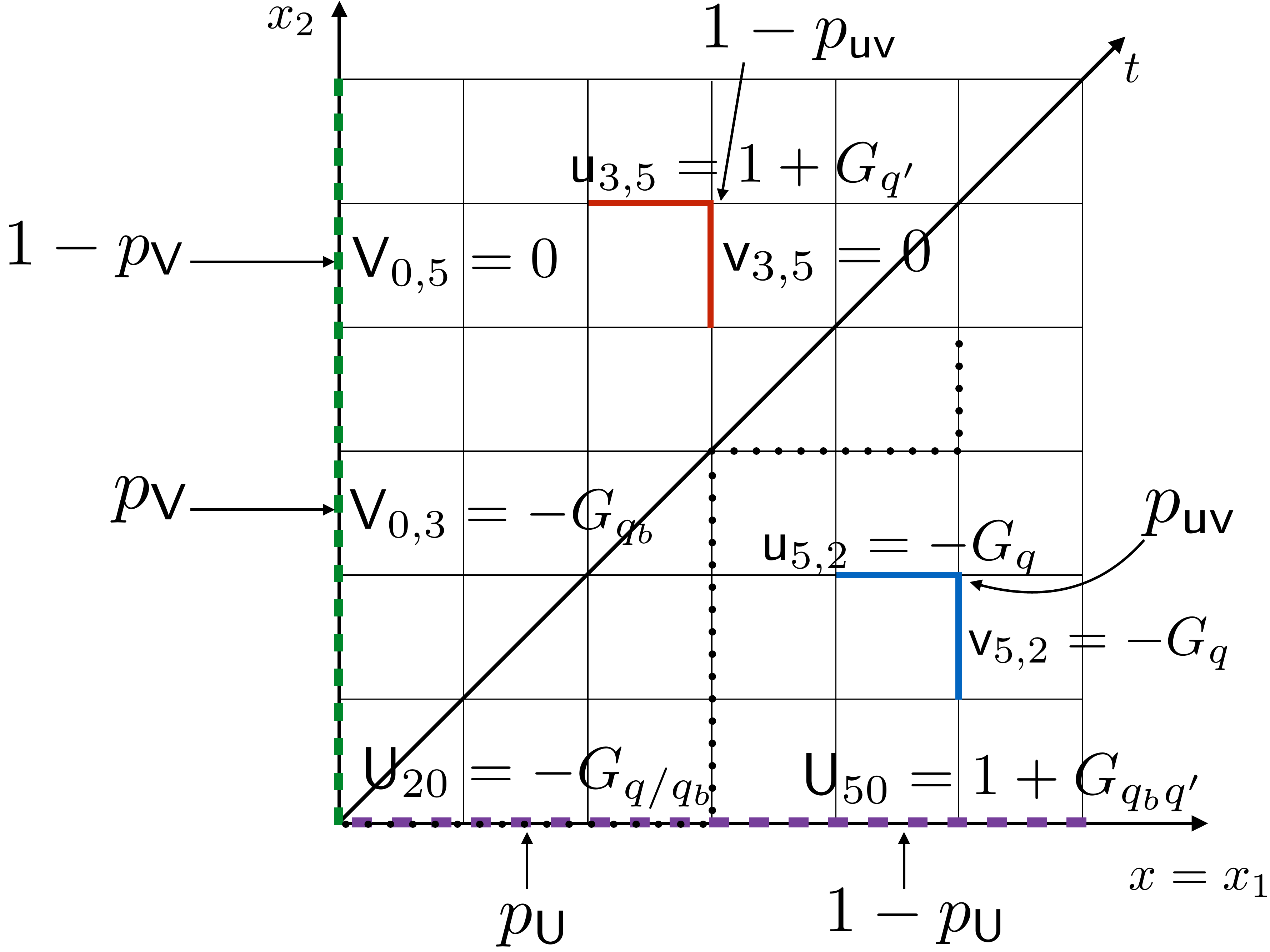}  } 
\caption{The Bernoulli-Geometric polymer with boundaries. In the bulk the couple of energies on the edges $(\su,\sv)$ are taken with probability $p_{\su \sv}$ (resp. $(1-p_{\su\sv})$) as $(\su,\sv) = (-G_q,-G_q)$ (blue edges above) (resp. $(\su,\sv) = (1+G_{q'},0)$ (red edges above)). The energies on the horizontal boundary $\sU$ (dashed-purple) are taken with probability $p_{\sU}$ (resp. $1- p_{\sU}$) as $\sU=-G_{q/q_b}$ (resp. $\sU=1+G_{q_b q'}$). The energies on the vertical boundary $\sV$ (dashed-green) are taken with probability $p_{\sV}$ (resp. $1- p_{\sV}$) as $\sV=-G_{q_b}$ (resp. $\sU=1+G_{q_b q'}$). The dotted line represents a possible polymer path from $(x_1,x_2) = (0,0)$ to $(x_1,x_2) = (5,4)$.}
\label{fig:Bernoulli-Geometric-Boundary}
\end{figure}

\begin{definition}\label{Def:BGbound}
{\bf The BG polymer with boundaries} We now consider the BG model previously defined and change the distribution of energies on the boundaries of $\mathbb{N}^2$. In the model with boundaries the energy on the edges $\hat{\cal E}(e)$ are distributed as
\bea \label{T0EnergiesBoundary}
&& \hat{\cal E} \left((x_1,x_2) \to (x_1+1 ,x_2) \right) = \su_{x_1+1 , x_2}   \text{ if }  x_2 \geq 1   , \nn \\
&& \hat{\cal E} \left((x_1,x_2) \to (x_1 ,x_2+1) \right)  = \sv_{x_1 , x_2 +1}  \text{ if }  x_1 \geq 1   \ , \nn \\
&& \hat{\cal E} \left((x_1,0) \to (x_1+1 ,0) \right) = \sU_{x_1+1 , 0}   \ , \nn \\
&& \hat{\cal E} \left((0,x_2) \to (0,x_2+1) \right)  = \sV_{0 , x_2 +1}  \ .
\eea
Where here the random energies in the bulk $(\su_{x_1,x_2} , \sv_{x_1,x_2})|_{x_1 , x_2 \geq 1} \sim(\su, \sv) $ are distributed as before with parameters $0<q<1$ and $0<q'<1$, see (\ref{T0DistEnergyuv1}) and (\ref{T0DistEnergyuv2}). The random energies on the edges of $\JN^2$, $\sU_{x_1 \geq 1, 0}$ and $\sV_{0, x_2 \geq 1}$  are independent from the random energies in the bulk and from each other. They are distributed as $\sU_{x_1 , 0}  \sim \sU$ and $\sV_{0, x_2} \sim \sV$ where
\bea \label{T0DistEnergyUV1}
&& \sU \sim  (1- \zeta_{\sU}) (1+ G_{q_b q'})  -\zeta_{\sU} G_{q/q_b} \in \JZ \ ,  \nn \\
&& \sV \sim -\zeta_{\sV} G_{q_b}  \in \JZ_- \ .
\eea
Here $q<q_b<1$ is a new parameter, $G_{q_b}$, $G_{q/q_b}$ and $G_{q_b q'}$ are independent geometric RVs distributed as in (\ref{T0probaGeo}), while $\zeta_{\sU}$ and $\zeta_{\sV}$ are Bernoulli RVs with parameter $p_{\sU}$ and $p_{\sV}$ distributed as in (\ref{T0probaBernou}) with
\bea
p_{\sU} =  \frac{1 - q_b q'}{1-qq'} \quad, \quad p_{\sV} =  \frac{1 - q'}{1-q_b q'} \ .
\eea
We can also directly write the probability distribution of $\sU$ and $\sV$ as
\bea \label{T0DistEnergyUV2}
Proba(\sU = k_{\sU} \in \JZ ) =&& p_{\sU} \delta(k_{\sU} \leq 0 ) (1-q/q_b)(q/q_b)^{-k_{\sU}}  \nn \\ 
&&+ (1 - p_{\sU} ) \delta(k_{\sU} \geq 1 ) (1-q_b q')(q_bq')^{k_{\sU} -1}      \   ,   \nn  \\
Proba(\sV = k_{\sV} \in \JZ_- ) =&& p_{\sV} \delta(k_{\sV} \leq 0 )(1-q_b)(q_b)^{-k_{\sV}} + (1 - p_{\sV} ) \delta(k_{\sV} = 0 ) \ .
\eea
Given a random environment specified by a drawing of the bulk and edges random energies we are interested in the optimal energy to go from the origin $(0,0)$ to the point $(x_1,x_2)$
\bea \label{Eq:Intro:RecBGStat}
 \hat \sE_{x_1,x_2} = {\rm \min} \left\{ \hat {\cal E}(\pi) = \sum_{e \in \pi} \hat {\cal E}(e) , \pi : (0,0) \to (x_1,x_2)  \right\}  \ .
\eea
\end{definition}

\begin{definition}\label{Def:BGstat}
{\bf The BG polymer with stationary initial condition} We define a third version of the BG polymer. Following the recursion equation (\ref{T0rec}), we define the DP optimal energy $\check \sE_t(x)$ for $t \geq -1$ and $x\in \JZ$ as 
\bea \label{EcheckrecTime}
&& \check \sE_{t}(x) = {\rm \min} \left( \check \sE_{t-1}(x-1) + \su_{t}(x) ,  \check \sE_{t-1}(x) + \sv_{t}(x) \right)  \quad \text{ for } t \geq 1 
\eea
and with the initial condition $\check \sE_t(0) = 1$ and 
\bea
 \check \sE_0(x)- \check \sE_{-1}(x-1) =  \sU(x) \quad , \quad  \check \sE_0(x)- \check \sE_{-1}(x) =  \sV(x) \quad \text{ for } x  \in \JZ \ .
\eea
Where $(\sU(x))_{x\in \JZ}$ and $(\sV(x))_{x\in \JZ}$ are two sets of iid RVs distributed as $\sU(x) \sim \sU$ and $\sV(x) \sim \sV$ with $\sU,\sV$ distributed as (\ref{T0DistEnergyUV1}), while the RVs $(\su_t(x),\sv_t(x))$ are distributed as before (\ref{T0DistEnergyuv1}).

\end{definition}

\subsection{Stationarity and reversibility properties} \label{SubSecOverStat}

In this section we now state the stationarity properties of the models previously defined. These properties will be shown rigorously in Sec.~\ref{sec:IBeta} and Sec.~\ref{sec:T0}\footnote{throughout the paper we will pay attention to emphasize the degree of rigor with which each result is shown, and in particular only fully rigorous results will be stated as Propositions}. Let us first define the notion of down-right paths.

\begin{definition} \label{Def:DRPath} 
A down-right path of length $N \in \JN^*$ on $\mathbb{Z}^2$ is as sequence of vertices of $\mathbb{Z}^2$ $(x_1(i) , x_2(i))_{i = 0 , \cdots , N}$ such that jumps are either downward: $(x_1(i+1) , x_2(i+1)) =(x_1(i) , x_2(i))  - (0,1) $, or are to the right: $(x_1(i+1) , x_2(i+1)) =(x_1(i) , x_2(i))  +(1,0) $. The set of edges crossed by the path $\pi_{dr}$ is $\{ (x_1(i) , x_2(i) ) \to (x_1(i +1) , x_2(i+1))  , i =0 , \cdots , N-1 \}$.
\end{definition}

\subsubsection{{\bf Stationarity and reversibility in the IB polymer with boundaries and stationary initial condition}} \label{SubSecOverStatIB}

Let us introduce, for $x_1,x_2 \geq 0$ and $(x_1,x_2) \neq (0,0)$, the ratios of partition sum on the horizontal and vertical edges leading to $(x_1,x_2)$ in the model with boundaries:
\bea \label{defUV}
&& \hat U_{x_1,x_2} := \frac{\hat Z_{x_1,x_2}}{ \hat Z_{x_1-1,x_2}}  \quad , \quad   \hat V_{x_1,x_2} := \frac{\hat Z_{x_1,x_2}}{ \hat Z_{x_1,x_2-1}} \ .
\eea
We will refer to these RVs as living on the edges of $\mathbb{N}^2$: $\hat U_{x_1,x_2}$ (resp. $\hat V_{x_1,x_2}$) is thought of as living on the horizontal (resp. vertical) edge leading to $(x_1,x_2)$. Note that on the boundaries these ratios coincide with the boundary weights in the IB polymer with boundaries: $\hat U_{x_1,0} = U_{x_1,0}$ and $\hat V_{0,x_2} = V_{0,x_2}$. Similarly in the model with stationary initial condition we define, for $t\geq 0$ and $x\in \JZ$:
\bea \label{defUVnew}
&& \check U_{t} (x) := \frac{\check Z_{t} ( x) }{\check Z_{t-1} ( x-1)}  \quad , \quad  \check V_{t} (x) := \frac{ \check Z_{t} ( x)}{  \check Z_{t-1} ( x)} \ .
\eea

The following four properties hold:

\begin{proposition}\label{prop:StatIBbound} 
{\bf Stationarity property of the IB polymer with boundaries}
For all down-right path on $\mathbb{N}^2$, the RVs $\hat U_{x_1,x_2}$ and $\hat V_{x_1,x_2}$ that live on the edges crossed by the down-right path are independent and distributed as $\hat U_{x_1,x_2} \sim U$ and $ \hat V_{x_1,x_2} \sim V$ with $U$ and $V$ distributed as in (\ref{Statio-Inv-Beta2}). In particular, since each edge of $\mathbb{N}^2$ belongs to at least one down-right path, the RVs $\hat U_{x_1,x_2}$ and $\hat V_{x_1,x_2}$ are all distributed as $U$ and $V$ in (\ref{Statio-Inv-Beta2}).
\end{proposition}

\begin{proposition}\label{prop:StatIBStat}
 {\bf Stationarity property of the IB polymer with stationary initial condition}
The process $(\check U_t(x) , \check V_t(x))_{t \in \JN , x \in \JZ}$ is stationary: $\forall t \in \JN$ {\it fixed}, the RVs $(\check U_t(x))_{x \in \JZ}$ and  $(\check V_t(x))_{x \in \JZ}$ are independent and distributed as $\check U_t(x) \sim U$ and $\check V_t(x) \sim V$ with $U$ and $V$ distributed as in  (\ref{Statio-Inv-Beta2}).
\end{proposition}

\begin{proposition}
{\bf Reversibility of the stationary process}\label{prop:RevIB}
Considering a finite time interval of duration $T \in \JN^*$ and the time-reversed coordinates and time reversed process variables defined by
\bea \label{time-reversed-coord}
&& t_R = T-t  -1 \quad , \quad x_R = -x   \\
&& \check U_{t_R}^R(x_R ) = \check U_{t = T -t_R} (x = -x_R +1) \quad , \quad \check V_{t_R}^R(x_R ) =\check V_{t = T -t_R} (x = -x_R)  \label{time-reversed-coord2} \ ,
\eea
we have the identity in law
\bea \label{reversibilityUV}
\left(\check U_t (x) , \check V_t(x) \right)_{t = 0, \dots , T ; x \in \mathbb{Z}} \sim \left(\check U^R_{t_R} (x_R) , \check V^R_{t_R}(x_R) \right)_{t_R = 0, \dots , T ; x_R \in \mathbb{Z}}  \ .
\eea
\end{proposition}

\begin{proposition}
{\bf Equivalence between models with boundaries and stationary initial condition} \label{prop:IBEquiv} We have
\bea 
(\check Z_t(x))_{(t,x) \in \JN^2} \sim (\hat Z_t(x))_{(t,x) \in \JN^2} \ .
\eea
\end{proposition}
The model with boundary conditions can thus be seen as an efficient way to study the model with stationary initial condition in the upper-right quadrant of $\mathbb{Z}^2$.

\subsubsection{{\bf Stationarity and reversibility in the BG polymer with boundaries and stationary initial condition}} \label{SubSecOverStatBG}

Conversely, let us introduce in the BG polymer with boundaries, for $x_1,x_2 \geq 0$ and $(x_1,x_2) \neq (0,0)$, the differences of optimal energies on the horizontal and vertical edges leading to $(x_1,x_2)$:
\bea \label{defUVT0}
&& \hat \sU_{x_1,x_2} :=  \hat \sE_{x_1,x_2} - \hat \sE_{x_1-1,x_2} \quad , \quad  \hat \sV_{x_1,x_2} :=  \hat \sE_{x_1,x_2} -  \hat \sE_{x_1,x_2-1}  \ .
\eea
And similarly, in the model with stationary initial condition, for $t\geq 0$ and $x\in \JZ$:
\bea \label{defUVnewT0}
&& \check \sU_{t} (x) = \check \sE_{t}(x)-\check \sE_{t-1}(x-1)  \quad , \quad  \check \sV_{t} (x) = \check \sE_{t}(x)-\check \sE_{t-1}(x) \ .
\eea

The following four properties hold:

\begin{proposition}
{\bf Stationarity property of the BG polymer with boundaries} \label{prop:StatBGbound} 
For all down-right path on $\mathbb{N}^2$, the RVs $\hat \sU_{x_1,x_2}$ and $\hat \sV_{x_1,x_2}$ that live on the edges crossed by the down-right path are independent and distributed as $\hat \sU_{x_1,x_2} \sim \sU$ and $ \hat V_{x_1,x_2} \sim \sV$ with $\sU$ and $\sV$ distributed as in (\ref{T0DistEnergyUV1}). In particular, since each edge of $\mathbb{N}^2$ belongs to at least one down-right path, the RVs $\hat \sU_{x_1,x_2}$ and $\hat \sV_{x_1,x_2}$ are all distributed as $\sU$ and $\sV$ in (\ref{T0DistEnergyUV1}).
\end{proposition}

\begin{proposition}
{\bf Stationarity property of the model with stationary initial condition}\label{prop:StatBGStat}
The process $(\check \sU_t(x) , \check \sV_t(x))_{t \in \JN , x \in \JZ}$ is stationary: $\forall t \in \JN$ {\it fixed}, the RVs $(\check \sU_t(x))_{x \in \JZ}$ and  $(\check \sV_t(x))_{x \in \JZ}$ are independent and distributed as $\check \sU_t(x) \sim \sU$ and $\check \sV_t(x) \sim \sV$ with $\sU$ and $\sV$ distributed as in  (\ref{T0DistEnergyUV1}).
\end{proposition}

\begin{proposition}
{\bf Reversibility of the stationary process}\label{prop:RevBG}
Considering a finite time interval of duration $T \in \JN^*$ and the time-reversed coordinates (\ref{time-reversed-coord}), the time reversed process is defined as
\bea \label{TRprocessT0}
\check \sU_{t_R}^R(x_R ) = \check \sU_{t = T -t_R} (x = -x_R +1) \quad , \quad \check \sV_{t_R}^R(x_R ) =\check \sV_{t = T -t_R} (x = -x_R)  \ ,
\eea
and we have the identity in law
\bea \label{T0reversibilityUV}
\left(\check \sU_t (x) , \check \sV_t(x) \right)_{t = 0, \dots , T ; x \in \mathbb{Z}} \sim \left(\check \sU^R_{t_R} (x_R) , \check \sV^R_{t_R}(x_R) \right)_{t_R = 0, \dots , T ; x_R \in \mathbb{Z}}  \ .
\eea
\end{proposition}

\begin{proposition}
{\bf Equivalence between models with boundaries and stationary initial condition}\label{prop:BGEquiv} We have
\bea
(\check \sE_t(x))_{(t,x) \in \JN^2} \sim (\hat \sE_t(x))_{(t,x) \in \JN^2} \ .
\eea
\end{proposition}

\subsection{Quenched free-energy in point to point models without boundaries} \label{SubSecOverFE}
Using the stationary properties stated above, we obtain in Sec.\ref{subsec:proof} asymptotic results for the mean quenched free-energy/optimal energy in the direction $(s_1,s_2)\in \JR_+^2$ in the point to point IB/BG polymer. These quantities are defined as
\bea \label{DefIntroFE}
f_{{\rm IB}}(s_1,s_2):= \lim_{N \to \infty} \frac{ - \overline{\log Z_{Ns_1, Ns_2}}}{N} \quad , \quad \sff_{{\rm BG}}(s_1,s_2) := \lim_{N \to \infty} \frac{\overline{\sE_{Ns_1,Ns_2}}}{N} \ .
\eea
Where here and throughout the paper the overline $\overline{()}$ denotes the average over the random environment. For a fixed direction $(s_1,s_2) \in \JR_+^2$ and bulk parameters ($\gamma,\beta)$/$(q,q')$, our results involves the solution of a saddle-point equation for a boundary parameter $\lambda=\lambda^*(s_1,s_2)$/$q_b = q_b^*(s_1,s_2)$.  Based on some unproven `natural' assumptions of convexity and regularity for $f_{{\rm IB}}(s_1,s_2)$ and $\sff_{{\rm BG}}(s_1,s_2)$ we obtain:

\bea \label{FEresultBIS}
 && f_{{\rm IB}}(s_1,s_2) = s_1 \left( -\psi (\beta +\lambda^* )+ \psi (\gamma -\lambda^* ) \right)  + s_2 \left(  -\psi (\beta +\lambda^* )+\psi (\lambda^* ) \right)    \   ,    \\
 && 0= s_1 \left( -\psi'(\beta +\lambda^* )- \psi'(\gamma -\lambda^* ) \right)  + s_2 \left(  -\psi'(\beta +\lambda^* )+\psi'(\lambda^* ) \right)   \   ,    \label{FEresultBISii}
\eea
where $\psi = \Gamma'/\Gamma$ is the diGamma function and with the condition $0<\lambda^*(s_1,s_2)<\gamma$ for the IB polymer, and
\bea \label{T0decomp11BIS}
&& \sff_{{\rm BG}}(s_1,s_2) = -\frac{  \left(q-\left(q_b^*\right)^2 q'\right)}{\left(q-q_b^*\right) \left(q_b^* q'-1\right)} s_1+\frac{q_b^*  \left(q'-1\right)}{\left(q_b^*-1\right) \left(q_b^* q'-1\right)} s_2   \   ,      \\
&& \left(q_b^*-1\right)^2 \left( q  \left(q_b^* \right)^2 \left(q'\right)^2+\left(q^2-4 q_b^* q+\left(q_b^*\right)^2\right) q'+q \right)  s_1 -\left(q-q_b^*\right)^2 \left(q'-1\right) \left(\left(q_b^*\right)^2 q'-1\right) s_2 = 0      \    ,  \label{T0decomp11BISii}
\eea
with the condition $q<q_b^*(s_1,s_2)<1$ for the BG polymer.

 Note that while (\ref{FEresultBISii}) is a transcendental equation for $\lambda^*(s_1,s_2)$, (\ref{T0decomp11BISii}) is a quartic equation for $q_b^*(s_1,s_2)$, which can be solved explicitly using radicals. These results cannot be considered as mathematical theorems since they rely on unproven assumptions (which could likely be proven by other means). Still, their derivation is very close to a mathematical proof. We note that the result (\ref{FEresultBIS}) for $f_{{\rm IB}}(s_1,s_2) $ coincides with the result obtained in Eq.(79)-(81) in \cite{usIBeta}\footnote{there $c_\varphi = f_{{\rm IB}}(1/2 + \varphi , 1/2 - \varphi)$ for $\varphi \in ]-1/2,1/2[$ and the equivalent of $\lambda^*$ there is the saddle-point parameter $k_{\varphi} = \gamma/2 + \lambda^*$} using non-rigorous replica calculations, and the above result thus gives a close to rigorous confirmation of one conjecture of \cite{usIBeta}.

{\bf Optimal angles} \\
Of interest are the {\it optimal angles} $\varphi_{{\rm opt}} \in ]-1/2,1/2[$, the `angles' for which the mean quenched free-energy/optimal energy {\it per unit length} in the direction  $\varphi \in ]-1/2,1/2[$, defined as
\bea
f^{{\rm p.u.l.}}_{{\rm IB}}(\varphi) = f_{{\rm IB}}(1/2+\varphi , 1/2- \varphi) \quad , \quad \sff^{{\rm p.u.l.}}_{{\rm BG}}\varphi) = \sff_{{\rm IB}}(1/2+\varphi , 1/2- \varphi)  \ ,
\eea
are maximum. These quantities are non-trivial in these anisotropic models and we obtain the explicit formulas
\bea \label{EqOverviewOptimalAngles}
\varphi_{{\rm opt}}^{{\rm IB}} = -\frac{1}{2} \frac{\psi'(\beta + \gamma/2)}{\psi'(\gamma/2)} \leq 0 \quad , \quad \varphi_{{\rm opt}}^{{\rm BG}}=-\frac{\left(\sqrt{q}-1\right)^2 q'}{2 \left(\sqrt{q} q'-1\right)^2}  \leq 0 \ .
\eea
These angles would correspond to the mean direction chosen by the polymer for a point to line polymer problem. The formula for $\varphi_{{\rm opt}}^{{\rm IB}}$ was already given in Eq.(83) of \cite{usIBeta}.

\subsection{Convergence of point to point models to their stationary state} \label{SubSecOverConv}

Finally, based on the upon results, we {\it conjecture} that the following limits in law holds: $\forall (L_u , L_v) \in \JN^2$ and $(s_1,s_2) \in \JR_+^2$
\bea \label{OverviewConvIB}
&& \lim_{N \to \infty}  \left( \frac{Z_{Ns_1 +x_1 , Ns_2+ x_2} }{Z_{Ns_1, Ns_2} } \right)_{0\leq x_1 \leq L_u,0\leq x_2 \leq L_v} \sim \left( \hat Z_{x_1,x_2} \right)_{0\leq x_1 \leq L_u,0\leq x_2 \leq L_v}  \\
&&  \lim_{N \to \infty}  \left( \sE_{Ns_1 +x_1 , Ns_2+ x_2} - \sE_{Ns_1 , Ns_2} \right)_{0\leq x_1 \leq L_u,0\leq x_2 \leq L_v} \sim \left( \hat \sE_{x_1,x_2} \right)_{0\leq x_1 \leq L_u,0\leq x_2 \leq L_v} \label{OverviewConvBG}
\eea  
where the left hand sides of these limits involve the point to point partition sum/optimal energy in the IB/BG polymer, and the right hand sides involve the corresponding quantities in the models with boundaries with boundary parameters $\lambda = \lambda^*(s_1,s_2)$ and $q_b = q_b^*(s_1,s_2)$, the solutions of the equations (\ref{FEresultBISii}) and (\ref{T0decomp11BISii}).

\subsection{Outline and some additional results not presented here}

The outline of the remaining of this manuscript is as follows. In Sec.~\ref{sec:IBeta} and \ref{sec:T0} we prove the stationarity and reversibility properties of the Inverse-Beta and Bernoulli-Geometric polymers of Sec.~\ref{SubSecOverStat}, and discuss the connections between our work and previous works. In Sec.~\ref{subsec:FEINMODELWITHBOUND} we obtain results for the asymptotic mean quenched free-energy and mean optimal energy in the IB and BG polymers with boundaries, and using these results we obtain in Sec.~\ref{subsec:proof} the corresponding formulas (\ref{FEresultBIS})-(\ref{T0decomp11BIS}) for the point to point models. In Sec.~\ref{subsec:EquivaAngBou} we discuss the conjectures for the convergence of both models to their stationary measure (\ref{OverviewConvIB})-(\ref{OverviewConvBG}). In Sec.~\ref{subsec:Fluctuations} we briefly discuss the nature of the fluctuations of the free-energy in the models with boundaries. Finally in Sec.~\ref{Sec:Num} we perform some simulations of the BG polymer and check our result and conjecture (\ref{T0decomp11BIS}) and (\ref{OverviewConvBG}) for this newly introduced model.

\section{Finite-temperature model: stationary measure of the Inverse-Beta polymer} \label{sec:IBeta}

In this section we show the stationarity properties of the IB polymer of Sec.~\ref{SubSecOverStatIB}. We follow closely the approach developed by Sepp\"{a}l\"{a}inen for the case of the Log-Gamma polymer \cite{Seppalainen2012} and adapt it to the Inverse-Beta polymer. We also discuss the connection between out work and previous works.

\subsection{Stationary property of the model with boundaries} 

We begin by showing the stationarity property Prop.~\ref{prop:StatIBbound} of the IB model with boundaries (see Def.~\ref{Def:IBbound}). First, note that in the bulk of $\JN^2$, i.e. for $x_1,x_2 \geq 1$, the partition sum $\hat Z_{x_1 , x_2}$ satisfies the bulk recursion
\bea \label{Statio-Inv-Beta1}
\hat Z_{x_1 , x_2} = u_{x_1,x_2} \hat Z_{x_1-1, x_2} + v_{x_1,x_2} \hat Z_{x_1 , x_2-1} \quad \text{for} \hspace{0.1cm} x_1,x_2 \geq 1  \ .
\eea
This implies that the vertical and horizontal ratios of partition sums $\hat U_{x_1,x_2}$ and $\hat V_{x_1,x_2}$ defined in (\ref{defUV}) satisfy the following recursion relation, valid for $x_1,x_2 \geq 1$,
\bea \label{inductionUV}
&& \hat U_{x_1,x_2} = \phi^{(1)} ( \hat U_{x_1,x_2-1} , \hat V_{x_1-1,x_2} , W_{x_1,x_2} )    \quad , \quad \hat V_{x_1,x_2} = \phi^{(2)} ( \hat U_{x_1,x_2-1} , \hat V_{x_1-1,x_2} , W_{x_1,x_2} )  \  ,
\eea 
where $\phi^{(i)}$ denotes the $i^{th}$ component of the image of the {\it stationarity-reversibility map} $\phi$ that we now define.

\begin{definition} \label{stationarityMapDef}
The stationarity-reversibility map is the function $\phi: (U,V,W) \in (\JR^*)^3 \to (U',V',W') \in (\JR^*)^3$ defined by
\bea \label{stationarityMap}
U' = W  + (W+1) \frac{U}{V}  \quad , \quad V' = W \frac{V}{U} + W +1   \quad , \quad  W' = \frac{U(V-1) }{U+V}   \   .   \ 
\eea
\end{definition}
It has the following properties:

\begin{proposition} \label{Prop:stationarityMap:Stationarity}
{\bf Stationarity}
If $(U,V,W)$ are three independent RVs distributed as in (\ref{Statio-Inv-Beta2}) and (\ref{Statio-Inv-Beta3}), then $(U',V',W'):=\phi(U,V,W)$ are three independent RVs distributed as in (\ref{Statio-Inv-Beta2}) and (\ref{Statio-Inv-Beta3}). 
\end{proposition}

\begin{proposition} \label{Prop:stationarityMap:Reversibility}
{\bf Reversibility} $\phi$ is an involution, i.e. $\phi \circ \phi = Id$. 
\end{proposition}

\begin{figure}
\centerline{\includegraphics[width=9cm]{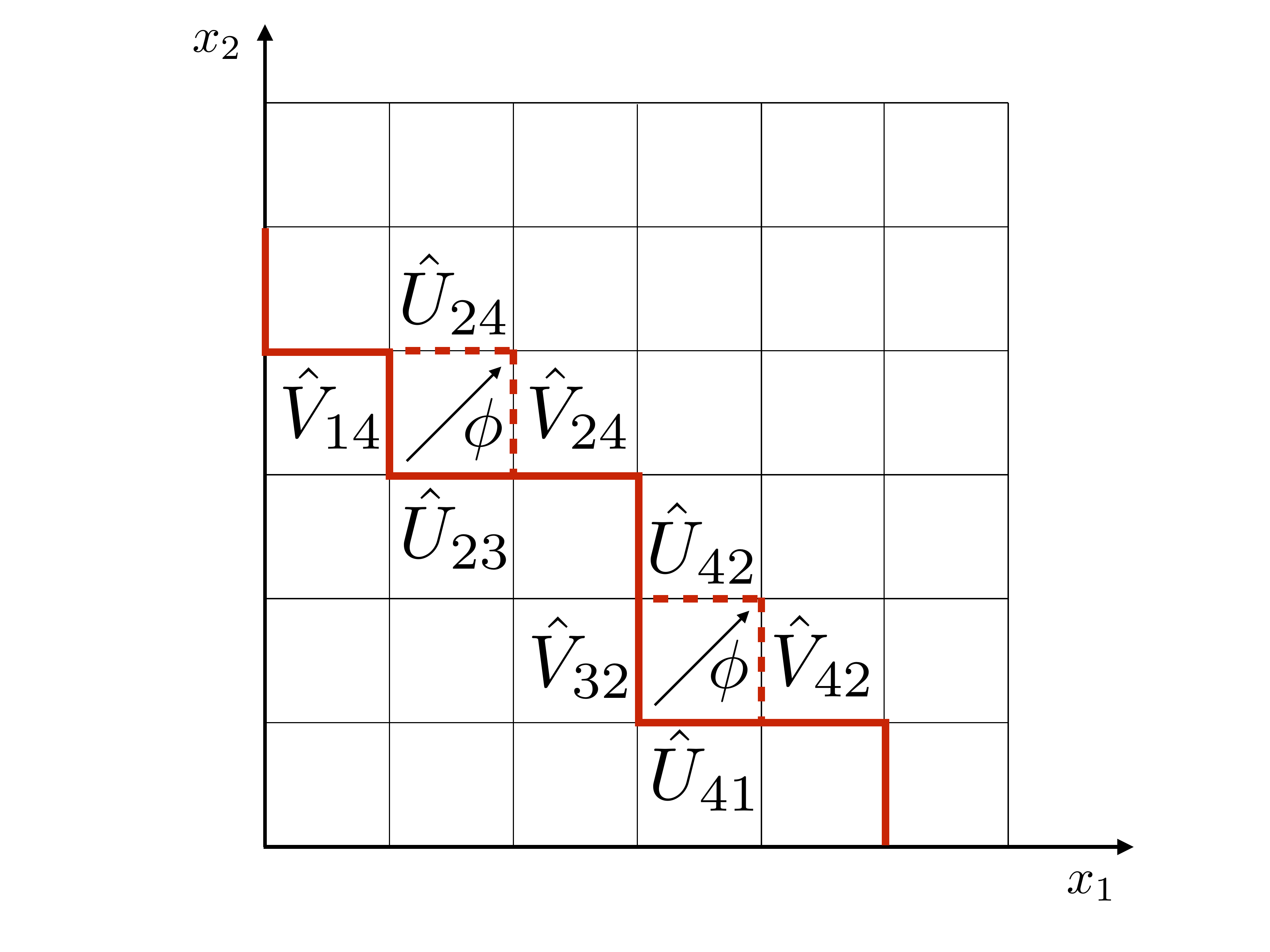}} 
\caption{Red: a down-right path on $\mathbb{N}^2$. Red-dashed: a new possible down-right path obtained from the first one by transforming two down-left corners into two top-rights ones (resulting from two down-left to top-right transformations, see Def.~\ref{Def:DLTR}). The RVs $\hat U_{x_1 , x_2} $ and $\hat V_{x_1,x_2}$ on the new down-right path are either the same as for the first path, or obtained from those on the first path using the stationarity-reversibility map $\phi$ through the induction (\ref{inductionUV}).}
\label{fig:drpathN2}
\end{figure}

These properties are proved in Appendix.~\ref{app:Stationarity}. Based on the above properties of $\phi$, the stationarity property of the model with boundary conditions Prop.~\ref{prop:StatIBbound} is proved by induction on the set of down-right paths (see Def.~\ref{Def:DRPath}) on $\JN^2$. We first need a definition: 

\begin{definition}\label{Def:DLTR}
{\bf Down-left to top-right transformation on down-right paths} 
A down-right path $\pi_{dr2}$ is a `down-left to top-right' (henceforth: DLTR) transformed down-right path of a down-right path $\pi_{dr2}$ if $\pi_{dr2}$ can be obtained from $\pi_{dr1}$ by a transformation where edges of $\pi_{dr1}$ forming a down-left corner, i.e. of the form $(x_1,x_2) \to (x_1 , x_2 -1) \to (x+1 , x_2-1)$), are replaced in $\pi_{dr2}$ by the two edges forming the corresponding top-right corner $(x_1,x_2) \to (x_1+1 , x_2) \to (x_1+1 , x_2-1) $ (see Fig.~\ref{fig:drpathN2}).
\end{definition}

Let us now give the proof of the stationarity property Prop.~\ref{prop:StatIBbound} of the model with boundary conditions. First, note that the stationarity property is trivially true for the down-right paths that follow exactly the boundaries of $\mathbb{N}^2$ (since on these down-right paths the RVs $\hat U_{x_1,x_2}$ and $\hat V_{x_1,x_2}$ are just the random Boltzmann weights on the boundaries $U_{x_1,0}$ and $V_{0,x2}$ which are independent and distributed as (\ref{Statio-Inv-Beta2})). Let us now suppose that the stationarity property Prop.~\ref{prop:StatIBbound} is true for a down right path $\pi_{dr1}$ such that $\pi_{dr1}$ contains one couple of edges of the form $(x_1,x_2) \to (x_1 , x_2 -1) \to (x+1 , x_2-1)$ (i.e. it contains two edges forming a down-left corner, see Fig.~\ref{fig:drpathN2}). The vertical edge then carries the RV $\hat V_{x_1,x_2}$ and the horizontal edge carries the RV $\hat U_{x_1+1 , x_2-1}$. Applying the induction (\ref{inductionUV}) on this couple of edges, we obtain the couple of RVs $(\hat U_{x_1+1,x_2} , \hat V_{x_1+1,x_2})$. These RVs, complemented by the other RVs $\hat U_{x_1',x_2'}$ and $\hat V_{x_1',x_2'}$ that live on $\pi_{dr1}$ and were left untouched by this induction, now live on a down right path $\pi_{dr2}$ defined such that the edges visited by $\pi_{dr2}$ are exactly those visited by $\pi_{dr1}$ except for the couple of edges $(x_1,x_2) \to (x_1 , x_2 -1) \to (x+1 , x_2-1)$ that is replaced by $(x_1,x_2) \to (x_1+1 , x_2) \to (x_1+1 , x_2-1) $ (see Fig.~\ref{fig:drpathN2}). Using the stationarity property of $\phi$  Prop.~\ref{Prop:stationarityMap:Stationarity} one concludes that those RVs satisfy the stationarity property Prop.~\ref{prop:StatIBbound}. Hence the DLTR transformation on down-right paths conserves the stationarity property, and we will generally think of the variables $(U,V)$ (resp. $(U',V')$) in (\ref{stationarityMap}) as living on down-left (resp. top right) corners (see Fig.~\ref{fig:drpathN2}). Finally, since any down-right path on $ \mathbb{N}^2$ can be obtained from a down-right path that follows exactly the edges of $\mathbb{N}^2$ by a sequence of DLTR transformations, the stationarity property holds for any down-right path on $\JN^2$. In this sense, the stationarity property of the model with boundaries can be thought of as a {\it propagation of boundary conditions}.

\subsection{Stationarity property of the model with stationary initial condition} \label{subsec:IBetaSM}

\begin{figure}
\centerline{\includegraphics[width=8.5cm]{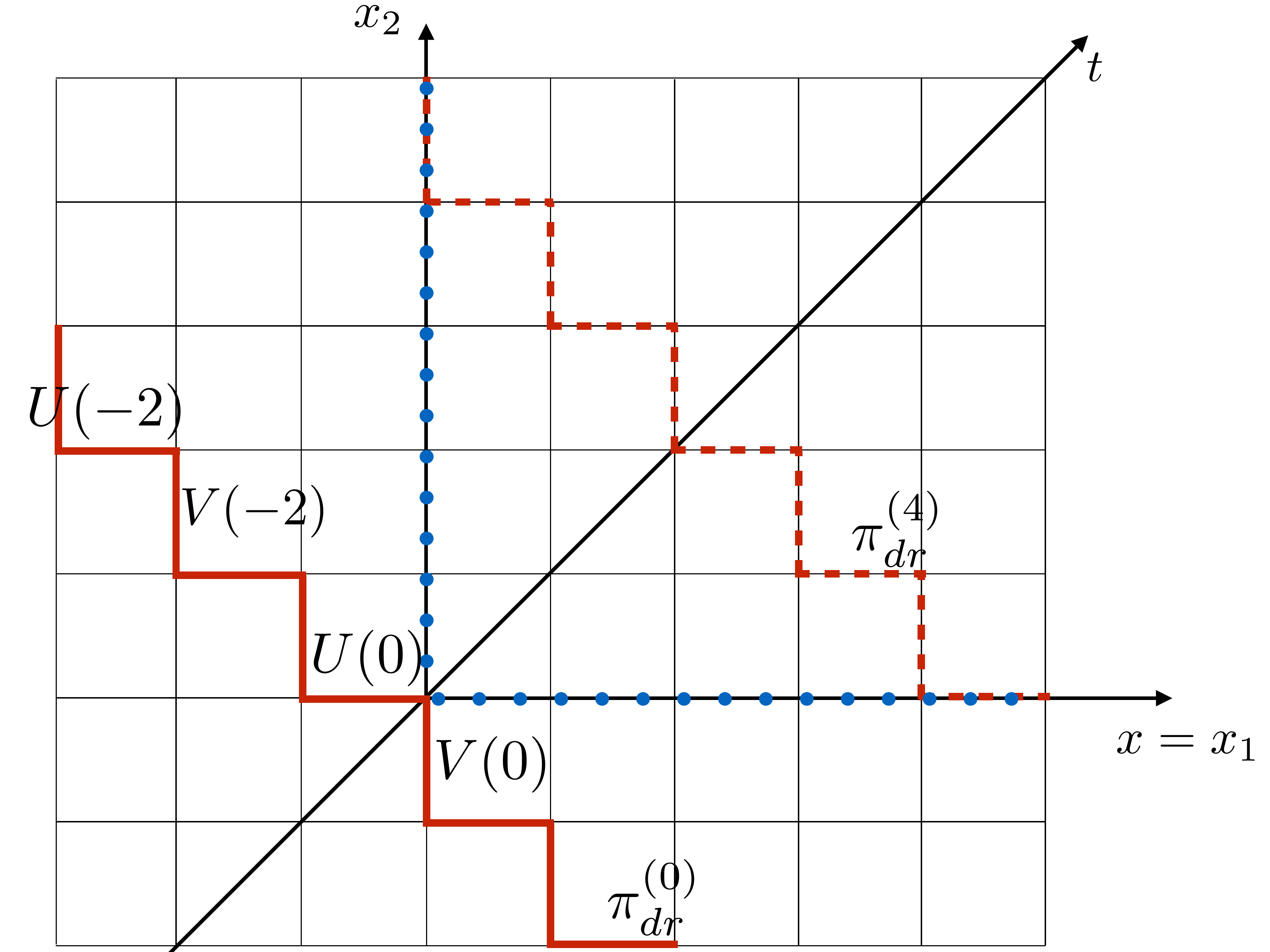}  \includegraphics[width=8.5cm]{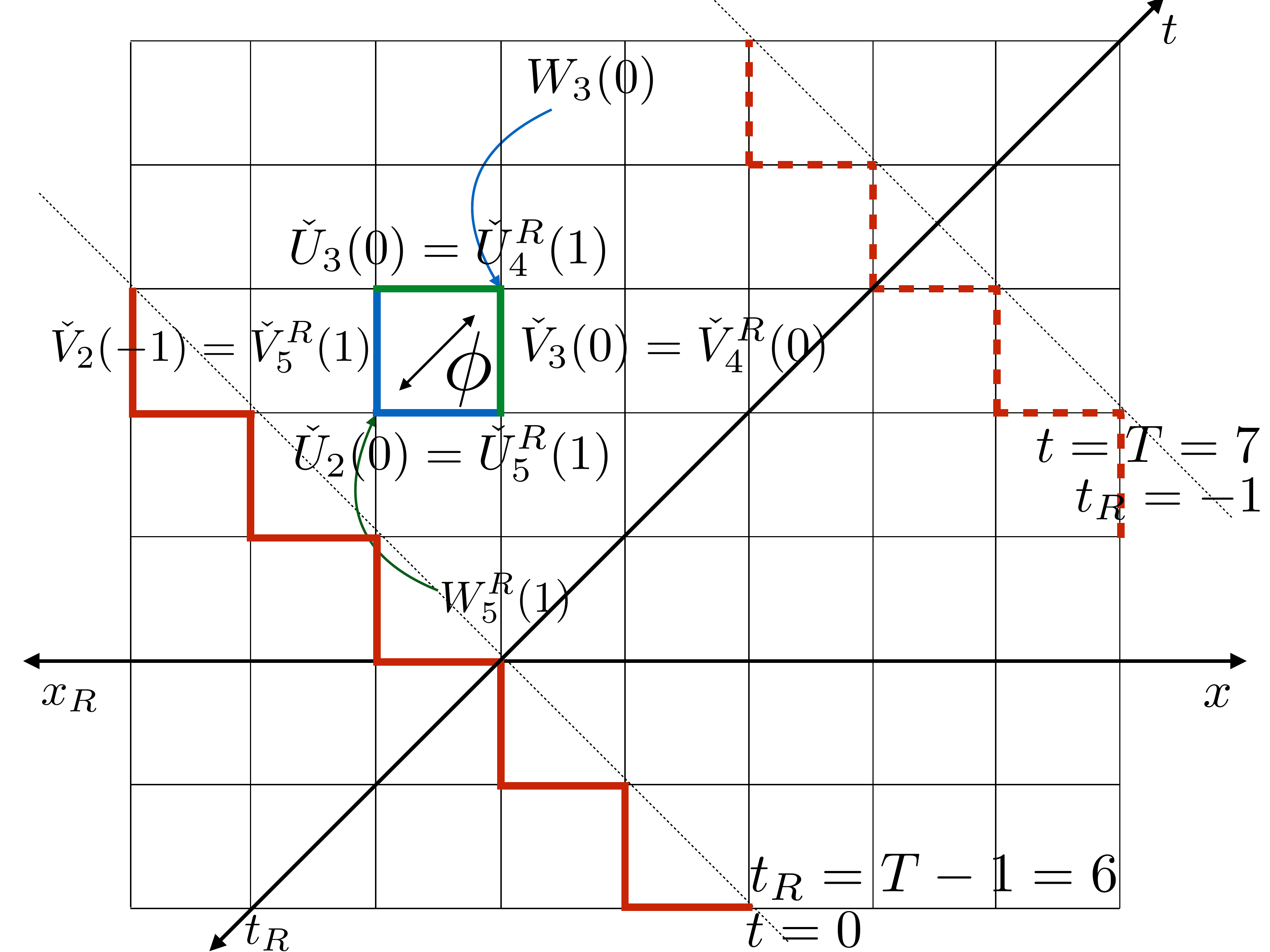}} 
\caption{Left: Stationary measure of the Inverse-Beta polymer with stationary initial condition. The initial down right path $\pi_{dr}^{(0)}$ on which the initial condition is defined carries RVs $\check U_{t=0}(x) = U(x) $ and $\check V_{t=0}(x) = V(x) $ for which the stationarity property holds. Any down-right paths obtained from $\pi_{dr}^{(0)}$ by down-left to top-right transformations then carries RVs $\check U_{t}(x)$ and $\check V_{t}(x)$ such that the stationarity property holds. These include e.g. all down-right paths $\pi_{dr}^{(t)}$ for $t \geq 0$ (such as $\pi_{dr}^{(4)}$ in dashed-red above) and the boundaries of $\mathbb{N}^2$ (in dotted blue above). Right: Illustration of the symmetry between the forward and time-reversed process for $T=7$. In the time evolution of the forward process, the RVs $\check U_2(0)$ and $\check V_2(-1)$ (on the blue edges above) are transformed using $\phi$ into the RVs $\check U_3(0)$ and $\check V_3(0)$ (on the green edges above) using the random Boltzmann weight $W_3(0)$. From this evolution one stores, using $\phi$ the random Boltzmann weights $W_5^R(1)$ later used in the time evolution of the backward process where the RVs $\check U_4^R(1)$ and $\check V_4^R(0)$ (on the green edges above) are evolved using $\phi$ into the RVs $\check U_5^R(1)$ and $\check V_5^R(1)$.}
\label{fig:drpathZ2}
\end{figure}

We now consider the IB with stationary initial condition defined in Def.~\ref{def:IBstat}. Similarly as before, the horizontal and vertical ratios of partition sums $\check U_t(x)$ and $\check V_t(x)$ (defined in (\ref{defUVnew})) satisfy the following recursion equation, valid for $t\geq 0$ and $x \in \JZ$
\bea \label{inductionUVnew}
\check U_{t+1}(x)  =  \phi^{(1)}(\check U_{t}(x) ,\check V_{t}(x-1)   , W_{t+1}(x))   \quad , \quad\check V_{t+1}(x)  =  \phi^{(2)}(\check U_{t}(x) ,\check V_{t}(x-1)   , W_{t+1}(x))     \    .  
\eea
In this model, the stationary initial condition (\ref{IBStat:CI}) is designed to provide an initial down right path on $\mathbb{Z}^2$,  
\bea
\pi_{dr}^{(0)} = \{ (x_1 ,x_2) = (m , -m) \to (m , -m-1)  \to (m+1 , -m-1)         , m \in \mathbb{Z} \}
\eea
on which the variable $\check U_{t}(x)$ and $\check V_{t}(x)$ defined in (\ref{defUVnew}) are all independent and distributed as in (\ref{Statio-Inv-Beta2}), (see Fig.~\ref{fig:drpathZ2}). Starting from this initial down-right path and successively applying DLTR transformations as described previously, one obtains the following improved (compared to Prop.~\ref{prop:StatIBStat}) stationarity property.

\begin{proposition}
{\bf Improved Stationarity property of the model with stationary initial condition} On each down-right path $\pi_{dr}$ on $\mathbb{Z}^2$ that can be obtained from $\pi_{dr}^{(0)}$ by a sequence of DLTR transformations, the variables $\check U_{t}(x)$ and $\check V_{t}(x)$ that live on $\pi_{dr}$ are independent and distributed as in (\ref{Statio-Inv-Beta2}).
\end{proposition}

For concreteness let us highlight some down-right paths that can be obtained from $\pi_{dr}^{(0)}$ using DLTR transformations (see left of Fig.~\ref{fig:drpathZ2}) and prove the properties Prop.~\ref{prop:StatIBStat} and Prop.~\ref{prop:IBEquiv}. These includes 

1) For all $t\geq 1$ the down-right path
\bea
\pi_{dr}^{(t)} = \{ (x_1 ,x_2) = (t+m , -m) \to (t+m , -m-1)  \to (t+m+1 , -m-1)         , m \in \mathbb{Z} \}
\eea 
In particular this implies the stationarity property Prop.~\ref{prop:StatIBStat}. Note that this shows that in the model with stationary initial condition, for all $t\geq 1$, the RVs $\{ \check Y_t(x) := \check Z_{t}(x+1)/\check Z_{t}(x)  , x \in \mathbb{Z} \}$ are iid and distributed as $U/V$ with $U$ and $V$ distributed as in (\ref{Statio-Inv-Beta2}). This stationarity property for the $\check Y_t(x)$ variables has the advantage of only involving partition sums at the same time coordinate $t$. It is trivially implied by the stronger property of stationarity of $\check U_{t}(x)$ and $\check V_{t}(x)$ on down-right paths and we will focus on the latter in the following.

 2) The boundary of $\mathbb{N}^2$, which is itself a down-right path, can also be obtained from $\pi_{dr}^{(0)}$. This shows that on the boundaries of $\JN^2$, the partition sums in the IB polymer with stationary initial condition $\check Z_{x_1,x_2}$ and in the model with boundary conditions $\hat Z_{x_1,x_2}$ are equivalent in law. Since the partition sums in these models in the remaining of $\JN^2$ are uniquely determined by their values on the boundaries and by the random BWs in the bulk of $\JN^2$, which coincide in both models, we obtain Prop.~\ref{prop:IBEquiv}, i.e. $(\check Z_{x_1,x_2})_{(x_1,x_2) \in \JN^2} \sim (\hat Z_{x_1,x_2})_{(x_1,x_2)\in \JN^2}$.

\medskip

{\it Remarks}
\begin{itemize}
\item{The condition $\check Z_0(0) = 1$ in the initial condition (\ref{IBStat:CI}) is arbitrary and could be replaced by any other constant or RV as long as it is independent of the variable $U(x)$ and $V(x)$. The equality in law between the model with stationary initial condition in the upper right quadrant and the model with boundary conditions then more generally reads $(\check Z_{x_1,x_2}/\check Z_{0,0})_{(x_1,x_2) \in \JN^2} \sim (\hat Z_{x_1,x_2})_{(x_1,x_2)\in \JN^2}$.}
\item{Here we have thus obtained a family (indexed by $\lambda$) of stationary measures for the Inverse-Beta polymer. These correspond to discrete random walks at fixed $t$ as a function of $x$ for the free energy $-\log \check Z_t(x)$. We will see in the following that these random walks have generally a non-zero drift, except in the `equilibrium case' $\lambda = \gamma/2$. This discrete stationary measure is thus a natural generalization of the stationary measure of the continuum DP, or equivalently of the $1$ dimensional KPZ equation \cite{HuseHenleyFisher,ForsterNelsonStephen}. Note also that as in the continuum case, the stationary measure only concerns quotients of partition sums/differences of free-energies and the one-point distribution of $\check Z_t(0)$ is not stationary. Hence the full process $(\check Z_t(x) )_{t = 0, \dots , T ; x \in \mathbb{Z}}$, which can be equivalently parametrized by the couple $\left(\check Z_t (x=0) , ( \check U_t(x) ,  \check V_t(x)  )_{t = 0, \dots , T ; x \in \mathbb{Z} }\right)$ is not stationary, but the process we are studying however $ ( \check U_t(x) ,  \check V_t(x)  )_{t = 0, \dots , T ; x \in \mathbb{Z}} $, is a marginal of the latter and is stationary.}
\end{itemize}

\subsection{Reversibility of the stationary measure: detailed balance property} \label{subsec:IBetaDB}

We now discuss the reversibility of the stationary process. We first study reversibility at the level of a DLTR transformation on down-right paths, then at the level of the process $(\check U_t(x) , \check V_t(x))_{x \in \JZ}$ and prove the property Prop.~\ref{prop:RevIB}.

\subsubsection{{\bf At the level of a down-left to top-right transformation}}

We now show a detailed-balance property for the stationary measure, namely that, if $(U,V,W)$ are distributed as in (\ref{Statio-Inv-Beta2}) and (\ref{Statio-Inv-Beta3}) and $(U',V',W') = \phi(U,V,W)$, then the PDF of the couples of couples of RVS $P((U',V') ; (U,V) )$ is symmetric by exchange $(U,V)  \leftrightarrow (U',V')$. Indeed, let us consider $((U',V') ; (U,V) )$ fixed and note $P_{stat}(U,V,W) = P_U(U) P_V(V) P_W(W)$ the stationary PDF of the triplet of RVs in (\ref{stationarityMap}) (see (\ref{Statio-Inv-Beta3}) and (\ref{Statio-Inv-Beta2}) for the expressions of $ P_U(U)$, $ P_V(V)$ and $ P_W(W)$). We have
\bea 
P((U',V') ; (U,V) ) && = \int dW   \delta( U'- \phi^{(1)}(U,V,W) ) \delta( V'- \phi^{(2)}(U,V,W) )  P_{stat}(U,V , W) \nn \\
&& = \int dW \int dU'' dV'' dW''   \delta(U'- \phi^{(1)}(U,V,W) ) \delta( V'- \phi^{(2)}(U,V,W) ) \label{formalDB1}  \\
&&   \quad \quad \quad \quad \quad \quad \delta^{(3)}( (U,V,W) - \phi(U'',V'',W'') ) P_{stat}(U'',V'' , W'') \nn \\ 
&& =  \int dW \int dU'' dV'' dW''   \delta(U'- U'' ) \delta( V'-  V'' )  \label{formalDB2}  \\
&& \quad \quad \quad \quad \quad \quad  \delta^{(3)}( (U,V,W) - \phi(U'',V'',W'') ) P_{stat}(U'',V'' , W'')  \nn  \\
&& =  \int dW''  \delta( U - \phi^{(1)}(U'',V'',W'') )  \delta( V - \phi^{(2)}(U'',V'',W'') )  P_{stat}(U'',V'' , W'')   \nn \\
\Longrightarrow P((U',V') ; (U,V) ) && =  P((U,V) ; (U',V') )  \label{formalDB}
\eea 
which is the desired detailed balance property. Here we have successively used that $\phi$ preserves the PDF $P_{stat}(U,V,W)$ (in (\ref{formalDB1})) and that $\phi$ is an involution (in (\ref{formalDB2})). This property can also be rewritten in the more usual form, using that $P((U',V') ; (U,V) )  = P((U',V') | (U,V) ) P_{stat}(U,V)$, with $P_{stat}(U,V) = P_U(U) P_V(V)$ the stationary PDF of the couple of RVs $(U,V)$,
\bea \label{DBIbeta}
\frac{ P((U',V') | (U,V) )}{ P((U,V) | (U',V') )} = \frac{P_{stat}(U',V')}{P_{stat}(U,V)} \ .
\eea
From a more pragmatical point of view, the above detailed balance property can also be proven using direct calculations. One easily obtains from (\ref{stationarityMap}) and (\ref{Statio-Inv-Beta2}) that
\bea
P((U',V') | (U,V) ) =  \frac{\Gamma(\gamma+\beta)}{\Gamma(\gamma) \Gamma(\beta)}\frac{ \left(V \left(-\frac{U-V U'}{V U'+V}\right)^{\beta } \left(\frac{U+V}{V U'+V}\right)^{\gamma }\right)}{\left(V U'-U\right)}  \delta \left(V'-\frac{U' V}{U}\right) \theta(U>0 ) \theta(V>1) \ ,
\eea
and Eq. (\ref{DBIbeta}) can then directly be checked.

\subsubsection{{\bf At the level of the full space-time process}}

We now give two proofs of the property Prop.~\ref{prop:StatIBStat}, with the first only relying on the detailed balance property (\ref{DBIbeta}) and which will be useful for the BG polymer case. We remind the reader that on a finite time interval $t\in [0 , T] $ with $T \in \JN^*$, the time-reversed coordinates are defined as (see (\ref{time-reversed-coord})) $t_R = T-t  -1 $ and $ x_R = -x $. 
The stationary forward process is defined by drawing a random environment $(W_t(x))_{t = 1 , \dots ,T  , x \in \mathbb{Z} }$ according to (\ref{Statio-Inv-Beta3}), an initial condition $(\hat U_{t=0} (x) , \hat V_{t=0}(x))$ according to the stationary measure (\ref{Statio-Inv-Beta2}), and let it deterministically evolve using (\ref{inductionUVnew}). The time-reversed process was defined for $t_R \in [0,T]$ in (\ref{time-reversed-coord2}) as
\bea
\check U_{t_R}^R(x_R ) = \check U_{t = T -t_R} (x = -x_R +1) \quad , \quad \check V_{t_R}^R(x_R ) =\check V_{t = T -t_R} (x = -x_R)  \ .
\eea
Let us first comment on this definition. First note the shift by one unity of the $x$ coordinate in the definition of $\check U_{t_R}(x_R )$ compared to $\check V_{t_R}^R(x_R )$. The reason for this is that, in the forward evolution, $\phi$ mixes up RVs $(U,V)$ living on edges leading to different vertices (forming a down-left corner) and creates RVs $(U',V')$ living on edges leading to the same vertex (forming a top-right corner) (see (\ref{inductionUVnew})). In the time-reversed process the $U'$ and $V'$ RVs are then reinterpreted as living on edges leading to different vertices (forming a down-left corner in the $(t_R,x_R)$ coordinates) whereas the RVs $U$ and $V$ live on edges leading to the same vertex (forming a top-right corner in the $(t_R,x_R)$ coordinates). The shift by one unity of the $t$ coordinate in the definition of $\check U_{t_R}(x_R )$ and $\check V_{t_R}^R(x_R )$ (compared to (\ref{time-reversed-coord})) ensures that the final values at $t=T$ of the forward process are initial values at $t_R=0$ for the backward process. This is illustrated on the right of Fig.~\ref{fig:drpathZ2}. Introducing these notations permits to rewrite the detailed balance condition (\ref{formalDB}) as
\bea \label{DB11}
P\left((\check U_{t+1}(x) , \check V_{t+1}(x)  ) , (\check U_{t}(x),\check V_{t}(x-1)  ) \right) && = P\left((\check U_{t}(x),\check V_{t}(x-1)  ),(\check U_{t+1}(x) , \check V_{t+1}(x)  ) \right) \nn \\
&&  = P\left((\check U^R_{t_R +1}(x_R),\check V^R_{t_R+1}(x_R)  ),(\check U^R_{t_R}(x_R) , \check V^R_{t_R}(x_R-1)  ) \right) \ 
\eea
(here we used that the process is homogeneous and stationary). Using inductively (\ref{DB11}) (and using that the measure is stationary and that the RVs $\check U_{t}(x)$ and $\check V_{t}(x) $ at different position $x$ are independent) shows the equality in law stated in Prop.~\ref{prop:RevIB} between the forward and time-reversed process. Another way to understand this reversibility property is to explicitly construct a random environment in which the reversed process performs a forward evolution. In this case we use the stronger (compared to the detailed balance property (\ref{formalDB})) property of reversibility of $\phi$ Prop.~\ref{Prop:stationarityMap:Reversibility}:
\begin{enumerate}
\item{Start from a drawing of a random environment $(W_t(x))_{t = 1 , \dots ,T  , x \in \mathbb{Z} }$ distributed as in (\ref{Statio-Inv-Beta3}) and of the variables $(\check U_{t=0} (x) , \check V_{t=0}(x))$ distributed according to the stationary measure (\ref{Statio-Inv-Beta2}).}
\item{Evolve  $(\check U_{t} (x) , \check V_{t}(x))$ from $t=0$ to $t=T$ according to (\ref{inductionUVnew}). At each time step, store also a new disorder RV as, for $1 \leq t_r \leq T$,
\bea
W^R_{t_R}(x_R)  = \phi^{(3)}( \check U_{t }(x ) ,  \check V_{t}(x-1) , W_{t+1}(x) )|_{t=T-t_R , x= -x_R +1} \ .
\eea
}
\item{Then, using that $\phi$ is an involution (Prop.~\ref{Prop:stationarityMap:Reversibility}) shows that the backward process satisfies 
\be \label{inductionUVrev}
\check U^R_{t_R+1}(x_R)  =  \phi^{(1)}(\check U^R_{t_R}(x) ,\check V^R_{t_R}(x-1)   , W^R_{t_R+1}(x))   \quad , \quad\check V^R_{t_R+1}(x)  =  \phi^{(2)}(\check U^R_{t_R}(x) ,\check V^R_{t_R}(x_R-1)   , W^R_{t_R+1}(x_R))   \ , 
\ee
that is, the backward process satisfies a forward evolution in the random environment $W^R_{t_R}(x_R)$, which is, using the properties of $\phi$, a legitimate Inverse-Beta random environment (i.e. the $W^R_{t_R}(x_R)$ are independent and distributed as (\ref{Statio-Inv-Beta3}) and are independent of the stationary initial condition $(\check U^R_{t_T=0} (x_R) , \check V^R_{t_R=0}(x_R))$).
}
\end{enumerate}
This shows in a more constructive fashion that the backward process in indistinguishable from a forward process and that the equality in law (\ref{reversibilityUV}) holds. This procedure is illustrated on the right of Fig.~\ref{fig:drpathZ2}. Note finally that if the RVs in the reversed process are interpreted as quotients of time-reversed partition sums $\check Z^R_{t_R}(x_R)$, we must have by definition
\bea
&& \check U^R_{t_R}(x_R) = \frac{\check Z^R_{t_R} (x_R) }{\check Z^R_{t_R-1}  (x_R-1)}  = \check U_{T-t_R} (-x_R +1)  =  \frac{ \check Z_{T-t_R} (-x_R +1) }{\check Z_{T-t_R-1} (-x_R )}   \   ,  \nn \\
&& \check V^R_{t_R}(x_R) = \frac{\check Z^R_{t_R} (x_R) }{\check Z^R_{t_R-1} (x_R)} = \check V_{T-t_R} (-x_R) = \frac{\check Z_{T-t_R} (-x_R ) }{\check Z_{T-t_R-1} (-x_R )}  \   ,   \ 
\eea
and an appropriate definition of $\check Z^R_{t_R}(x_R)$ is thus
\bea \label{BackwardZ}
\check Z^R_{t_R}(x_R) := \frac{1}{\check Z_{T-t_R -1} (-x_R) } \ .
\eea
Alternatively one can multiply this definition by a constant term as $\check Z^R_{t_R}(x_R) :=\check Z_{T -1} (0)/ \check Z_{T-t_R -1} (-x_R) $ to ensure the initial condition $\check Z^R_{t_R}(0)=1$ as well. In this case one has in law $ (\check Z_t(x) )_{t = 0, \dots , T ; x \in \mathbb{Z}} \sim (\check Z^R_{t_R}(x_R) )_{t_R = 0, \dots , T ; x_R \in \mathbb{Z}}$.

\subsection{Relation to other models} \label{subsec:relation}

In this section we explicitly consider the implication of our results for the Log-Gamma and Stric-Weak polymers, two exactly solvable models of DPs on $\JZ^2$ that can be obtained as limits of the IB polymer. We will not discuss here the $0$ temperature limits $(\gamma , \beta) \to (0,0)$, whose discussion is reported to Sec.~\ref{subsec:T0relation}. We will use here the language of polymers with boundaries to discuss the stationary measures.

\subsubsection{{\bf Limit to the Log-Gamma polymer}}

In \cite{usIBeta} it was shown that the point-to-point partition-sum of the Inverse-Beta polymer (without boundaries) converges to the partition-sum of the Log-Gamma polymer as
\bea
\lim_{\beta \to \infty} \frac{Z_{x_1,x_2}}{\beta^{x_1+x_2}} = Z^{LG}_{x_1,x_2} \ , 
\eea
where the limit holds in law and $Z_{x_1,x_2}^{LG}$ is the partition sum of the Log-Gamma polymer. The latter is defined as in Def.~\ref{Def:ptopIB} but in this case the random variables are distributed as $u^{LG} = v^{LG}$ and $(u^{LG})^{-1}$ is distributed as a Gamma distribution with parameter $\gamma >0$\footnote{Here $u^{LG} = v^{LG}$ means that the random Boltzmann weights can equally well be interpreted as living on the vertices of the square lattice}. At the level of the random Boltzmann weights the convergence in law reads
\bea
 \left( \frac{u}{\beta} , \frac{v}{\beta} \right) \sim \left( \frac{1- Beta(\gamma,\beta)}{\beta Beta(\gamma,\beta)} , \frac{1}{ \beta Beta(\gamma,\beta) } \right) \sim_{\beta \to \infty} (u^{LG} , v^{LG})  \sim \frac{(1,1)}{Gamma(\gamma)} .
\eea

In the same way, using (\ref{Statio-Inv-Beta2}), (\ref{Statio-Inv-Beta3}), a stationary Log-Gamma polymer with boundaries is obtained as
\bea \label{LogGamma}
&& \hat Z^{LG}_{x_1,x_2} = \lim_{\beta \to \infty} \frac{\hat Z_{x_1,x_2}}{\beta^{x_1+x_2}}   \  ,   \nn \\
&& U^{LG} = \lim_{\beta \to \infty} \frac{U}{\beta} \sim \left(Gamma(\gamma-\lambda) \right)^{-1}    \   ,   \nn \\
&& V^{LG} = \lim_{\beta \to \infty} \frac{V}{\beta} \sim \left(Gamma(\lambda)\right)^{-1}   \   ,    \nn \\
&& W^{LG} = \lim_{\beta \to \infty} \frac{W}{\beta} \sim \left(Gamma(\gamma) \right)^{-1}   \   ,   
\eea
(all these limits hold in law). Here $\hat Z^{LG}_{x_1,x2}$ is the partition sum of the Log-Gamma polymer with boundaries defined as for the IB polymer with boundaries (see Def~\ref{Def:IBbound}) with random BWs distributed as $U_{x_1,0} \sim U^{LG}$, $V_{0,x_2} \sim V^{LG}$ and $u_{x_1,x_2} = v_{x_1,x_2} \sim W^{LG}$. This is the same model as introduced in \cite{Seppalainen2012} and our results of stationarity in the IB polymer imply the results Lemma 3.2 and Theorem 3.3 of \cite{Seppalainen2012}.

\subsubsection{{\bf Limit to the Strict-Weak polymer}}

In \cite{usIBeta} it was shown that the point-to-point partition-sum of the Inverse-Beta polymer converges to the partition-sum of the Strict-Weak polymer without boundaries as
\bea
\lim_{\gamma \to \infty} \gamma^{x_1} Z_{x_1,x_2} = Z^{SW}_{x_1,x_2} \ , 
\eea
where the limit holds in law and $Z_{x_1,x_2}^{SW}$ is the partition sum of the Strict-Weak polymer. It is defined as in Def.~\ref{Def:ptopIB} but in this case the random variables are distributed as $v^{SW} =1$ and  $u^{SW}$ is distributed with a Gamma distribution of parameter $\beta >0$. At the level of the random Boltzmann weights the convergence in law reads
\bea \label{IBtoSW3}
 ( \gamma u ,  v) \sim \left( \frac{\gamma(1 - Beta(\gamma , \beta))   }{Beta(\gamma , \beta)} , \frac{1}{Beta(\gamma , \beta)} \right)\sim_{\gamma \to \infty} (u^{SW} , v^{SW}) \sim \left( Gamma(\beta)  , 1\right)   \   .
\eea
A stationary Strict-Weak polymer with boundaries is similarly obtained as, using (\ref{Statio-Inv-Beta2}) and (\ref{Statio-Inv-Beta3}),
\bea \label{StrictWeak}
&&\hat Z^{SW}_{x_1,x_2} =  \lim_{\gamma \to \infty}  \gamma^{x_1} \hat Z_{x_1,x_2}     \   ,   \nn \\ 
&& U^{SW} = \lim_{\gamma \to \infty} \gamma  U \sim Gamma(\beta+\lambda)   \   ,    \nn \\
&& V^{SW} = \lim_{\gamma \to \infty} V\sim  \left( Beta(\lambda, \beta) \right)^{-1}\nn   \  ,      \\
&& W^{SW} = \lim_{\gamma \to \infty}  \gamma W \sim Gamma(\beta)   \  ,   
\eea
(all these limits hold in law). Here $\hat Z^{SW}_{x_1,x_2}$ is the partition sum of the stationary Strict-Weak polymer with boundaries defined as for the IB polymer with boundaries with random BWs distributed as $U_{x_1,0} \sim U^{SW}$, $V_{0,x_2} \sim V^{SW}$, $u_{x_1,x_2} \sim W^{LG}$ and $v_{x_1,x_2} =1$. It satisfies stationarity and reversibility properties inherited from those of the IB polymer (see Sec.~\ref{SubSecOverStatIB}). We note that this stationary Strict-Weak polymer with boundaries differs from the one considered in \cite{StrictWeak1}. Indeed, the admissible paths considered in \cite{StrictWeak1} differ from ours, and so does the stationarity property there obtained which involve ratios of partition sums slightly different from ours (see Definition 6.1, Proposition 6.2 and Lemma 6.3 in \cite{StrictWeak1}). While these two stationary process are different, we note that the ESPs that underly them are different incarnations of the Beta-Gamma algebra of RVs.

\section{$0$ temperature model: Stationary measure of the Bernoulli-Geometric polymer} \label{sec:T0}

In this section we obtain the stationarity properties of the BG polymer with boundary conditions and stationary initial condition stated in Sec.~\ref{SubSecOverStatBG} and discuss the link between our results and previous results on other models. Thanks to the notations we used, the proof of these properties will be (almost) completely analogous to the finite temperature case and we will thus give much less details in this section.

\subsection{Stationarity properties of the Bernoulli-Geometric polymer}

Let us first focus on the case of the BG polymer with boundaries defined in Def.~\ref{Def:BGbound}. In the bulk of $\JN^2$, the optimal energy in the BG polymer with boundaries satisfies the following recursion equation  
\bea \label{T0hatErec}
\hat \sE_{x_1,x_2} = {\rm \min} \left(  \hat \sE_{x_1-1,x_2} + \su_{x_1,x_2} ,   \hat \sE_{x_1,x_2-1} + \sv_{x_1,x_2} \right)  \text{ for } (x_1,x_2) \in (\JN^*)^2  \ .
\eea
This implies the bulk recursion equation for the horizontal and vertical differences of optimal energies (see (\ref{defUVT0}))
\be \label{T0recsUsV}
\hat \sU_{x_1,x_2} =   \phi_{T=0}^{(1)}\left(\hat \sU_{x_1,x_2-1} , \hat \sV_{x_1-1,x_2} ,  \su_{x_1,x_2} , \sv_{x_1,x_2}  \right)   \quad , \quad \hat \sV_{x_1,x_2}=  \phi_{T=0}^{(2)}\left(\hat \sU_{x_1,x_2-1} , \hat \sV_{x_1-1,x_2} ,  \su_{x_1,x_2} , \sv_{x_1,x_2}  \right)    \ . 
\ee
where we have introduced the $T=0$ {\it stationarity map} that we now define.

\begin{definition}
The $T=0$ {\it stationarity map} $\phi_{T=0}$ is the function $\phi_{T=0} : (\sU ,\sV  , \su, \sv) \in \JZ^4 \to (\sU' , \sV') \in \JZ^2$ defined by
\bea \label{T0StationarityMap}
&& \sU' = {\rm min}\left( \su , \sv + \sU - \sV \right)    \quad , \quad \sV' = {\rm min}\left( \su+\sV-\sU , \sv \right)  = \sU' + \sV - \sU   \   .
\eea
\end{definition}
It has the following properties (below and throughout the rest of the paper $\perp$ means `independent of'):

\begin{proposition} \label{prop:T0statMap:Stat}
{\bf Stationarity} If $\sU$, $\sV$, $\su$ and $\sv$ are RVs distributed as in (\ref{T0DistEnergyuv1}) and (\ref{T0DistEnergyUV1}) with $\sU \perp \sV \perp (\su , \sv)$, then the RVs $\sU'$, $\sV'$ in (\ref{T0StationarityMap}) are distributed as in (\ref{T0DistEnergyUV1}) with $\sU' \perp \sV'$.
\end{proposition}

\begin{proposition}
{\bf Detailed balance} \label{prop:T0statMap:DB}
If $\sU$, $\sV$, $\su$ and $\sv$ are RVs distributed as in (\ref{T0DistEnergyuv1}) and (\ref{T0DistEnergyUV1}) with $\sU \perp \sV \perp (\su , \sv)$ and $\sU'$ and $\sV'$ are given by (\ref{T0StationarityMap}), then
\bea \label{T0DB}
Proba\left( (\sU' , \sV') = (k_{\sU'} , k_{\sV'})  ,   (\sU , \sV) = (k_{\sU} , k_{\sV}) \right) =Proba\left( (\sU' , \sV') = (k_{\sU} , k_{\sV})  ,   (\sU , \sV) = (k_{\sU'} , k_{\sV'}) \right)   \   .
\eea
\end{proposition}

These two properties are proved in Appendix~\ref{app:T0Stationarity}. Thanks to the existence of these properties and of the induction (\ref{T0recsUsV}), the stationarity properties of the BG polymer with boundaries (and similarly of the BG polymer with stationary initial condition) then easily follow as in the previous section by induction on down-right paths.

{\it Remarks}
\begin{itemize}
\item{Note that contrary to the stationarity-reversibility map $\phi$ defined for the Inverse-Beta polymer in Def.~\ref{stationarityMapDef}, the stationarity map of the $\phi_{T=0}$ model is not an involution. We were not able to extend as before $\phi_{T=0}$ to an involution $\tilde \phi_{T=0} : (\sU ,\sV  , \su, \sv) \to (\sU' , \sV' , \su', \sv')$ that conserves the PDF and the independence of $\sU$, $\sV$ and of the couple $(\su,\sv)$. We believe this is related to the fact that the recursion equation (\ref{T0recsUsV}) `loses some memory', in the sense that if $\su_{x_1,x_2}$ in (\ref{T0recsUsV}) is too large, its value cannot be inferred from the sole knowledge of $\sU_{x_1,x_2}$ and $\sV_{x_1,x_2}$. Nevertheless, we were still able to prove the detailed balance property (\ref{T0DB}), which is sufficient to prove the reversibility property of the stationary process Prop.~\ref{prop:RevBG} in the BG polymer with stationary initial condition as for the IB polymer with stationary initial condition: the only difference is that we do not have the explicit construction of the random environment in which the backward process (\ref{TRprocessT0}) satisfies a forward evolution equation. }
\item{Sets of random variables satisfying a stationarity property similar to the one of the stationarity map (\ref{stationarityMap}) have played over the years an important role in the theory of queuing systems since they also provide in this framework models with an exact solvability property. The first occurrence of a property of this type in this context is due to Burke for the case of exponentially distributed RVs \cite{Burke}. Since then such properties have been designated as Burke properties. Examples of sets of RVs for which Burke properties have been shown notably include systems of Geometric variables \cite{OConnell2005} and more recently mixture of Bernoulli and Geometric variables very similar to the ones considered here \cite{Martin2009}. The exact solvability property studied in \cite{Martin2009} does not however seem trivially connected to the one studied here. From the technical point of view we note that it involves $4$ independent Geometric RVs (while our property involves $5$), and more conceptually the model studied in \cite{Martin2009} naturally corresponds  to a problem of first passage percolation, while our model interpolates between problems of first and last passage percolation (see Sec.~\ref{subsec:T0relation}).}
\end{itemize}

\subsection{Relation to other models} \label{subsec:T0relation}

Let us now discuss the relations between this model and other known models. We discuss this in the framework of the model with boundaries in order to obtain the stationary measure of the limiting model as well. In the following we will only study the limits at the level of the random energies $(\su,\sv,\sU,\sV)$. Each limit can be used to define a model equivalent to the BG polymer with boundaries (see Def.~\ref{Def:BGbound}) with different distributions of energies in the bulk and on the boundaries and a stationarity property on down-right paths.

\subsubsection{{\bf $q' \to 0$ limit: last passage percolation with geometric waiting times}}

An isotropic limit of the model is obtained by letting $q' \to 0$. In this case the random energies that enters into the definition of the model with boundaries are distributed as
\bea \label{gLPP}
&& \su^{gLPP} = \sv^{gLPP} = -G_q \ ,  \nn \\
&& \sU^{gLPP} = - G_{q/q_b} \ , \nn \\
&& \sV^{gLPP} = -G_{q_b} \ .
\eea
This model exactly corresponds to geometric last passage percolation as e.g. studied in \cite{Johansson2000} for the case without boundaries (note that $\su^{gLPP} = \sv^{gLPP}$ implies that the bulk random energies can be interpreted as living on the vertices of $\mathbb{N}^2$). Indeed, note that while the random energies in the Bernoulli-Geometric polymer can generally be both positive and negative, in this limit the energies are always negative and the energy-minimization problem can be reinterpreted as a maximization problem of the last passage time. The latter is given by ${\sf T}_{x_1 ,x_2} := - {\sf E}_{x_1,x_2} = {\rm \max} \left\{ \sum_{e \in \pi} {\sf t}_e , \pi : (0,0) \to (x_1,x_2)  \right\}$, where the random waiting times on the edges are the opposite of the random energies, ${\sf t}_e := - {\cal E}(e) \geq 0$. This model was denoted Geo-LPP in Fig.~\ref{fig:spaceofmodels}.

\subsubsection{{\bf $q \to 0$ limit: a first passage percolation problem with Geometric waiting times}}

An anisotropic limit is obtained by letting $q \to 0$ with $q'$ fixed. We obtain
\bea \label{bgFPP}
&& (\su^{bgFPP} ,\sv^{bgFPP})  =  ((1-\xi_{\su \sv}) (1+G_{q'}) , 0) \ ,  \nn \\
&& \sU^{bgFPP} = (1- \xi_{\sU})(1+G_{q_b q'}) \ , \nn \\
&& \sV^{bgFPP} = - \xi_V G_{q_b}  \ .
\eea
with now $p_{\su \sv} = 1-q'$, $p_{\sU} = 1-q_b q'$ and $p_{\sV} = \frac{1-q'}{1-q_b q'}$. Note that in this limit the energies on the (bulk) edges are either $0$ (for vertical edges) or positive. Note also that we can replace the bulk energies on horizontal edges by a simple geometric RV since we have the equality in law $(1-\xi_{\su \sv}) (1+G_{q'}) \sim G_{q'}$. In this limit the optimal energy ${\sE}_{x_1,x_2}$ is thus always the sum of positive terms and the model is naturally interpreted as a model of first passage percolation. Here the first passage time is ${\sf T}_{x_1 ,x_2} :=  + {\sf E}_{x_1,x_2} = {\rm min} \left\{ \sum_{e \in \pi} {\sf t}_e , \pi : (0,0) \to (x_1,x_2)  \right\}$, where the random waiting times on the edges are equal to the random energies, ${\sf t}_e := + {\cal E}(e) \geq 0$. This model was denoted Anisotropic Geo-FPP in Fig.~\ref{fig:spaceofmodels}. This model was already studied in the language of queuing system in \cite{OConnell2005} where the authors obtained an analogue Burke property and also showed that the model could be solved exactly using the RSK correspondence.

\subsubsection{{\bf Continuous limit of the Bernoulli-Geometric polymer and $T=0$ limit of the Inverse-Beta polymer}}

We now discuss the exponential/continuous limit. It is obtained by letting $\epsilon \to 0^+$ with
\bea
q = 1- \gamma' \epsilon \quad, \quad q' = 1 - \beta' \epsilon  \quad , \quad q_b = 1- (\gamma' - \lambda') \epsilon   \   ,
\eea
where $\gamma' , \beta' >0$ and $0< \lambda' < \gamma'$ (to ensure $q_b>q$) are three parameters. In this limit the energies have to be rescaled by $\epsilon$ and converge in law to exponentially distributed random variables as
\bea \label{T0expWeight}
&& (\su^{B-Exp} , \sv^{B-Exp}) = \epsilon (\su , \sv) \to_{ \epsilon \to 0} ( (1- \zeta_{\su \sv} ) E_{\beta'} - \zeta_{\su \sv} E_{\gamma'},-\zeta_{\su \sv} E_{\gamma'} , )  \nn \\
&& \sU^{B-Exp} =   \epsilon \sU \to_{\epsilon \to 0} (1- \zeta_{\sU}) E_{\beta' + \lambda'} - \zeta_{\sU} E_{\gamma' - \lambda'} \nn \\
&& \sV^{B-Exp} =   \epsilon \sV \to_{\epsilon \to 0} - \zeta_{\sV}E_{\lambda'}
\eea
where $\zeta_{\su \sv}$, $\zeta_{\sU} $ and $\zeta_{\sV}$ are Bernoulli RVs with parameters $p_{\su \sv} =  \frac{\beta'}{\beta' + \gamma'}$, $p_{\sU}  =\frac{\beta' + \lambda'}{\beta' + \gamma'}$ and $p_{\sV}  =\frac{\beta' }{\beta' + \lambda'}$ and $E_{\gamma'}$, $E_{\beta'}$, $E_{\gamma' - \lambda'}$, $E_{\lambda'}$ and  $E_{\beta' + \lambda'}$ denote exponentially distributed RVs. Let us recall here that the PDF of an exponentially distributed RV is
\bea
E_\alpha \sim Exp(\alpha) \quad,  \quad p(E_\alpha ) =  \alpha e^{ -\alpha E_{\alpha}} \ .
\eea
The optimal energy in this model has to be scaled accordingly as
\bea
\hat \sE^{B-Exp}_{x_1,x_2} = \lim_{\epsilon \to 0}  \epsilon \hat{\sE}_{x_1 ,x_2},
\eea
and the results previously obtained in the BG polymer with boundaries also apply to this model using the now exponentially distributed weights (\ref{T0expWeight}). We call this model the Bernoulli-Exponential polymer with boundaries (denoted as Bernoulli-Exp in Fig.~\ref{fig:spaceofmodels}). This model can also be obtained from the IB polymer with boundaries using $\gamma = \epsilon \gamma'$, $\beta= \epsilon \beta'$, $\lambda = \epsilon \lambda'$ and scaling the energies as
\bea \label{IBtoBE}
&& (\su^{B-Exp} , \sv^{B-Exp} ) = -\epsilon (\log u ,\log v) \to_{\epsilon \to 0} ( (1- \zeta_{\su \sv} ) E_{\beta'} - \zeta_{\su \sv} E_{\gamma'} , -\zeta_{\su \sv} E_{\gamma'} , ) \ ,    \nn \\
&&  \sU^{B-Exp} = -\epsilon \log U \to_{\epsilon \to 0} (1- \zeta_{\sU}) E_{\beta' + \lambda'} - \zeta_{\sU} E_{\gamma' - \lambda'}  \ , \nn \\
&&  \sV^{B-Exp} = -\epsilon \log V \to_{\epsilon \to 0} - \zeta_{\sV}E_{\lambda'}   \   , \nn \\  
&& \hat \sE^{B-Exp}_{x_1,x_2}  = \lim_{\epsilon \to 0}  - \epsilon \log \hat Z_{x_1 ,x_2}   \  .
\eea
Here the convergence in law of the logarithm of the random Boltzmann weights of the Inverse-Beta polymer to a mixture of Bernoulli and exponential distributions was shown in \cite{usIBeta}. We refer the reader to \cite{usIBeta} for the Bethe ansatz study of this polymer model (without boundary conditions) where the authors notably obtain the full distribution of the optimal energy and show Tracy-Widom GUE asymptotic limit. The Bernoulli-Exponential polymer with boundaries has stationarity properties inherited from the stationarity properties of the IB polymer with boundaries, and was first introduced using the limit (\ref{IBtoBE}). {\it The definition of the Bernoulli-Geometric polymer with boundaries Def.~\ref{Def:BGbound} was found by trial and error as a discretization of the Bernoulli-Exponential polymer with boundaries which conserves these stationarity properties} (see in particular Appendix~\ref{app:T0Stationarity}). 

{\bf Isotropic limit: Exponential last passage percolation} \\
Note that the Bernoulli-Exponential polymer (\ref{T0expWeight}) admits an isotropic limit $\beta' \to \infty$ which converges to exponential last passage percolation: in this limit
\bea
 && \su^{eLPP} = \sv^{eLPP}  = - E_{\gamma'} \ ,    \nn \\
&& \sU^{eLPP} =  - E_{\lambda'}  \ , \nn \\
 && \sV^{eLPP} =  -E_{\gamma' - \lambda'} \   .   
\eea
This model can also be obtained from the continuum limit ($ (q,q_b ) = (1- \epsilon \gamma' , q_b = 1- \epsilon \lambda') $, $ \epsilon \to 0$) of geometric last passage percolation (\ref{gLPP}), or also directly as the zero-temperature limit ($(\gamma , \lambda)= \epsilon (\gamma', \lambda')$ with $\epsilon \to 0$) of the Log-Gamma polymer (\ref{LogGamma}), and was denoted Exp-LPP in Fig.~\ref{fig:spaceofmodels}. The first occurrence of this stationary model in the literature was in the language of queuing system and is due to Burke in \cite{Burke}. Here again $\su^{eLPP} = \sv^{eLPP}$ implies that the random energies can be interpreted as living on the sites of $\mathbb{N}^2$.

{\bf Anisotropic limit: anisotropic Exponential first passage percolation} \\
One can also consider an anisotropic limit $ \gamma' \to \infty$ of the Bernoulli-Exponential polymer (\ref{T0expWeight}) to obtain a first passage percolation problem with exponential waiting times:
\bea
&& ( \su^{eFPP} , \sv^{eFPP} ) = ( E_{\beta'} , 0) \nn \\
&& \sU^{eFPP} = E_{\beta' + \lambda'} \nn \\
&& \sV^{eFPP}  = - \zeta_V E_{\lambda'}        \   .
\eea
This model can also be obtained from the continuum limit ($ (q',q_b ) = (1- \epsilon \beta' , q_b = 1- \epsilon \lambda') $, $ \epsilon \to 0$) limit of (\ref{gLPP}), or also as the zero temperature limit ($(\beta, \lambda) = \epsilon (\beta' , \lambda')$ with $\epsilon \to 0$) of the Strict-Weak polymer (\ref{StrictWeak}). As for its geometric counterpart (\ref{gLPP}) this model was studied in \cite{OConnell2005}. It was noted Anisotropic Exp-FPP in Fig.~\ref{fig:spaceofmodels}.

\section{Quenched free-energy, Angle-Boundary parameter equivalence and convergence to the stationary state of point to point models}\label{sec:FE}

In this section we first obtain in Sec.~\ref{subsec:FEINMODELWITHBOUND} preliminary results on the mean optimal energy in the BG polymer with boundaries and the mean free-energy in the IB polymer with boundaries. In Sec.~\ref{subsec:proof} we use these results to obtain the mean quenched optimal energy/free-energy in models without boundaries (Eq.~(\ref{FEresultBIS}) and (\ref{T0decomp11BIS})). In Sec.~\ref{subsec:EquivaAngBou} we will discuss the convergence of point to point models to their stationary state (see Sec.~\ref{SubSecOverConv}). Finally in Sec.~\ref{subsec:Fluctuations} we will comment on free-energy fluctuations and optimal path properties in models with boundaries.

\smallskip

 Let us first collect here some definitions for the mean energies of bulk and boundaries random Boltzmann weights/energies in these models. Below and as before $u,v,U,V$ and $\su, \sv, \sU, \sV$ denote RVs distributed as in Def.~\ref{Def:IBbound} and Def.~\ref{Def:BGbound}. We define

\bea \label{freeEnergyBW}
&  f_U^{\gamma , \beta} (\lambda) :=-\overline{\log{U}} = -\psi (\beta +\lambda )+ \psi (\gamma -\lambda )  \quad  \quad &  \sff_\sU^{q, q'} (q_b) :=\overline{\sU} =  \frac{q_b^2 q'-q}{\left(q_b-q\right) \left(1-q_b q'\right)}   \  , \nn \\
&  \!\!\!\!\!\!\!\!\!\! f_V^{\gamma , \beta} (\lambda):=-\overline{\log{V}} = -\psi (\beta +\lambda )+\psi (\lambda )  \quad  \quad &  \sff_\sV^{q, q'} (q_b):=\overline{\sV} =  - \frac{1-q'}{1-q_b q'}  \frac{q_b}{1-q_b}   \   ,   \nn \\
&  \!\!\!\!\!\!\!\!\!\!\!\!\!\!\!\!\!\!\!\!\!\!\!\!\!\!\!\!\!\!   f_u^{\gamma , \beta} := -\overline{\log u} =-\psi (\beta)+ \psi (\gamma  )  \quad  \quad &  \sff_\su^{q, q'} := \overline{\su} = \frac{q'-q}{(1-q)(1-q')}    \   ,     \nn \\
&  \!\!\!\!\!\!\!\!\!\!\!\!\!\!\!\!\!\!\!\!   f_v^{\gamma , \beta}:=-\overline{\log{v}} = -\psi (\beta +\gamma )+\psi (\gamma )  \quad  \quad &  \sff_\sv^{q, q'}:=\overline{\sv} = - \frac{1-q'}{1-q q'}  \frac{q}{1-q} \  , 
\eea
where $\psi = \frac{\Gamma'}{\Gamma}$ is the diGamma function. A key property of models with boundaries, that will notably play a crucial role in the remaining of this section, is that boundaries are attractive.

Indeed it follows from the fact that $\psi$ is strictly increasing and concave that $f_U^{\gamma , \beta} (\lambda) \leq f_u^{\gamma , \beta}  $ (equality for $\lambda \to  0$), $f_V^{\gamma , \beta} (\lambda) \leq f_v^{\gamma , \beta}  $, (equality for $\lambda \to  \gamma$). Furthermore, when $\lambda \to 0$ (resp. $ \lambda \to \gamma$), $ f_V^{\gamma , \beta} (\lambda) \to - \infty$ (resp.  $ f_U^{\gamma , \beta} (\lambda) \to - \infty$) and the vertical (resp. horizontal) boundary becomes infinitely attractive. Note also that $ f_V^{\gamma , \beta}(\lambda)$ increases with $\lambda$ while $ f_U^{\gamma , \beta}(\lambda)$ decays with $\lambda$. Finally note that for $0< \lambda < \gamma/2$ (resp.  $\gamma/2< \lambda < \gamma$), $f_V^{\gamma , \beta}(\lambda) < f_U^{\gamma , \beta}(\lambda)$ (resp. $f_U^{\gamma , \beta}(\lambda) < f_V^{\gamma , \beta}(\lambda)$) and the vertical (resp. horizontal) boundary is the most attractive. Both boundaries have the same mean energy for $\lambda = \gamma/2$, i.e. $f_U^{\gamma , \beta}(\gamma/2) = f_V^{\gamma , \beta}(\gamma/2)$, a special case referred to as the equilibrium situation in the rest of the paper. Similarly, note that for $q<q_b<1$,  $ \sff_\sU^{q, q'} (q_b)  < \sff_\su^{q, q'}$,  $ \sff_\sV^{q, q'} (q_b)  < \sff_\sv^{q, q'}$. Note also that $ \sff_\sU^{q, q'} (q_b)$ increases when $q_b$ increases with $\sff_\sU^{q, q'} (q_b)  \to_{q_b \to q^+} -\infty$ and $\sff_\sU^{q, q'} (q_b)  \to_{q_b \to 1^-} \sff_\su^{q, q'}$, while $ \sff_\sV^{q, q'} (q_b)$ decays when $q_b$ increases with $\sff_\sV^{q, q'} (q_b)  \to_{q_b \to q^+} \sff_\sv^{q, q'} $ and $\sff_\sV^{q, q'} (q_b)  \to_{q_b \to 1^-}  - \infty$. Finally $ \sff_\sU^{q, q'} (q_b)  <  \sff_\sV^{q, q'} (q_b)$ for $q_b < \sqrt{q}$, $ \sff_\sU^{q, q'} (q_b)  > \sff_\sV^{q, q'} (q_b)$ for $q_b > \sqrt{q}$ and in the `equilibrium case' $q_b = \sqrt{q}$ we have $ \sff_\sU^{q, q'} (\sqrt{q}) =  \sff_\sV^{q, q'} (\sqrt{q})$.

%  \begin{figure}
% \centerline{\includegraphics[width=5.5cm]{FreeEnergy} \quad \quad  \quad \quad  \includegraphics[width=5.5cm]{FreeEnergyT0}} 
% \caption{Mean value of bulk and boundary random energies in the IB (with $\gamma=\beta=1$, left) and BG (with $q=1/4$ and $q'=1/2$, right) polymers with boundaries (\ref{freeEnergyBW}). In the equilibrium cases (vertical lines) the boundaries are equally attractive.}
% \label{fig:meanvalueofenergies}
% \end{figure}  

\subsection{Free-energy in models with boundaries } \label{subsec:FEINMODELWITHBOUND}

{\bf Bernoulli-Geometric polymer}

Let us first focus on the Bernoulli-Geometric polymer with boundaries defined in Def.~\ref{Def:BGbound} and write the optimal energy $\hat \sE_{x_1,x_2}$ for $(x_1 , x_2) \in \mathbb{N}^2$ as, 
\bea
\hat \sE_{x_1,x_2} =\sum_{i =0}^{x_1} \hat \sU_{i , 0} + \sum_{ j = 0}^{x_2}  \hat \sV_{x_1 , j} \ .
\eea
Note that this decomposition does not follow a down-right path and the variables $\hat \sU_{x_1,x_2}$ and $\hat \sV_{x_1,x_2}$ in the two sums are correlated. Each one however is distributed as $\hat \sU_{i , 0} \sim  \sU$ and $\hat \sV_{x_1, j} \sim \sV$ as in (\ref{T0DistEnergyUV1}). Hence we obtain, $\forall (x_1,x_2) \in \mathbb{N}^2$,
\bea \label{T0meanEnergy}
\overline{ \hat \sE_{x_1,x_2} } = x_1 \sff_{\sU}^{q,q'}(q_b) + x_2 \sff_{\sV}^{q,q'}(q_b) \ ,
\eea 
where $ \sff_{\sU}^{q,q'}(q_b)$ and $ \sff_{\sV}^{q,q'}(q_b)$ were given in (\ref{freeEnergyBW}). In particular the mean optimal energy in the direction $(s_1,s_2)$ is, for $(s_1 , s_2) \in \mathbb{R}_+^2$,
\bea \label{T0FEB}
\hat \sff_{{\rm BG}}(s_1,s_2 , q_b) := \lim_{N \to \infty} \frac{1}{N} \overline{ \hat \sE_{ N s_1, N s_2} } = s_1 \sff_{\sU}^{q,q'}(q_b) + s_2 \sff_{\sV}^{q,q'}(q_b) \ .
\eea
We can also consider the mean optimal energy per-unit-length in a direction $-1/2<\varphi<1/2$ as $\hat \sff^{{\rm p.u.l.}}_{{\rm BG}}(\varphi , q_b) := \lim_{t \to \infty} \frac{1}{t} \overline{ \hat \sE_{ t} (x = (1/2 +\varphi) t )} =  \hat \sff_{{\rm BG}}(1/2 + \varphi ,1/2 - \varphi, q_b) $, with conversely $\hat \sff_{{\rm BG}}(s_1,s_2 , q_b) =  (s_1+s_2) \hat \sff^{{\rm p.u.l.}}_{{\rm BG}}( \frac{s_1-s_2}{2(s_1+s_2)} , q_b)$. Note that from (\ref{T0FEB}), it is clear that the mean optimal energy per-unit-length $\hat \sff^{{\rm p.u.l.}}_{{\rm BG}}(\varphi,q_b)$ is linear in $\varphi$. Furthermore, note that in the special case $q_b = \sqrt{q}$ (referred to as the equilibrium case earlier), $\sff_{\sU}^{q,q'}(q_b)= \sff_{\sV}^{q,q'}(q_b)$ and $\hat \sff^{p.u.l.}_{{\rm BG}}(\varphi,q_b)$ does not depend on $\varphi$. We will come back to this point later.

{\bf Inverse-Beta polymer}\\
In the same way, in the case of the Inverse-Beta polymer with boundaries, $\forall (x_1 , x_2) \in \mathbb{N}^2$, $-\log \hat Z_{x_1,x_2} =- \sum_{i =0}^{x_1} \log \hat U_{i , 0} - \sum_{ j = 0}^{x_2}  \log \hat V_{x_1 , j}$. We thus have $- \overline{ \log \hat Z_{x_1,x_2}} =x_1 f_{U}^{\gamma,\beta}(\lambda) + x_2 f_{V}^{\gamma,\beta}(\lambda)$, implying that the mean quenched free-energy in the direction $(s_1,s_2) \in \JR_+^2$ is
\bea \label{FEB}
\hat f_{{\rm IB}}(s_1,s_2 , \lambda) := - \lim_{N \to \infty} \frac{1}{N} \overline{  \log \hat Z_{N s_1 , N s_2} } =s_1 f_{U}^{\gamma,\beta}(\lambda) + s_2 f_{V}^{\gamma,\beta}(\lambda)    \ .
\eea
And the free-energy per-unit-length in the direction $\varphi \in ]-1/2,1/2[$, $\hat f^{{\rm p.u.l.}}_{{\rm IB}}(\varphi , \lambda) := - \lim_{t \to \infty} \frac{1}{t} \overline{  \log \hat Z_{t (1/2+\varphi) , t(1/2  - \varphi)} } = \hat f_{{\rm IB}}(1/2+\varphi , 1/2 - \varphi , \lambda )$ with conversely $ \hat f_{{\rm IB}}(s_1,s_2 , \lambda) = (s_1+s_2)  \hat{f}^{{\rm p.u.l.}}_{{\rm IB}}( \frac{s_1-s_2}{2(s_1+s_2)} , \lambda)$. As before, note from (\ref{FEB}) that $\hat f^{{\rm p.u.l.}}_{{\rm IB}}(\varphi , \lambda)$ is generally linear in $\varphi$, with the special case that it is constant in the equilibrium situation $\lambda = \gamma/2$.

\subsection{Free-energy in models without boundaries}  \label{subsec:proof}
\subsubsection{{\bf Bernoulli-Geometric polymer}}

\begin{figure}
\centerline{\includegraphics[width=7.5cm]{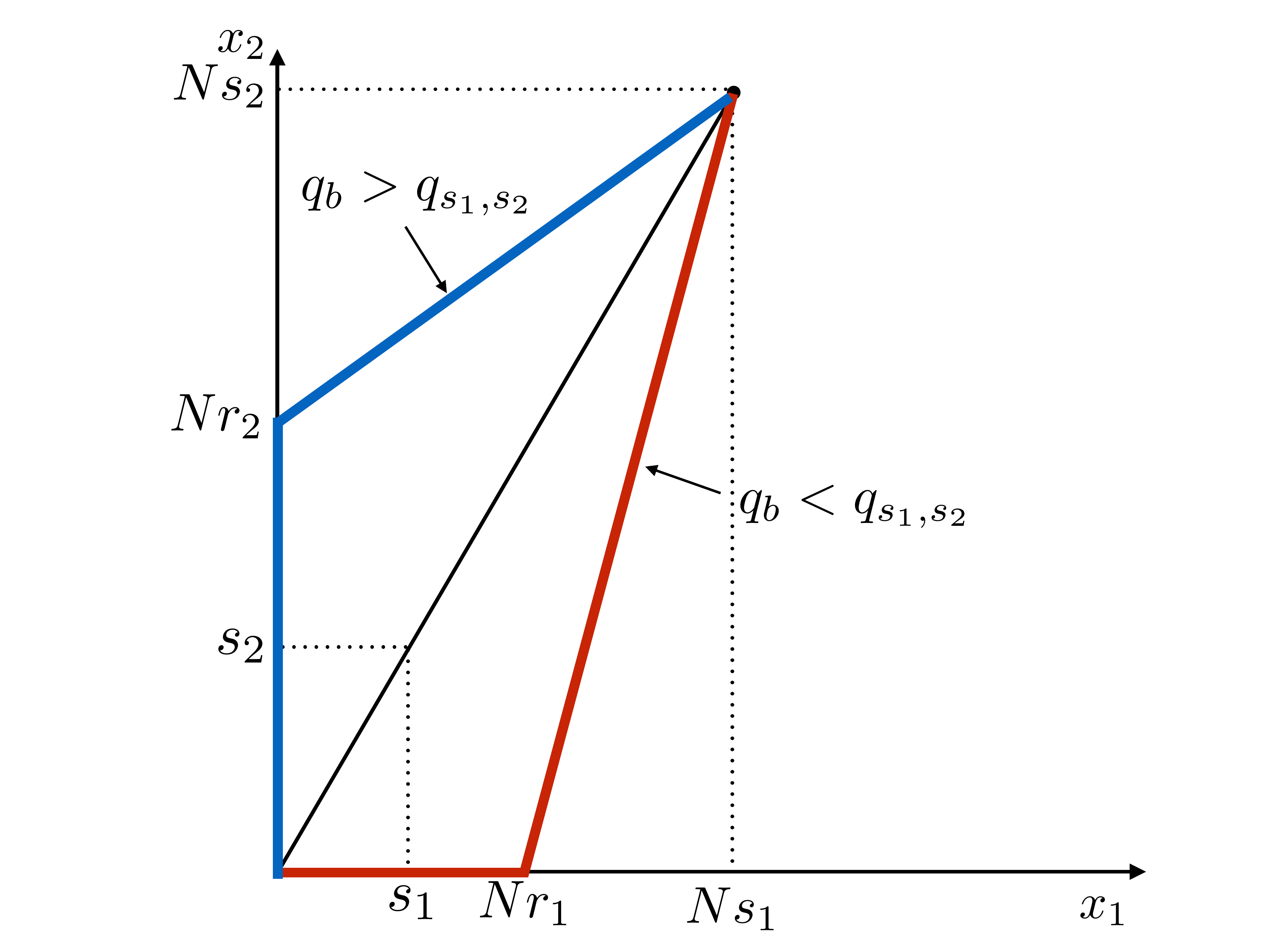} } 
\caption{Cartoon of the notations used in the derivation of (\ref{T0decomp10}). At large $N$ in a  fixed direction $(s_1,s_2)$ and varying the boundary parameter $q_b$, the optimal polymer path sticks to the horizontal (resp. vertical) boundary for $N r_1$ (resp. $Nr_2$) steps for $q_b < q_{s_1,s_2}$ (resp. for $q_b > q_{s_1,s_2}$). We show that $r_1$ decreases from $s_1$ to $0$ (resp. $r_2$ increases from $0$ to $s_2$) when $q_b$ increases from $q$ to $q_{s_1,s_2}$ (resp. from $q_{s_1,s_2}$ to $1$). We show that $q_{s_1 s_2} = q_b^*(s_1,s_2)$ the solution of the saddle-point equation $\frac{\partial}{\partial q_b} \hat \sff_{{\rm BG}}(s_1,s_2, q_b )|_{q_b = q_b^*(s_1,s_2)}   =  0$ and that for this boundary parameter in the direction $(s_1,s_2)$ we have the identity $\sff_{{\rm BG}}(s_1,s_2) =  \hat  \sff_{{\rm BG}}(s_1,s_2,q_b^*(s_1,s_2) )$. }
\label{fig:picproof}
\end{figure}

The first part of this section is devoted to the derivation of the formula (\ref{T0decomp11BIS}) for $\sff(s_1,s_2 )$. We believe it to be rather instructive and the main ideas are summarized in Fig.~\ref{fig:picproof}. Furthermore we introduce in this derivation several elements which will be important in Sec.~\ref{subsec:EquivaAngBou}. The ideas used in this derivation are close in spirit to those used in \cite{Seppalainen2013} for the Log-Gamma polymer, however the proof in \cite{Seppalainen2013} cannot be straightforwardly adapted to our problem since it relied on the $x_1 \leftrightarrow x_2$ symmetry that is absent in our anisotropic models\footnote{although we note that this was indeed accomplished in \cite{Emrah} for the case of inhomogeneous last passage percolation models with on site geometric or exponential waiting times}. 

\smallskip

{\bf Derivation of a formula for $\sff_{{\rm BG}}(s_1,s_2 )$} \\
Let us now consider again the Bernoulli-Geometric model with boundaries defined in Def.~\ref{Def:BGbound}. $\forall (x_1 ,x_2) \in (\JN^*)^2$ we write the decomposition
\bea \label{T0decomp1}
\hat \sE_{x_1,x_2} = {\rm min}\left\{ {\rm min}_{i \in [1 , x_1]} \left( \sum_{j=1}^{i} \hat \sU_{j,0} + \sv_{i,1} + \sE^{i,1}_{x_1,x_2}  \right)  ,  {\rm min}_{i \in [1 , x_2]} \left( \sum_{j=1}^{i} \hat \sV_{0,j} + \su_{1,i} + \sE^{1,i}_{x_1,x_2}  \right)    \right\}   \   ,
\eea
where we have introduced $\forall (x_1 ,x_2  , x_1' , x_2' ) \in (\JN^*)^4 $ with $x_1' \leq x_1 $ and $x_2'  \leq x_2$, the minimal energy to go from $(x_1',x_2')$ to $(x_1,x_2)$
\bea
\hat \sE^{x_1',x_2'}_{x_1,x_2} = {\rm \min} \left\{ \hat {\cal E}(\pi) = \sum_{e \in \pi} \hat {\cal E}(e) , \pi : (x_1',x_2') \to (x_1,x_2)  \right\}  \ .
\eea
Note that an up-right path from $(x_1',x_2')$ to $(x_1,x_2)$ cannot pass upon an edge on the boundary of $\mathbb{N}^2$. Hence the random energies encountered along the way are only of the bulk type and thus $\hat \sE^{x_1',x_2'}_{x_1,x_2} $ corresponds to an optimal energy in a model without boundaries. More precisely we have the equality in law, using the statistical translational invariance of the disorder,
\bea \label{T0decomp2}
\hat \sE^{x_1',x_2'}_{x_1,x_2} \sim \sE_{x_1-x_1' , x_2-x_2'} \ ,
\eea
where here $\sE_{x_1,x_2}$ denotes the optimal energy in the point to point Bernoulli-Geometric model as defined in Def.~\ref{Def:ptopBG}. Using (\ref{T0decomp1}), the definitions (\ref{T0FEB}) and (\ref{DefIntroFE}) and the equality in law (\ref{T0decomp2}) we obtain, scaling $i \sim N r$ in (\ref{T0decomp1}),
\bea \label{T0decomp3}
&& \hat \sff_{{\rm BG}}(s_1,s_2, q_b) =s_1 \sff_{\sU}^{q,q'}(q_b) + s_2 \sff_{\sV}^{q,q'}(q_b)  \nn \\
&&  = {\rm min}\left\{ {\rm inf}_{0 \leq r \leq s_1} ( r  \sff_{\sU}^{q,q'}(q_b)  + \sff_{{\rm BG}}(s_1-r,s_2) )   ,  {\rm inf}_{0 \leq r \leq s_2} ( r  \sff_{\sV}^{q,q'}(q_b)  + \sff_{{\rm BG}}(s_1,s_2-r)  ) \right\} \ .
\eea
The goal is now to `invert' (\ref{T0decomp3}) to obtain $\sff_{{\rm BG}}(s_1,s_2)$. Let us fix $s_1 ,s_2 >0$ and study the properties of (\ref{T0decomp3}) as a function of $q_b\in [q,1]$. Note that in the limit $q_b \to 1$, $\sff_{\sV}^{q,q'}(q_b) \to -\infty$ while other quantities stay bounded. As a consequence $\hat \sff_{{\rm BG}}(s_1,s_2, q_b) \sim s_2 \sff_{\sV}^{q,q'}(q_b)$ and the minimum in the right hand side of (\ref{T0decomp3}) is attained in the second ${\rm inf}$ with $r \to s_2$. Conversely, in the limit $q_b \to q$, $\sff_{\sU}^{q,q'}(q_b) \to -\infty$ while other quantities stay bounded, and thus $\hat \sff_{{\rm BG}}(s_1,s_2, q_b) \sim s_1 \sff_{\sU}^{q,q'}(q_b)$ and the minimum in the right hand side of (\ref{T0decomp3}) is attained in the first ${\rm inf}$ with $r \to s_1$. From this it is clear that there exists a constant $q_{s_1,s_2} \in ]q_b , 1[$ such that
\bea\label{T0decomp4}
s_1 \sff_{\sU}^{q,q'}(q_b) + s_2 \sff_{\sV}^{q,q'}(q_b)  = && \theta(q_{s_1,s_2} -q) {\rm inf}_{0 \leq r \leq s_1} ( r  \sff_{\sU}^{q,q'}(q_b)  + \sff_{{\rm BG}}(s_1-r,s_2) ) \nn \\
&& + \theta(q-q_{s_1,s_2} ){\rm inf}_{0 \leq r \leq s_2} ( r  \sff_{\sV}^{q,q'}(q_b)  + \sff_{{\rm BG}}(s_1,s_2-r)  )         .
\eea 
Let us implicitly define two functions $r_1 :  q_b \in [q , q_{s_1,s_2}] \to r_1(q_b) \in [0,s_1]$ and $r_2 :  q_b \in [q_{s_1,s_2} , 1] \to r_2(q_b) \in [0,s_2]$ such that
\bea \label{T0decomp5}
s_1 \sff_{\sU}^{q,q'}(q_b) + s_2 \sff_{\sV}^{q,q'}(q_b)  = && \theta(q_{s_1,s_2} -q_b) ( r_1(q_b)  \sff_{\sU}^{q,q'}(q_b)  + \sff_{{\rm BG}}(s_1-r_1(q_b),s_2) ) \nn \\
&& + \theta(q_b-q_{s_1,s_2} ) ( r_2(q_b)  \sff_{\sV}^{q,q'}(q_b)  + \sff_{{\rm BG}}(s_1,s_2-r_2(q_b))  ) .
\eea
They satisfy $r_1(q_b) \to_{q_b \to q}  s_1$, $r_2(q_b) \to_{q_b \to 1} s_2$ and are such that
\bea \label{T0decomp6}
&& \forall q_b \in ]q , q_{s_1,s_2}[  \quad , \quad \sff_{\sU}^{q,q'}(q_b) - \partial_1 \sff_{{\rm BG}}(s_1-r_1(q_b) ,s_2) ) = 0   \   ,   \nn \\
&& \forall q_b \in ]q_{s_1,s_2} , 1[  \quad , \quad \sff_{\sV}^{q,q'}(q_b) - \partial_2 \sff_{{\rm BG}}(s_1 ,s_2 -r_2(q_b)) ) = 0 \ .
\eea
Differentiating these equations with respect to $q_b$, we obtain
\bea \label{T0decomp7}
&& \forall q_b \in ]q , q_{s_1,s_2}[  \quad ,  \quad ( \sff_{\sU}^{q,q'})'(q_b) + \partial_1^2 \sff_{{\rm BG}}(s_1-r_1(q_b) ,s_2) ) r_1'(q_b) = 0    \   ,   \nn \\
&& \forall q_b \in ]q_{s_1,s_2} , 1[  \quad , \quad  (\sff_{\sV}^{q,q'})'(q_b) + \partial_2^2 \sff_{{\rm BG}}(s_1 ,s_2 -r_2(q_b)) ) r_2'(q_b)  = 0   \  .
\eea
From this and using the fact that $\sff_{\sU}^{q,q'}(q_b)$ (resp. $\sff_{\sV}^{q,q'}(q_b)$) is strictly increasing (resp. decreasing) as a function of $q_b$ and assuming that $\sff(s_1 ,s_2) ) $ is a strictly convex function, we obtain that $r_1(q_b)$ (resp. $r_2(q_b)$) should be strictly decreasing (resp. increasing) on $]q,q_{s_1,s_2}[$ (resp. $]q_{s_1,s_2},1[$). Note now that $\hat \sff_{{\rm BG}}(s_1,s_2, q_b) = s_1 \sff_{\sU}^{q,q'}(q_b) + s_2 \sff_{\sV}^{q,q'}(q_b)$ is a concave function of $q_b$ on $]q,1[$ with a single maximum at some $q_b^* \in ]q ,1[$. In particular $s_1 \sff_{\sU}^{q,q'}(q_b) + s_2 \sff_{\sV}^{q,q'}(q_b)$ is not constant on any sub-interval and hence both $r_1(q_b)$ and $r_2(q_b)$ cannot be constant on any subinterval either. Combined with the fact that $r_1(q_b)$ is strictly decreasing, this shows that $r_1(q_b) >0$ $\forall q_b <q_{s_1,s_2}$. In the same way $r_2(q_b) >0$ $\forall q_b >q_{s_1,s_2}$. Let us now differentiate (\ref{T0decomp5}) with respect to $q_b$ for $q_b \neq q_{s_1,s_2}$ and use the saddle-point equation (\ref{T0decomp6}), we obtain
\bea \label{T0decomp8}
s_1 (\sff_{\sU}^{q,q'})'(q_b) + s_2 (\sff_{\sV}^{q,q'})'(q_b)  =   \theta(q_{s_1,s_2} -q_b) r_1(q_b)  (\sff_{\sU}^{q,q'})'(q_b)  + \theta(q_b-q_{s_1,s_2} ) r_2(q_b) ( \sff_{\sV}^{q,q'})'(q_b)  .
\eea
In particular this shows that $\frac{\partial}{\partial q_b}\hat \sff_{{\rm BG}}(s_1,s_2, q_b) = s_1 (\sff_{\sU}^{q,q'})'(q_b) + s_2 (\sff_{\sV}^{q,q'})'(q_b) $ is not $0$ $\forall q_b \neq q_{s_1,s_2}$. Since we know that $\frac{\partial}{\partial q_b}\hat \sff_{{\rm BG}}(s_1,s_2, q_b) = 0$ for $q_b = q_b^*$ we necessarily obtain
\bea \label{T0decomp9}
q_{s_1,s_2} = q_b^* \quad \text{and} \quad 0 = \lim_{q_b \to (q_b^*)^{-}}r_1(q_b)  (\sff_{\sU}^{q,q'})'(q_b) =   \lim_{q_b \to (q_b^*)^+} r_2(q_b)  (\sff_{\sV}^{q,q'})'(q_b)  \ .
\eea
And hence $\lim_{q_b \to (q_b^*)^{-}} r_1(q_b) = \lim_{q_b \to (q_b^*)^{+}}r_2(q_b) = 0$.

\smallskip

{\bf Final formula for $\sff_{{\rm BG}}(s_1,s_2) $} \\
Using finally by continuity (\ref{T0decomp5}) for $q_b \to q_b^*$ we obtain our final result for the optimal energy of the model without boundaries: $\forall (s_1 , s_2) \in \mathbb{R}_+^2$,
\bea \label{T0decomp10}
&& \sff_{{\rm BG}}(s_1,s_2) =\hat \sff_{{\rm BG}}(s_1,s_2, q_b = q_b^*(s_1,s_2))   \  , \nn \\
&& \frac{\partial}{\partial q_b} \hat \sff_{{\rm BG}}(s_1,s_2, q_b )|_{q_b = q_b^*(s_1,s_2)} = 0    \   .
\eea
Using (\ref{freeEnergyBW}) and (\ref{T0FEB}), these formal formulas are rewritten more explicitly in (\ref{T0decomp11BIS}).

\begin{figure}
\centerline{\includegraphics[width=6cm]{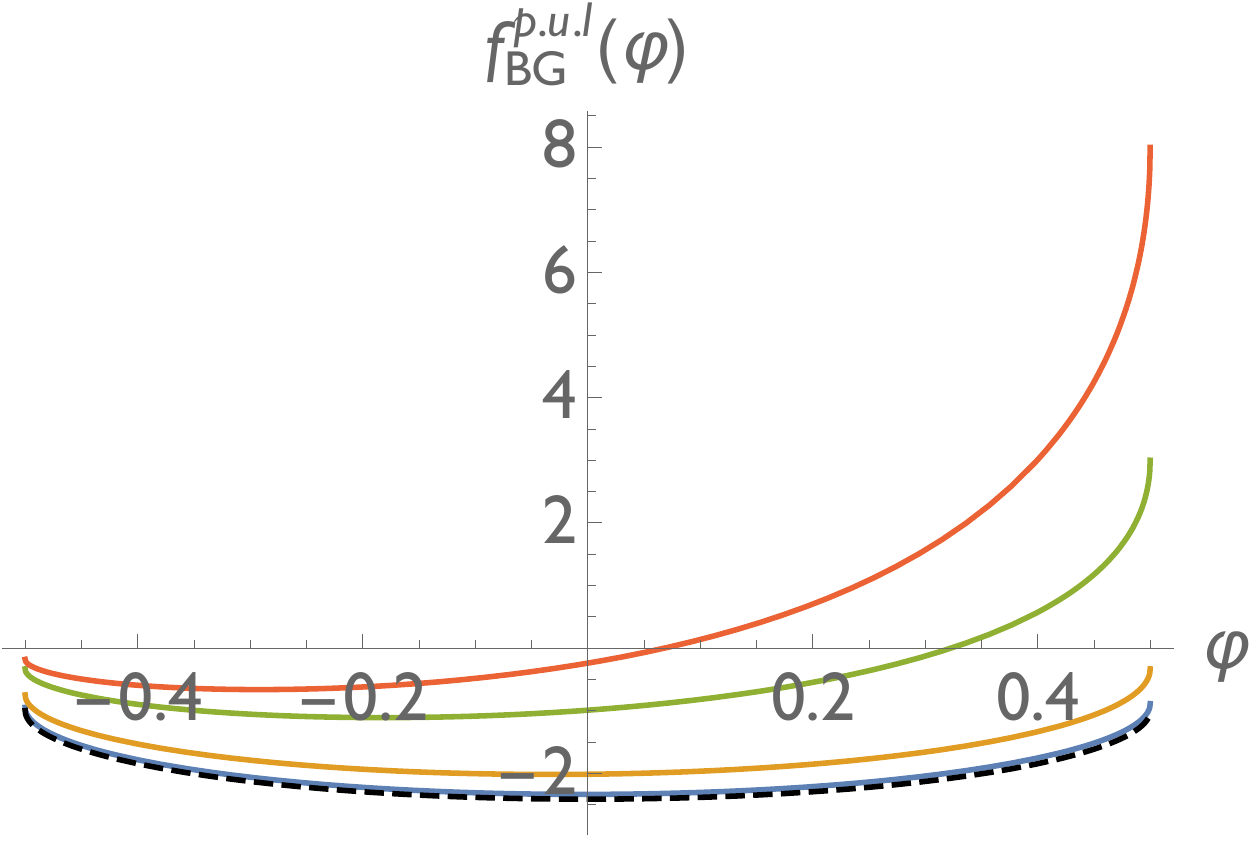} \quad \quad \includegraphics[width=6cm]{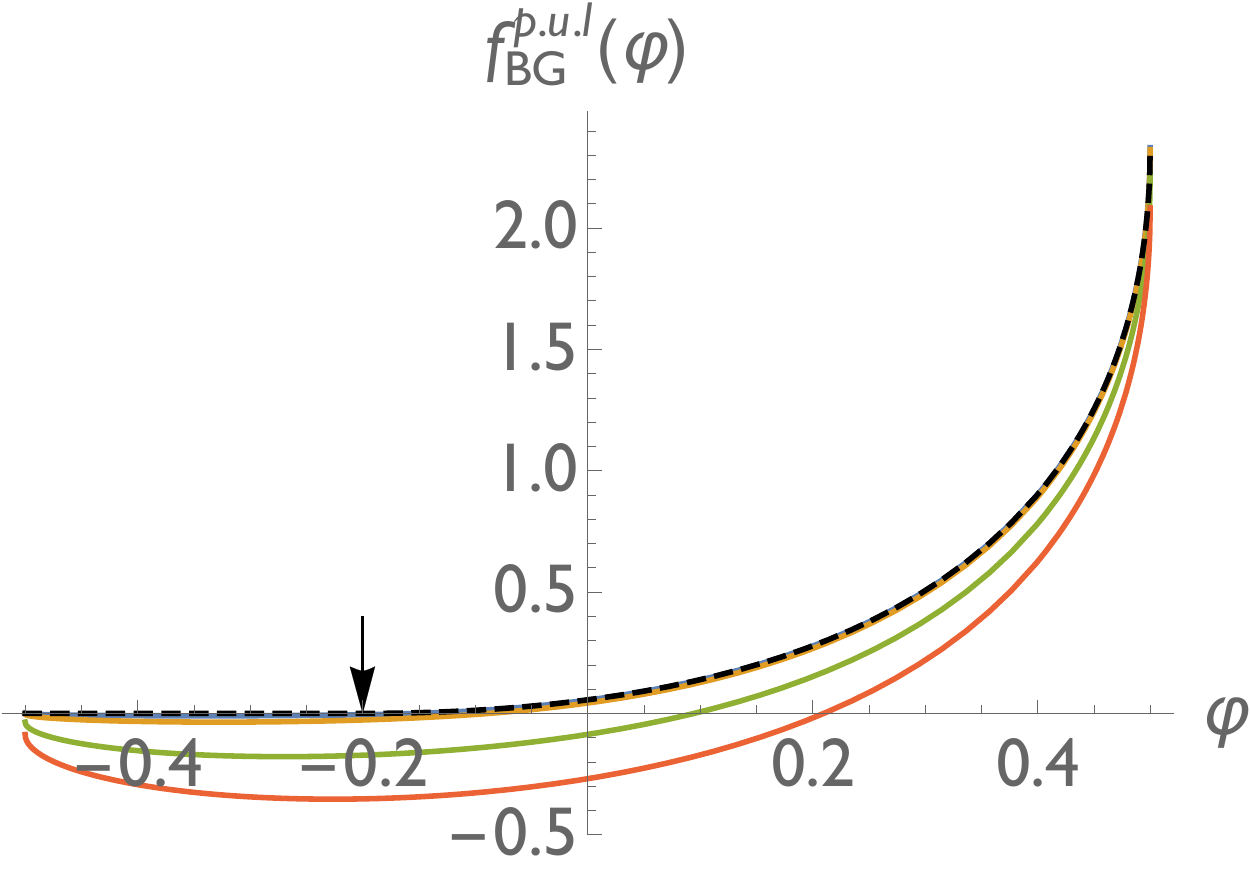} } 
\caption{Left: Optimal energy per-unit-length $\sff^{{\rm p.u.l.}}_{{\rm BG}}(\varphi)$ in the point to point BG polymer (\ref{T0decomp101}) for $q=0.5$ and $q'=0.1, 0.4, 0.8, 0.9$ (plain lines, blue, orange, green and red) and in the last passage percolation limit $q' \to 0$ (black dashed line) (\ref{JohannsonResult}). Right: Optimal energy per-unit-length $\sff^{{\rm p.u.l.}}_{{\rm BG}}(\varphi)$ in the point to point BG polymer (\ref{T0decomp101}) for $q'=0.7$ and $q'=0.001, 0.01, 0.1, 0.2$ (plain lines, blue, orange, green and red) and in the first passage percolation limit $q \to 0$ (black dashed line) (\ref{FPPresult}). The arrow indicates the percolation threshold of the $q\to 0$ limit $\varphi_{q' = 0.7} = -0.2$.}
\label{fig:FEFPPandLPP}
\end{figure}

{\bf Free-energy per-unit-length, optimal angle and the equilibrium case} \\
The free-energy per-unit-length in the direction $\varphi \in ]-1/2,1/2[$, $ \sff^{{\rm p.u.l.}}_{{\rm BG}}(\varphi)= \sff_{{\rm BG}}(1/2+\varphi,1/2-\varphi)$ is similarly given by
\bea \label{T0decomp101}
&& \sff^{{\rm p.u.l.}}_{{\rm BG}}(\varphi) = (1/2 + \varphi) \sff_\sU^{qq'}(q_b^*(\varphi)) + (1/2-  \varphi)\sff_\sV^{qq'}(q_b^*(\varphi))   \\
&&  0 = (1/2 + \varphi) \partial_{q_b}\sff_\sU^{qq'}(q_b^*(\varphi)) + (1/2 - \varphi) \partial_{q_b}\sff_\sU^{qq'}(q_b^*(\varphi)) \ . \label{T0decomp101bis}
\eea 
It is plotted in Fig.~\ref{fig:FEFPPandLPP} for various values of $q,q'$. $\sff^{{\rm p.u.l.}}_{{\rm BG}}(\varphi)$ reaches its minimum at the angle $\varphi_{{\rm opt}}^{{\rm BG}}$ such that, using the saddle-point structure, in (\ref{T0decomp101})-(\ref{T0decomp101bis})
\bea
\frac{\partial \sff^{{\rm p.u.l.}}_{{\rm BG}}(\varphi)}{\partial \varphi} = 0 =  \sff_\sU^{qq'}(q_b^*(\varphi_{{\rm opt}}^{{\rm BG}})) -\sff_\sV^{qq'}(q_b^*(\varphi_{{\rm opt}}^{{\rm BG}}))  \ .
\eea
And using (\ref{freeEnergyBW}) this shows that $\sff^{p.u.l.}_{{\rm BG}}(\varphi)$ is minimum when $q_b^*(\varphi) = \sqrt{q}$, the boundary parameter already referred to as the equilibrium boundary parameter. The optimal angle is thus obtained using (\ref{T0decomp101bis}) with $q_b = \sqrt{q}$ and one obtains (\ref{EqOverviewOptimalAngles}).

\smallskip

{\bf Last-Passage-Percolation limit} \\
In the isotropic case $q' = 0$ case we easily obtain from the above formulas, using $\hat \sff_{{\rm BG}}(s_1,s_2,q_b)|_{q'=0} = \frac{  q}{q-q_b} s_1+\frac{q_b }{q_b -1} s_2 $ and that the quartic equation for $q_b^*$ in (\ref{T0decomp11BIS}) becomes a simpler quadratic equation, that
\bea \label{JohannsonResult}
 \sff_{{\rm BG}}(s_1,1)|_{q' = 0}=\frac{s_1  q+2 \sqrt{s_1  q}+q}{q-1}   \quad , \quad   \sff_{{\rm BG}}^{{\rm p.u.l.}}(\varphi)|_{q' = 0} =  \frac{(1-2 \phi ) \sqrt{\frac{2 q \phi +q}{1-2 \phi }}+q}{q-1}.   
\eea
This reproduces the already known result first obtained by Johansson using the RSK correspondence (see Theorem 1.1 in \cite{Johansson2000} with there $s_1 = \gamma$). The function $\sff_{{\rm BG}}^{{\rm p.u.l.}}(\varphi)|_{q'=0}$ is plotted in black-dashed on the left of Fig.~\ref{fig:FEFPPandLPP} for $q=0.5$.

\smallskip

{\bf First-Passage-Percolation limit} \\
As discussed in Sec.\ref{subsec:T0relation}, the $q \to 0$ limit of the model is a model of first passage percolation with Bernoulli-Geometric waiting times on horizontal edges only (see (\ref{gLPP})). Taking the limit $q \to 0$ of the above formulas is less straightforward than in the last-passage-percolation limit. Indeed in this limit $\hat \sff_{{\rm BG}}^{{\rm p.u.l.}}(s_1,s_2,q_b)|_{q=0} = -\frac{  q_b q'}{q_b q' -1} s_1+\frac{q_b  \left(q'-1\right)}{\left(q_b-1\right) \left(q_b q'-1\right)} s_2 $, and though $\hat \sff_{{\rm BG}}^{{\rm p.u.l.}}(s_1,s_2,q_b)|_{q=0}$ is still concave as a function of $q_b$, $\lim_{q_b \to 0} \hat \sff(s_1,s_2,q_b)|_{q=0} = 0 > -\infty$ and one of the important element in the derivation of (\ref{T0decomp10}) does not hold anymore. One can however repeat a similar derivation and obtain that, at fixed $q'$ and as a function of $\varphi$, $\hat \sff_{{\rm BG}}^{{\rm p.u.l.}}(\varphi,q_b)|_{q=0} = \hat \sff_{{\rm BG}}(1/2+\varphi,1/2 - \varphi,q_b)|_{q=0}$ reaches its maximum on $q_b = [0,1]$ at $q_b = 0$ for $\varphi \leq \varphi_{q'} :=  1/2 - q'$. In those cases $\sff_{{\rm BG}}^{{\rm p.u.l.}}(\varphi)|_{q =0} = \hat \sff_{{\rm BG}}^{{\rm p.u.l.}}(\varphi,0)|_{q=0}=0$. For $\varphi > \varphi_{q'}$ on the other hand $\hat \sff_{{\rm BG}}^{{\rm p.u.l.}}(\varphi,q_b)|_{q=0}$ reaches its maximum on $q_b = q_b^* \in ]0,1[$ at some $q_b^*$ solution of the quadratic equation $\partial_{q_b} \hat \sff_{{\rm BG}}^{{\rm p.u.l.}}(\varphi,q_b=q_b^*)|_{q=0} = 0$ and in those cases $\sff_{{\rm BG}}^{{\rm p.u.l.}}(\varphi)|_{q =0} = \hat \sff_{{\rm BG}}^{{\rm p.u.l.}}(\varphi,q_b^*)|_{q=0}>0$. Solving the resulting quadratic equation one obtains that $\sff_{{\rm BG}}^{{\rm p.u.l.}}(\varphi)|_{q =0}$ is given by the non-analytic form
\bea \label{FPPresult}
\sff_{{\rm BG}}^{{\rm p.u.l.}}(\varphi)|_{q =0} =\theta \left(\varphi - (\frac{1}{2} - q') \right)  \frac{\left(2 \sqrt{2} \sqrt{(1-2 \varphi ) q'}-2 q'+2 \varphi -1\right)  }{2 \left(q'-1\right)}   \geq  0 \ .
\eea
This formula can also easily be obtained by first solving explicitly the quartic equation in (\ref{T0decomp11BIS}) and then taking the limit $q \to 0$. A `natural' way to interpret this non-analytic behavior is the existence of a percolation threshold. Indeed, the optimal energy $E_{x_1,x_2}$ is equal to $0$ iff there exist a path from $(0,0)$ to $(x_1,x_2)$ such that all the Bernoulli variables $\xi_{\su \sv}$ on the horizontal edges encountered by the path are $0$ (which occurs for each edge with probability $q'$, see (\ref{gLPP})). When $(x_1,x_2) = t(1/2-\varphi , 1/2 + \varphi)$ with $t \to \infty$ and for $\varphi = -1/2$ it is trivial that this occurs with probability $1$, and an interesting question is whether there exist a critical angle $\varphi_c$ up to which this still occurs with probability $1$. In the region $\varphi \geq \varphi_{q'}$ this is clearly not the case since $\sff_{{\rm BG}}^{{\rm p.u.l.}}(\varphi)|_{q =0}  >0$. In the region $\varphi \leq \varphi_{q'}$, $\sff_{{\rm BG}}^{{\rm p.u.l.}}(\varphi)|_{q =0} =0$ and a natural guess would be $\varphi_c = \varphi_{q'}$, although we cannot simply rule out here the possibility that the optimal path encounters a non-extensive number (i.e. $o(t)$) of edges such that $\xi_{\su \sv} \neq 0$. Around $\varphi_{q'}$ we obtain, for $ \delta \varphi>0$, a quadratic behavior $\sff_{{\rm BG}}^{{\rm p.u.l.}}(\varphi_{q'} + \delta \varphi)|_{q =0} \simeq \frac{1}{4 q'} \frac{\delta \varphi^2}{1-q'} + O(\delta \varphi^3)$. The function $\sff_{{\rm BG}}^{{\rm p.u.l.}}(\varphi)|_{q=0}$ is plotted in black-dashed on the right of Fig.~\ref{fig:FEFPPandLPP} for $q'=0.7$.

\subsubsection{{\bf Inverse-Beta polymer}}

Let us now consider the IB polymer with boundaries defined in Def.~\ref{Def:IBbound}. $\forall (x_1 ,x_2) \in (\JN^*)^2$, we write,
\bea \label{decomp1}
\hat Z_{x_1,x_2} = \sum_{i=1}^{x_1}   \prod_{j=1}^i \hat{U}_{j,0}  v_{i,1} \hat Z^{i,1}_{x_1,x_2} + \sum_{i=1}^{x_2}   \prod_{j=1}^i \hat{V}_{0,j}  u_{1,i} \hat Z^{1,i}_{x_1,x_2} \ ,
\eea
where we have introduced $\forall (x_1 ,x_2  , x_1' , x_2' ) \in (\JN^*)^4$ with $x_1' \leq x_1 $ and $x_2'  \leq x_2$, the partition sum for polymers with starting point $(x_1',x_2')$ and $(x_1 , x_2) $, $\hat Z^{x_1',x_2'}_{x_1,x_2} = \sum_{\pi :(x_1',x_2') \to (x_1 , x_2) } \prod_{e \in \pi} \hat w (e)$. Since the Boltzmann weights taken into account in this partition sum are all bulk-type weights, we have the equality in law $\hat Z^{x_1',x_2'}_{x_1,x_2} \sim Z_{x_1-x_1',x_2-x_2'}$ where $ Z_{x_1,x_2}$ is the partition sum of the point to point IB polymer as defined in Def.~\ref{Def:ptopIB}. The decomposition in (\ref{decomp1}) expresses the partition sum  $\hat Z_{x_1,x_2}$ as a sum of $x_1+x_2$ positive terms. Hence we have the two inequalities
\bea \label{decomp2}
&&-\log \hat Z_{x_1,x_2} \geq   -  \log\left(  (x_1+x_2) {\rm max}\left\{ {\rm max}_{i \in [0,x_1] }  \prod_{j=1}^i \hat{U}_{0,j}  v_{1,i} \hat Z^{1,i}_{x_1,x_2} ,  {\rm max}_{i \in [0,x_2] } \prod_{j=1}^i \hat{V}_{j,0}  u_{i,1} \hat Z^{i,1}_{x_1,x_2} \right \} \right) \nn \\
&&  -\log \hat Z_{x_1,x_2}  \leq  -  \log \left(  {\rm max}\left\{ {\rm max}_{i \in [0,x_1] }  \prod_{j=1}^i \hat{U}_{0,j}  v_{1,i} \hat Z^{1,i}_{x_1,x_2} ,  {\rm max}_{i \in [0,x_2] } \prod_{j=1}^i \hat{V}_{j,0}  u_{i,1} \hat Z^{i,1}_{x_1,x_2} \right \} \right) \ .
\eea
Taking average values in (\ref{decomp2}), scaling $(x_1 , x_2) = N(s_1,s_2)$ and $ i  \sim N r$ with $N \gg 1$ and using the definitions (\ref{FEB}), (\ref{freeEnergyBW}) and (\ref{DefIntroFE}) we obtain
\bea \label{decomp3}
&& \hat f_{{\rm IB}}(s_1,s_2, \lambda) =s_1 f_{U}^{\gamma , \beta}(\lambda) + s_2 f_{V}^{\gamma , \beta}(\lambda)\nn \\
&&  = {\rm min}\left\{ {\rm inf}_{0 \leq r \leq s_1} ( r  f_{U}^{\gamma , \beta}(\lambda)+ f_{{\rm IB}}(s_1-r,s_2) )   ,  {\rm inf}_{0 \leq r \leq s_2} ( r  f_{V}^{\gamma , \beta}(\lambda)  + f_{{\rm IB}}(s_1,s_2-r)  ) \right\} \ .
\eea
Note that this equation has the exact same structure as the equation (\ref{T0decomp3}) relating the optimal energies in the Bernoulli-Geometric polymer with and without boundaries. Furthermore, the functions $ \hat f_{{\rm IB}}(s_1,s_2, \lambda)$, $f_{U}^{\gamma , \beta}(\lambda)$ and $ f_{V}^{\gamma , \beta}(\lambda) $ have similar analytical properties as a function of $\lambda$ than the mean optimal energies for the BG polymer (see (\ref{freeEnergyBW})). We can thus repeat the precedent derivation, and, using that $ \hat f_{{\rm IB}}(s_1,s_2, \lambda) $ is a concave function of $\lambda$ on $]0 , \gamma[$ with a unique maximum $\lambda^* \in ]0 , \gamma[$, we obtain for the point to point IB polymer, $\forall(s_1,s_2) \in \mathbb{R}_+^2$
\bea
&& f_{{\rm IB}}(s_1,s_2) = \hat f_{{\rm IB}}(s_1,s_2 , \lambda^*(s_1,s_2)) \nn \\
&& \partial_{\lambda} f_{{\rm IB}}(s_1,s_2 , \lambda)|_{\lambda =\lambda^*(s_1,s_2)} = 0  \ ,
\eea
with $\lambda^*(s_1,s_2) \in ]0 , \lambda[$. Using the formulas (\ref{freeEnergyBW}) and (\ref{FEB}) we obtain (\ref{FEresultBIS}). The derivation of the formula (\ref{EqOverviewOptimalAngles}) for the optimal angle $\varphi_{{\rm opt}}^{{\rm IB}}$ is identical to the BG polymer case.

\subsection{Convergence to the stationary measures} \label{subsec:EquivaAngBou}

We now discuss the conjectures (\ref{OverviewConvIB}) and (\ref{OverviewConvBG}) using heuristic arguments. We note that making rigorous and extending the picture discussed in this section is an active research area (focusing on the existence and characterization of so-called Busemann functions and stationary cocycles), see e.g. \cite{GeorgiouRassoulAghaSeppalainen} for problems of directed last passage percolation (including a discussion of the exactly solvable geometric case), \cite{DamronHanson} for undirected first passage percolation and \cite{Seppalainen2015} for the Log-Gamma polymer. We discuss the conjecture for the  BG polymer (\ref{OverviewConvBG}), the argument for the IB polymer being, at the level of rigor of this section, identical. In the following and until the end of the paper we will heavily use the notation $q_b^*(s_1,s_2)$ to denote the solution of the saddle-point equation (\ref{T0decomp10}), or alternatively the notation $q_b^*(\varphi)$ to denote the solution of the saddle-point equation (\ref{T0decomp101bis}).

\medskip

Let us thus again consider the optimal energy $\sE_{x_1,x_2}$ in the point to point BG polymer defined in Def.~\ref{Def:ptopBG}. Let us suppose that, given an arbitrary direction $(s_1,s_2) \in \mathbb{R}_+^2$ and fixing a total horizontal and vertical length $L_u \geq 1$ and $L_v \geq 1$, the difference of optimal energies in the rectangle delimited by the points $(N s_1 , N s_2) \to (N s_1 +L_u , N s_2)\to (N s_1 +L_u , N s_2 + L_v)\to (N s_1 , N s_2 + L_v)\to (N s_1 , N s_2 )$ converges to a well defined ensemble of RVs. That is
\bea \label{T0conv1}
(\sE_{Ns_1 + x_1 , Ns_2 +  x_2} - \sE_{Ns_1  , Ns_2 } )_{ 0 \leq x_1 \leq L_u , 0 \leq x_2 \leq L_v} \sim_{N \to \infty} (\tilde{\sE}_{x_1,x_2})_{ 0 \leq x_1 \leq L_u , 0 \leq x_2 \leq L_v} \ .
\eea
Where the $\tilde{\sE}_{x_1,x_2}$ are $O(1)$ RVs. It is clear that if the above convergence holds, the difference of horizontal and vertical energies $\tilde \sU_{x_1,x_2} := \tilde{\sE}_{x_1,x_2}-\tilde{\sE}_{x_1-1,x_2}$ and $\tilde \sV_{x_1,x_2} := \tilde{\sE}_{x_1,x_2}-\tilde{\sE}_{x_1,x_2-1}$ should be homogeneously distributed. In other words their distributions should be invariant by induction using the stationarity map $\phi_{T=0}$ (\ref{T0StationarityMap}). It is thus natural to identify the RVs $\tilde{\sE}_{x_1,x_2}$ with the optimal energies $\hat\sE_{x_1,x_2}$ in the BG polymer with boundaries. We however need to specify self-consistently the value of the boundary parameter $q_b$. To do so, let us evaluate the mean value $\overline{\tilde{\sE}_{x_1,x_2}}$ as
\bea \label{AngleFeEquiv1}
\overline{\tilde{\sE}_{x_1,x_2}} && = \overline{\sE_{Ns_1 + x_1 , Ns_2 +  x_2} }  - \overline{  \sE_{Ns_1  , Ns_2 }  }  \nn \\
&& \simeq N \sff_{{\rm BG}}( s_1 + x_1/N , s_2 + x_2/N) - N \sff_{{\rm BG}}( s_1 , s_2 ) \nn \\
&& \simeq x_1 \partial_1 \sff_{{\rm BG}}( s_1 , s_2) + x_2 \partial_2 \sff_{{\rm BG}}( s_1 , s_2)  \nn \\
&& \simeq  x_1 \sff_{\sU} ^{q,q' }(q_b^*(s_1,s_2)) + x_2  \sff_{\sV} ^{q,q' }(q_b^*(s_1,s_2))  \nn \\
&& \simeq  \overline{\hat{\sE}_{x_1,x_2}} \text{ if } q_b =q_b^*(s_1,s_2) \ ,
\eea
where we used the definition (\ref{DefIntroFE}), the result (\ref{T0decomp10}), the saddle-point equation in (\ref{T0decomp10}) to compute the derivatives $\partial_i \sff$ and (\ref{T0meanEnergy}). This calculation thus suggests that we have the equality in law, already given in (\ref{OverviewConvBG})
\bea \label{AngleFeEquiv2}
(\sE_{Ns_1 + x_1 , Ns_2 +  x_2} - \sE_{Ns_1  , Ns_2 } )_{ 0 \leq x_1 \leq L_u , 0 \leq x_2 \leq L_v} \sim_{N \to \infty}  (\hat{\sE}_{x_1,x_2})_{ 0 \leq x_1 \leq L_u , 0 \leq x_2 \leq L_v}   \text{ with } q_b =q_b^*(s_1,s_2) \ .
\eea
That is we relate the differences of energies in a specific direction at large length in the model without boundaries with the optimal energy in the model with boundaries with a specific boundary parameter. Note that this result fails if one starts to scale the length of the rectangles with $N$. This is obvious if one scales $L_u \sim N$, but the result is also expected to fail for the smaller scaling $L_u \sim N^{\frac{2}{3}}$. Indeed the exponent $2/3$ is the known rugosity exponent of directed polymer in $d=1+1$ and should correspond to the typical scale at which correlations between energy differences appear.

\medskip

\begin{figure}
\centerline{\includegraphics[width=7.5cm]{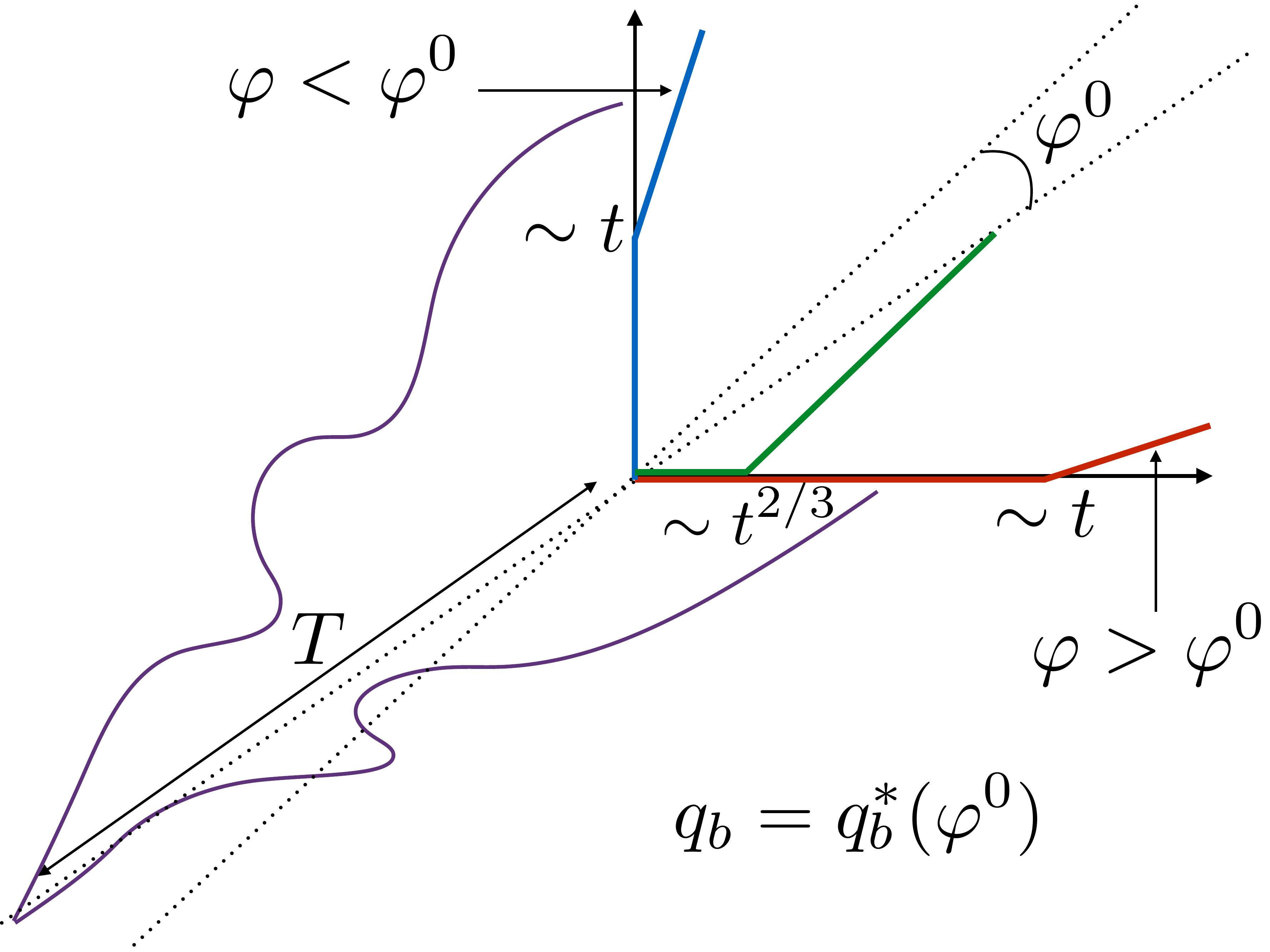} } 
\caption{Convergence to the stationary measure and equivalence between direction and stationarity parameter $q_b$ in the Bernoulli-Geometric polymer. The optimal energies in the model with boundaries with parameter $q_b$ are interpreted as differences of optimal energies for the point to point BG polymer (purple paths above) starting from infinity $T \gg 1$ in the direction $\varphi^0$ such that $q_b = q_b^*(\varphi^0)$. Optimal path properties in the BG polymer with boundaries strongly depend on the observed direction. For $\varphi > \varphi^0$ (resp. $\varphi < \varphi^0$) the optimal polymer path sticks to the vertical boundary (blue path) (resp. horizontal boundary (red path)) for a macroscopic amount of time $\sim t$. In the direction $\varphi = \varphi^0$, we conjecture following the work of Sepp\"{a}l\"{a}inen \cite{Seppalainen2012} that the optimal polymer path stays on the boundary for a time of order $t^{2/3}$ only (green path).}
\label{fig:EquivalenceAngleBoundary}
\end{figure}

Let us now reinterpret following this picture some properties of the IB model with boundaries. Following the convergence in law (\ref{AngleFeEquiv2}), the optimal energies $\overline{\hat{\sE}_{t}(x)}$ in the model with boundaries with parameter $q_b = q_b^*(\varphi^0)$ with $\varphi^0\in]-1/2,1/2[$, are thus interpreted as the difference of energies in the model without boundaries when the polymer starts from infinity in the direction with angle $\varphi^0$ (see Fig.~\ref{fig:EquivalenceAngleBoundary}). In this interpretation the linear dependence of the free-energy per-unit-length in the model with boundaries $\hat \sff_{{\rm BG}}^{{\rm p.u.l.}}(\varphi , q_b = q_b^*(\varphi^0) )$ as a function of $\varphi$ is natural since the model with boundaries is obtained by `zooming in' on a specific region of the model without boundaries in the direction $\varphi = \varphi^0$. One easily checks using calculations similar to those of (\ref{AngleFeEquiv1}) that $\hat \sff_{{\rm BG}}^{{\rm p.u.l.}}(\varphi , q_b =q_b^*(\varphi^0)) =  \sff_{{\rm BG}}^{{\rm p.u.l.}}(\varphi^0)+ (\varphi - \varphi^0) \partial_{\varphi} \sff_{{\rm BG}}^{{\rm p.u.l.}}(\varphi) |_{\varphi = \varphi^0} $. In particular, as we already saw, the direction of optimal energy for the model without boundaries $\varphi^0 = \varphi^{{\rm BG}}_{{\rm opt}}$ (such that $\partial_{\varphi} \sff_{{\rm BG}}^{{\rm p.u.l.}}(\varphi) |_{\varphi = \varphi^{{\rm BG}}_{{\rm opt}}} =0 $) corresponds to the equilibrium boundary parameter $q_b = \sqrt{q}$ for which the optimal energy in the model with boundaries $\hat \sff_{{\rm BG}}^{{\rm p.u.l.}}(\varphi , q_b)$ is constant: there the model with boundaries is obtained by `zooming in' on the region of optimal energy of the model without boundaries.

\subsection{{\bf A remark on optimal paths and energy fluctuations in models with boundaries}} \label{subsec:Fluctuations}

Let us now briefly discuss some asymptotic properties of the model with boundaries. For concreteness we will consider the Bernoulli-Geometric polymer but the discussion can be easily adapted to the Inverse-Beta case. We suppose that the boundary parameter $q_b$ corresponds to a direction $\varphi^0$ for which $q_b = q_b^*(\varphi^0)$, the solution of the saddle-point-equation (\ref{T0decomp101bis}). As we saw before in (\ref{AngleFeEquiv2}) the optimal energies $\hat{\sE}_{x_1,x_2} = \hat{\sE}_{t=(x_1+x_2)}(x=x_1)$ in the model with boundaries on a finite domain are naturally interpreted as the asymptotic limit of the difference of energies of the model without boundaries in the direction $\varphi^0$. That is, for $T \gg 1$ and $(t,x) \in \mathbb{N}^2$ fixed we have
\bea \label{eqinlaw10}
\hat \sE_t(x) \sim \sE_{T+t}((1/2 + \varphi_0) T + x) - \sE_{T}((1/2 + \varphi_0) T ) \ .
\eea
As such, asymptotic properties of optimal energies of the model with boundaries in a direction $\varphi$, $\hat \sE_t(x=(1/2 + \varphi)t)$ with $t \gg 1$, are to be interpreted with caution for the model without boundaries since (\ref{eqinlaw10}) is a priori valid only for $t$ fixed and $T \to \infty$. With this in mind, let us now discuss the properties of the energy fluctuations and of the optimal path in the model with boundaries. 

\medskip

Let us first comment on some elements that appeared in the proof of (\ref{T0decomp10}) in Sec.~\ref{subsec:proof}. There we defined, for a fixed direction $(s_1,s_2)=(1/2 + \varphi , 1/2- \varphi)$ and varying the boundary parameter $q_b$, two functions $r_1^{\varphi}(q_b)$ and $r_2^{\varphi}(q_b)$ (here we emphasize the dependence on $\varphi$ of these quantities by superscript). These two functions correspond to the average length divided by $t$ spend by the optimal polymer path on the horizontal (for $r_1^{\varphi}(q_b)$) or vertical (for $r_2^{\varphi}(q_b)$) boundary of $\mathbb{N}^2$. Hence for $q_b< q_b^*(\varphi)$ we saw that the polymer spends on average a macroscopic amount of his time ($\sim r_1^{\varphi}(q_b) t$) on the horizontal boundary, while for $q_b > q_b^*(\varphi)$ the polymer spends on average a macroscopic amount of his time ($\sim  r_2^{\varphi}(q_b) t$) on the vertical boundary.

\smallskip

Conversely, fixing now $q_b = q_b^*(\varphi^0)$ for some $\varphi^0 \in ]-1/2,1/2[$, and varying $\varphi$, for $\varphi > \varphi^0$ (resp. $\varphi < \varphi^0$), the optimal polymer path spends on average a macroscopic amount of time $ \sim \tilde r_1^{\varphi_0}(\varphi) t$ (resp. $ \tilde r_2^{\varphi_0}(\varphi) t$) on the horizontal (resp. vertical) boundary with $ \tilde r_1^{\varphi_0}(\varphi) = r_1^{\varphi}(q_b =  q_b^*(\varphi^0))$  (resp.  $ \tilde r_2^{\varphi_0}(\varphi) = r_2^{\varphi}(q_b =  q_b^*(\varphi^0))$). Hence, for $\varphi > \varphi_0$ (resp. $\varphi < \varphi_0$), the optimal energy $\hat \sE_t(x=(1/2 + \varphi)t)$ contains a sum of order $t$ terms of iid distributed RVs of the $\sU$ type (resp. of the $ \sV$ type) and one thus expects the fluctuations of $\hat \sE_t(x=(1/2 + \varphi)t)$ to scale as $\sqrt{t}$. Thus, in any direction $\varphi \neq \varphi^0$, one does not observe fluctuations of order $t^{1/3}$ as could have naively been expected from KPZ universality, the reason being that the polymer is then typically pinned by one of the two attractive boundaries for a macroscopic (i.e. of order $t$) amount of time.

\medskip

An important question is then to understand how the fluctuations of $\hat \sE_t(x=(1/2 + \varphi)t)$ scale with $t$ when $\varphi = \varphi^0$. In \cite{Seppalainen2012} Sepp\"{a}l\"{a}inen showed in the Log-Gamma polymer case that these fluctuations scale with the characteristic exponent $t^{1/3}$ as expected from KPZ universality, and a typical polymer path then only spend a time of order $t^{2/3}$ on one of the two boundaries. It is likely that the arguments presented in \cite{Seppalainen2012} could be adapted to our models. We will not prove it here and continue by assuming that the fluctuations of $\hat \sE_t(x=(1/2 + \varphi)t)$ are of order $t^{1/3}$ when $\varphi = \varphi^0$.

\medskip

Following the above remarks, it is clear that the direction $\varphi = \varphi^0$ is a special direction for the model with boundaries. It is the only direction for which the fluctuations of the optimal energy scale with the expected $t^{1/3}$ exponent. Moreover, the direction $\varphi = \varphi^0$ is the only direction for which the mean optimal energy in the model with boundaries $\hat \sff_{{\rm BG}}^{{\rm p.u.l.}}(\varphi^0 , q_b^*(\varphi^0))$ coincide with the mean optimal energy of the underlying model without boundaries $\sff_{{\rm BG}}^{{\rm p.u.l.}}(\varphi^0)$. From these two facts, it appears reasonable to conjecture that the equivalence in law (\ref{eqinlaw10}), which a-priori only holds for $t , x \ll T$, also holds for $t=O(T)$ with $x =(1/2 + \varphi^0)t + \hat x$ with $\hat x = O(1)$. That is, asymptotic properties of the model with boundaries with parameter $q_b = q_b^*(\varphi^0)$ reproduce those of the model without boundaries if one looks in the characteristic direction $\varphi = \varphi^0$.

\medskip

On the other hand for directions $\varphi \neq \varphi^0$ it is clear that asymptotic properties of the model with boundaries cannot be interpreted in the model without boundaries and the equivalence in law (\ref{eqinlaw10}) does not hold anymore. An example of such properties is as follows. In a direction $\varphi > \varphi^0$ we have
\bea
\hat \sE_t((1/2 + \varphi) t)  && = \hat \sE_t((1/2 + \varphi) t)-\hat \sE_t((1/2 + \varphi^0) t) + \hat \sE_t((1/2 + \varphi^0) t) \nn \\
&& = \sum_{ x = (1/2 + \varphi^0) t}^{ x = (1/2 + \varphi) t} \left(  \hat \sU_t(x) - \hat \sV_t(x) \right) + \hat \sE_t((1/2 + \varphi^0) t)  \ .
\eea
Subtracting the average values over disorder in the above equation, one gets that $\hat \sE_t((1/2 + \varphi) t)  - t \hat \sff^{{\rm p.u.l.}}_{{\rm BG}} (\varphi , q_b)$ is the sum of $t(\varphi - \varphi^0)$ independent centered RVs (note that these RVs indeed live on a down-right path $\pi_{dr}^{(t)}$) distributed as $\sU - \sV  -\overline{\sU} + \overline{\sV}$, and of another centered term $\hat \sE_t((1/2 + \varphi^0) t)  - t \hat \sff^{{\rm p.u.l.}}_{{\rm BG}}(\varphi^0 , q_b)$ whose fluctuations scale as $t^{1/3}$ (admitting the above discussion). Hence it is then clear that in the large time limit the fluctuations of $\hat \sE_t((1/2 + \varphi) t)$ are Gaussian distributed and we have the convergence in law
\bea
\frac{\hat \sE_t((1/2 + \varphi) t)  - t \hat \sff^{{\rm p.u.l.}}_{{\rm BG}}(\varphi , q_b)}{\sigma(\varphi) \sqrt{t}} \sim_{t \to \infty} \chi_{{\cal N}(0 , 1) }  \  ,
\eea
where here, $q_b = q_b^{*}(\varphi^0)$, $\sigma(\varphi) = \sqrt{ (\varphi - \varphi^0) \left(  \overline{\sU^2}^c + \overline{\sV^2}^c \right) } $ and $\chi{_{\cal N}(0 , 1)}$ is a RV distributed with a standard unit centered normal distribution.

\section{Numerical results for the zero-temperature model} \label{Sec:Num}

In this section we report results of numerical simulations of the point to point Bernoulli-Geometric polymer (without boundaries, see Def.~\ref{Def:ptopBG}) for three sets of parameters. For each set we have $q =0.5$ and we vary the anisotropy parameter: we consider an almost isotropic case $q'_1 =0.1$ and two strongly anisotropic cases $q'_2 = 0.8$ and $q'_3 = 0.9$. For each set we perform $2\times 10^5$ simulations of independent random environments of size $2048 \times 2048$. For each random environment we measure using a transfer matrix algorithm the optimal energy and horizontal and vertical energy differences $\sE_{t_j}((1/2 + \varphi_k)t_j )$, $\sU_{t_j}((1/2 + \varphi_k)t_j ) := \sE_{t_j}((1/2 + \varphi_k)t_j ) -\sE_{t_j-1}((1/2 + \varphi_k)t_j -1) $ and $\sV_{t_j}((1/2 + \varphi_k)t_j ):= \sE_{t_j}((1/2 + \varphi_k)t_j ) -\sE_{t_j-1}((1/2 + \varphi_k)t_j)$ for different times $t_j = 2^{ j +5}$ with $j =1 , \dots , 6$ (hence $t_1 = 64$ and $t_6 = 2048$) and different angle parameters $\varphi_k = - 0.4 + \frac{k-1}{10}$ with $k = 1 , \dots ,9$.

\medskip

\begin{figure}
\centerline{\includegraphics[width=6cm]{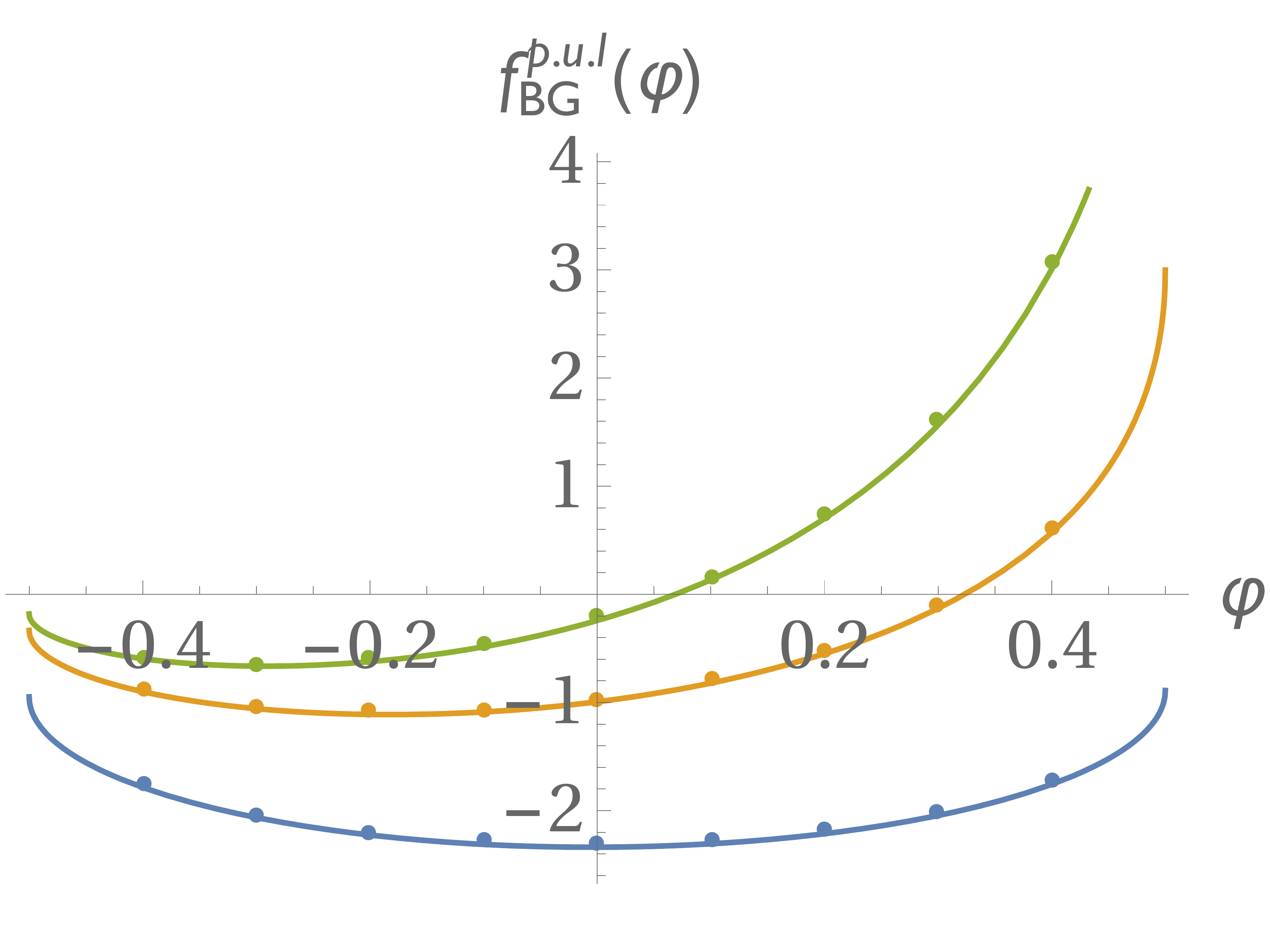} } 
\caption{Comparison between the exact result for the mean asymptotic optimal energy per-unit-length in the Bernoulli-Geometric polymer $\sff_{{\rm BG}}^{{\rm p.u.l.}}(\varphi)$ (see (\ref{T0decomp101})) for $q=0.5$ and $q' = q_1 = 0.1$ (blue line), $q' = q_2 = 0.8$ (orange line) and $q' = q_3= 0.9$ (green line) and the numerically obtained value $\overline{\sE_{t_6}((1/2 + \varphi_k)t_6 ) }/t_6$ for each set of parameters and each angle $\varphi_k$ (dots, same color code).}
\label{fig:numFE}
\end{figure}

We first compare in Fig.\ref{fig:numFE} for each set of parameters our exact result for the asymptotic value of the mean optimal energy per-unit-length (\ref{T0decomp101}) with the numerically obtained value $\overline{\sE_{t_j}((1/2 + \varphi_k)t_j ) }/t_j$ for $j=6$ (i.e. polymers of length $t=t_6 = 2048$ for each set of parameters and each angle $\varphi_k$. We obtain an excellent agreement.

\medskip

\begin{figure}
\centerline{\includegraphics[width=5.7cm]{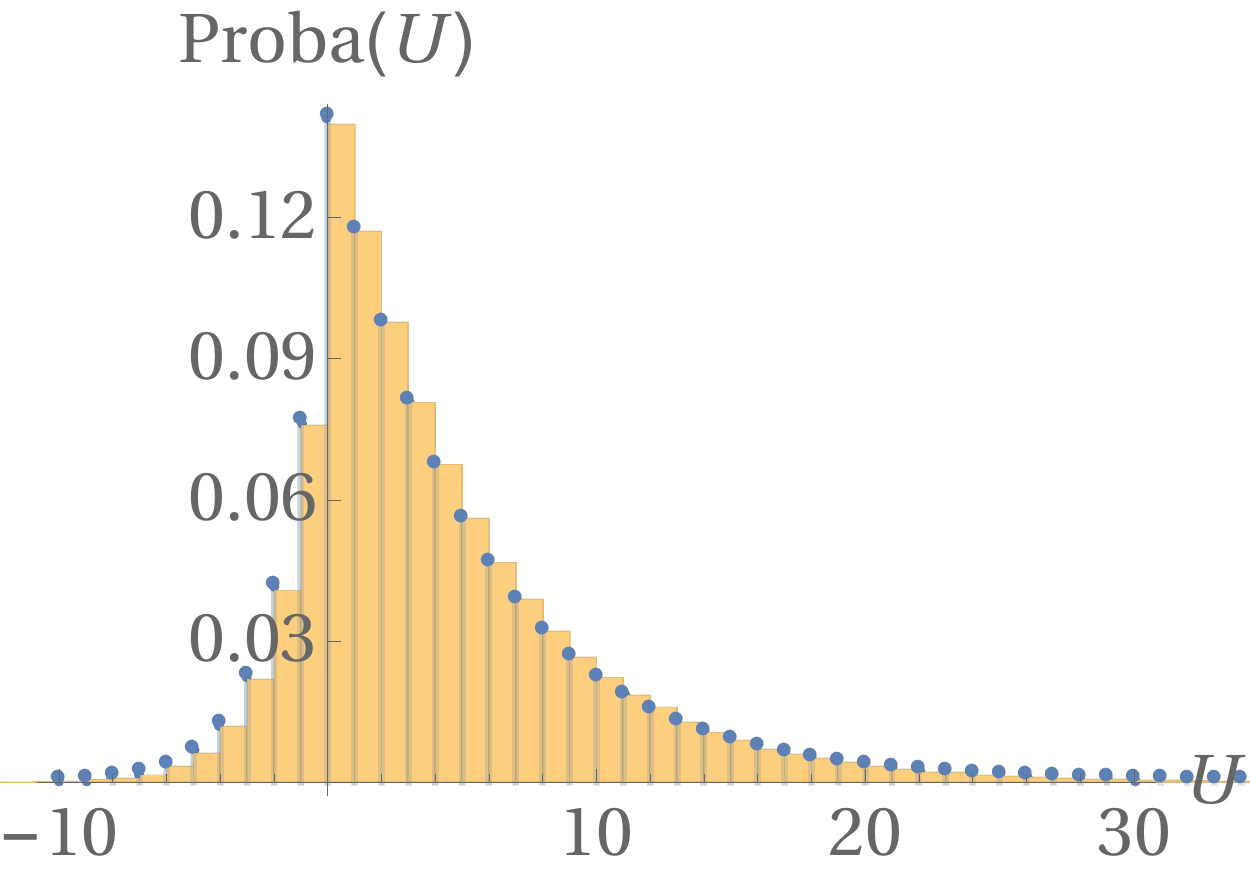} \includegraphics[width=5.7cm]{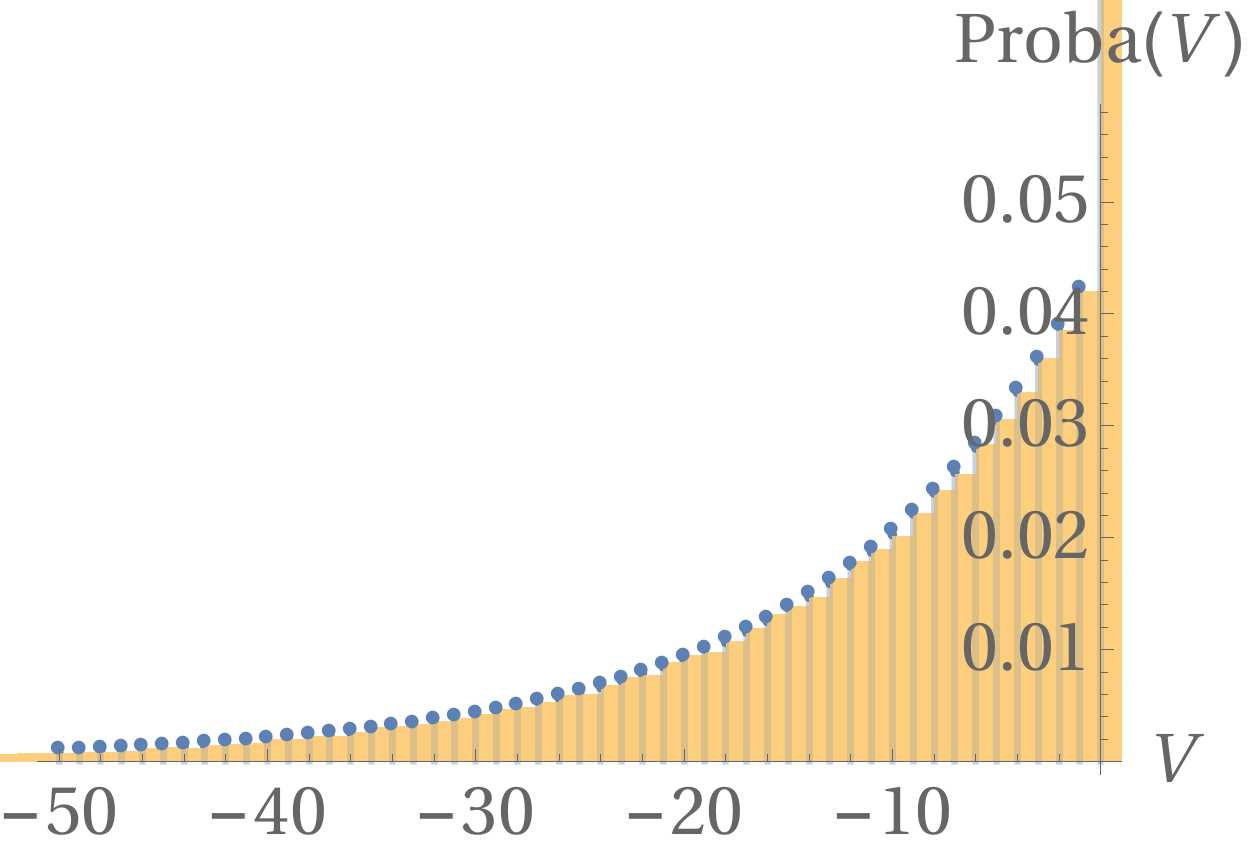} \includegraphics[width=5.7cm]{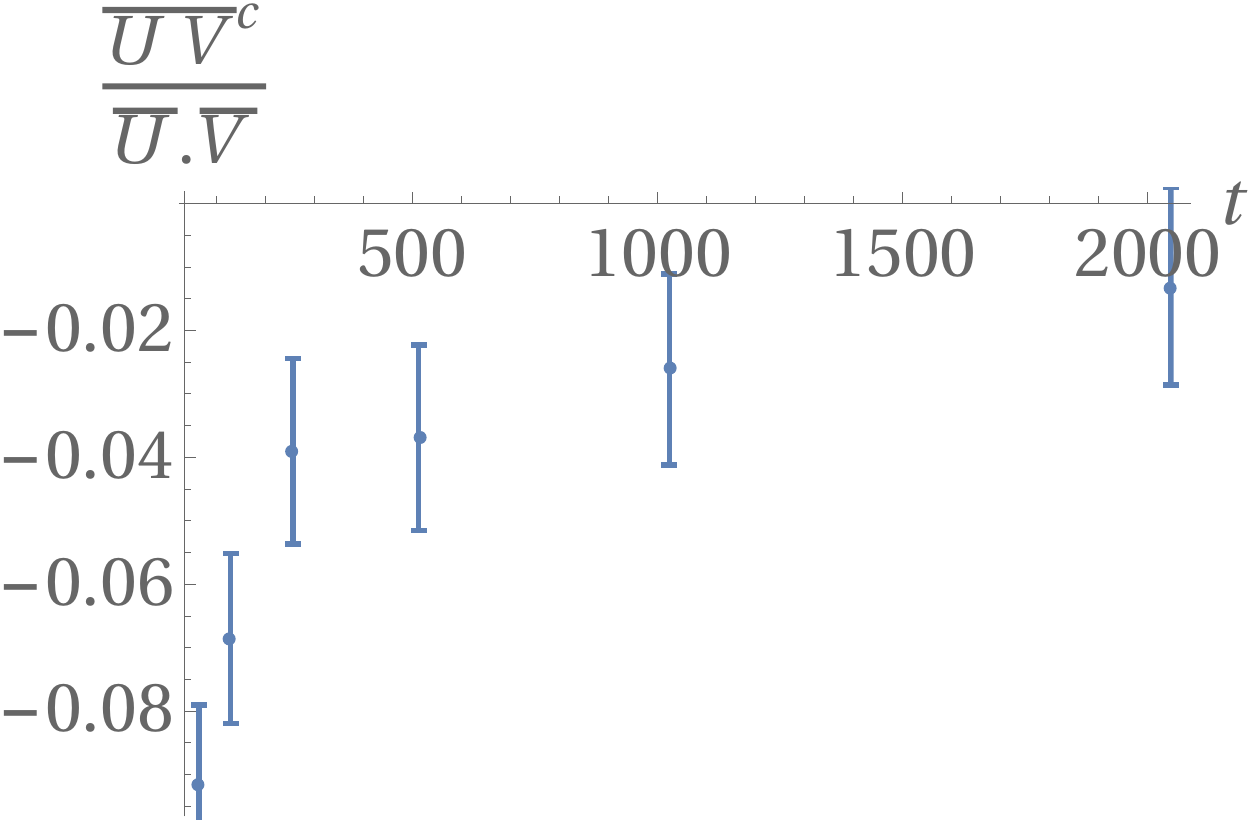} } 
\caption{Left: (resp. Middle:) Comparison between the numerically obtained PDF of $\sU_{t_6}((1/2 + \varphi_8)t_6)$ (resp. $\sV_{t_6}((1/2 + \varphi_8)t_6)$) in the simulations with parameters $q=0.5$, $q' = 0.9$ (yellow histogram) and the PDF of $\sU$ (resp. $\sV$) given in (\ref{T0DistEnergyUV2}) with $q_b=q_b^* \simeq 0.922824$ (blue dots). Right: numerically obtained normalized covariance $\frac{\overline{\sU_{t_j} ((1/2 + \varphi_8)t_j) \sV_{t_j} ((1/2 + \varphi_8)t_j)}^c}{\overline{\sU_{t_j} ((1/2 + \varphi_8)t_j)}  \times \overline{\sV_{t_j} ((1/2 + \varphi_8)t_j)}}$ in the simulations with parameters $q=0.5$, $q' = 0.9$ as a function of $t_j$ (blue dots). Error bars are $3-$sigma Gaussian estimates.}
\label{fig:num28}
\end{figure}

\begin{figure}
\centerline{\includegraphics[width=5.7cm]{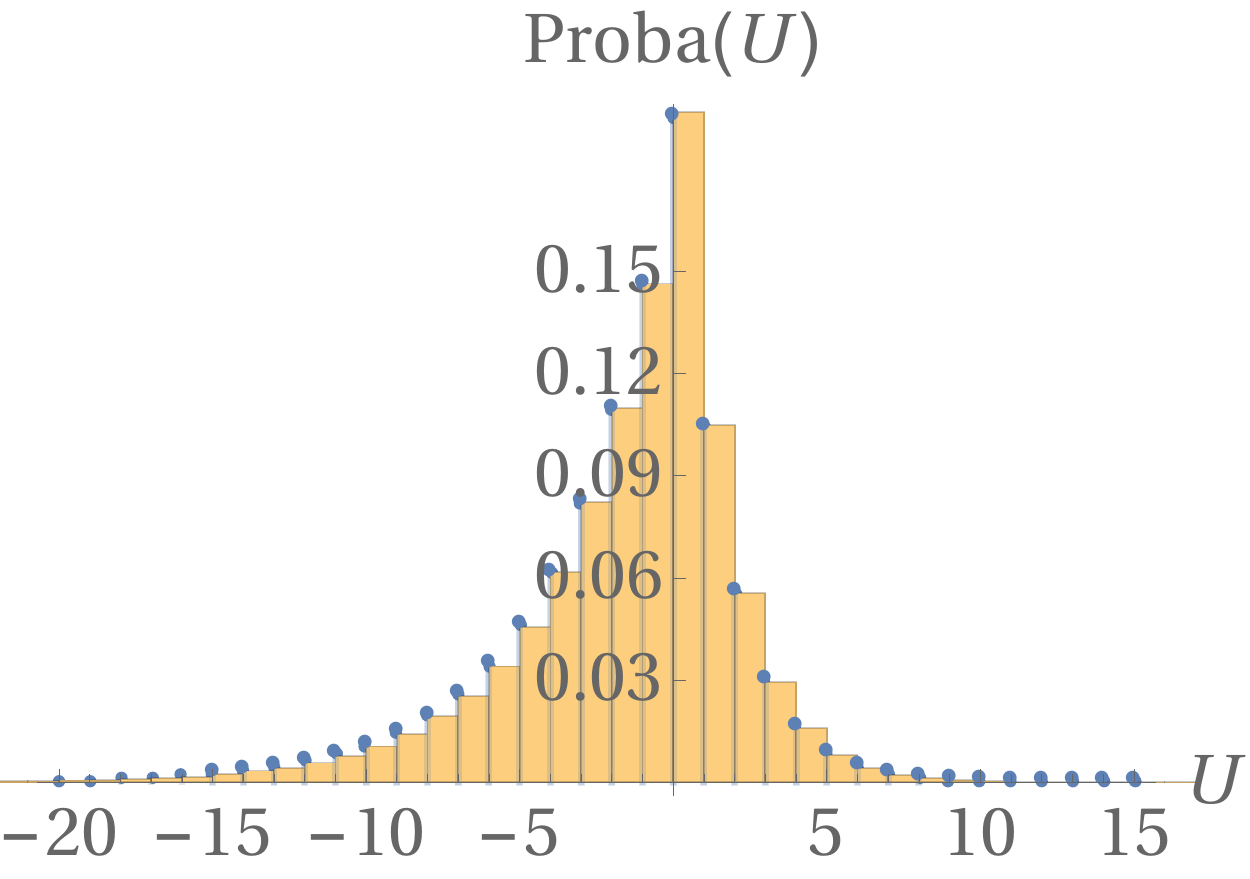} \includegraphics[width=5.7cm]{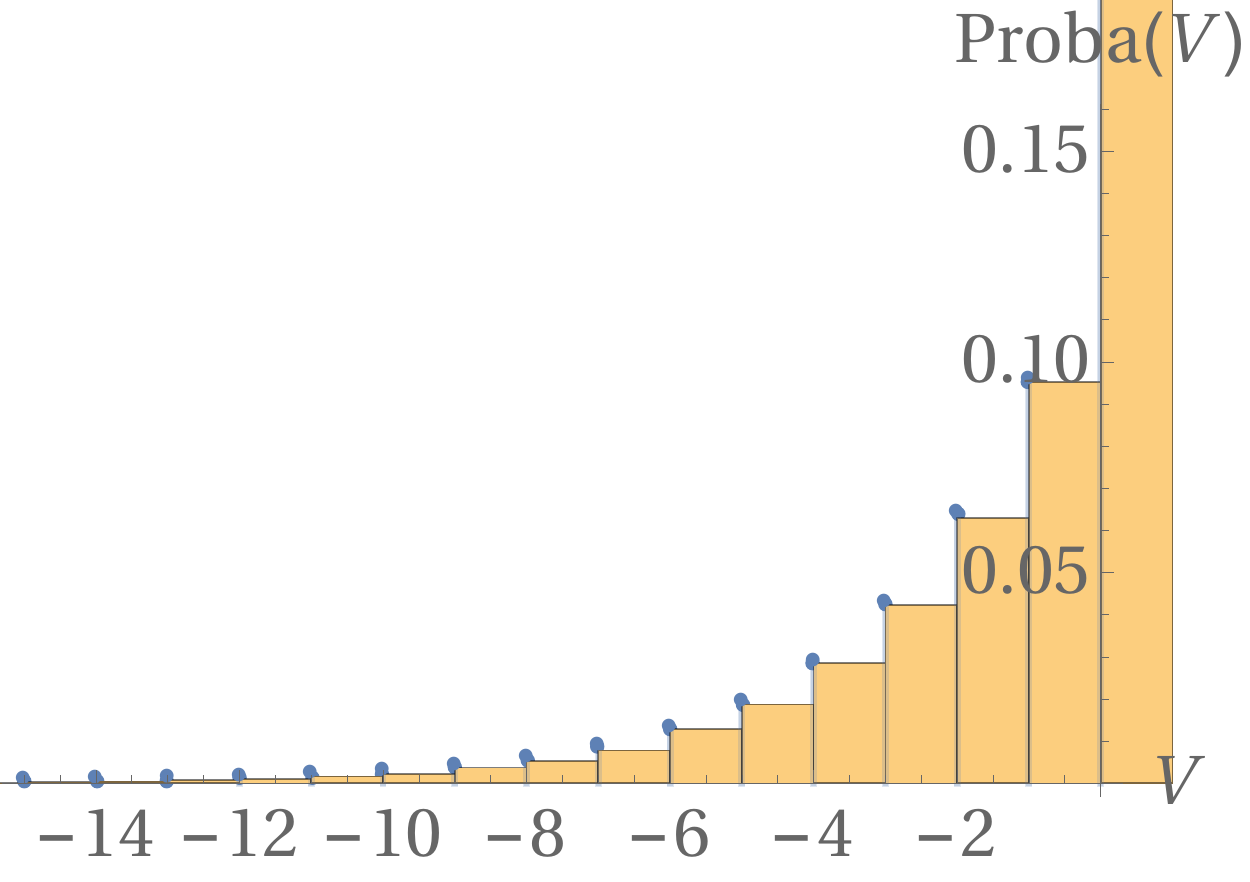} \includegraphics[width=5.7cm]{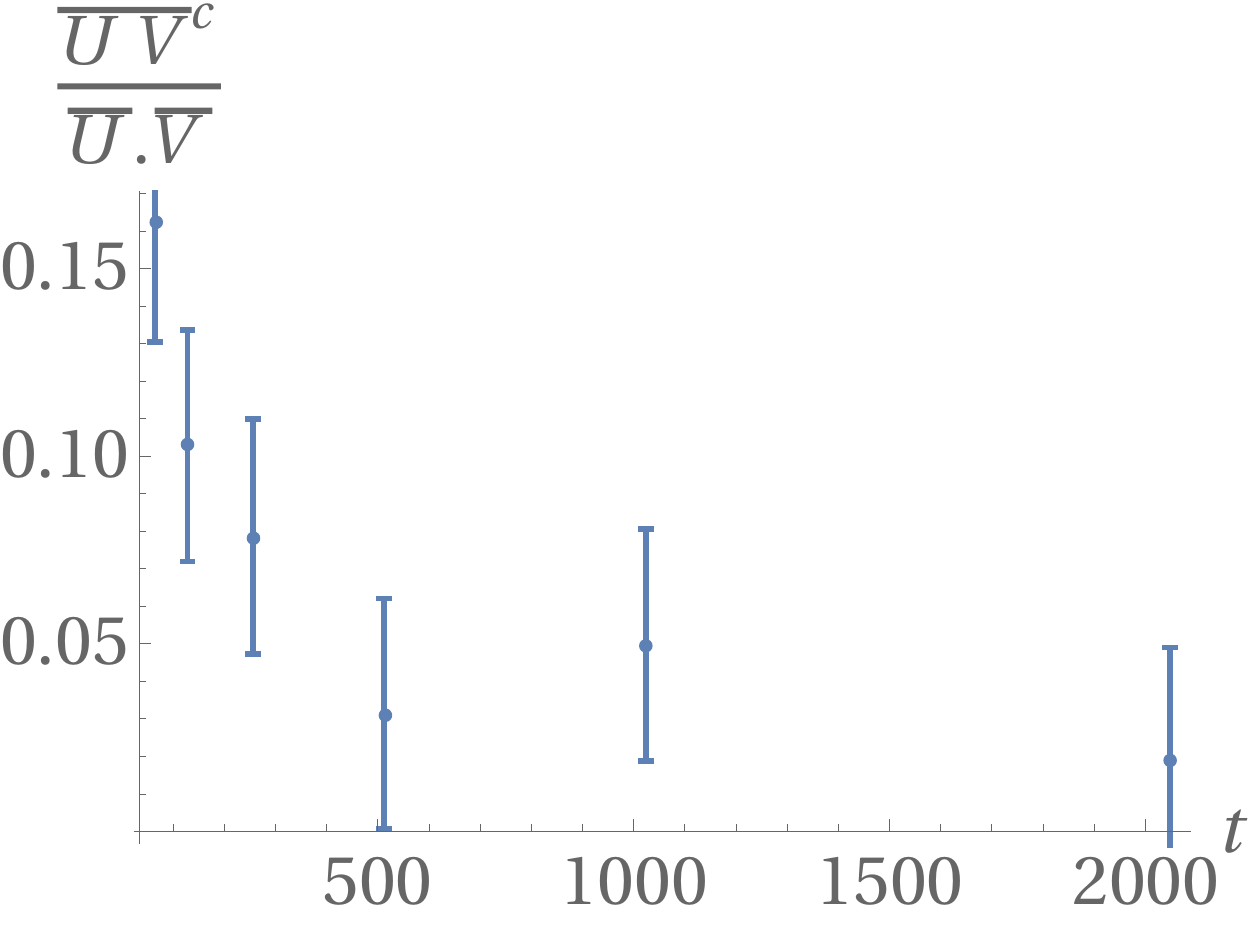} } 
\caption{Left: (resp. Middle:) Comparison between the numerically obtained PDF of $\sU_{t_6}((1/2 + \varphi_2)t_6)$ (resp. $\sV_{t_6}((1/2 + \varphi_2)t_6)$) in the simulations with parameters $q=0.5$, $q' = 0.8$ (yellow histogram) and the PDF of $\sU$ (resp. $\sV$) given in (\ref{T0DistEnergyUV2}) with $q_b=q_b^* \simeq 0.667665$ (blue dots). Right: numerically obtained normalized covariance $\frac{\overline{\sU_{t_j} ((1/2 + \varphi_2)t_j) \sV_{t_j} ((1/2 + \varphi_2)t_j)}^c}{\overline{\sU_{t_j} ((1/2 + \varphi_2)t_j)}  \times \overline{\sV_{t_j} ((1/2 + \varphi_2)t_j)}}$ in the simulations with parameters $q=0.5$, $q' = 0.8$ as a function of $t_j$ (blue dots). Error bars are $3-$sigma Gaussian estimates.}
\label{fig:num12}
\end{figure}

We then check our conjecture (\ref{OverviewConvBG}). The latter notably implies, combined with Prop.~\ref{prop:StatIBbound}, that the differences of horizontal and vertical energies in a given direction, $\sU_{t_j}((1/2 + \varphi_k)t_j )$ and $\sV_{t_j}((1/2 + \varphi_k)t_j )$, converge to independent random variables distributed as $\sU$ and $\sV$ in (\ref{T0DistEnergyUV2}), with for each $\varphi_k$ the boundary parameter $q_b$ chosen as $q_b = q_b^*(\varphi_k)$, the solution of the saddle-point equation (\ref{T0decomp11BISii}) with $(s_1,s_2) = (1/2-\varphi_k , 1/2+\varphi_k)$. In Fig.\ref{fig:num28} we analyze the numerical results for the set of parameters with $q' = q'_3 = 0.9$. We first obtain numerically the PDF of horizontal and vertical differences of optimal energies $\sU_{t_6}((1/2 + \varphi_8)t_6)$, $\sV_{t_6}((1/2 + \varphi_8)t_6)$ for an angle $\varphi = \varphi_8 = 0.3$ and polymers of length $t=t_6 = 2048$, and compare it with our asymptotic prediction (\ref{T0DistEnergyUV2}) (the appropriate boundary parameter is there found to be $q_b^* \simeq 0.922824$). We obtain an excellent agreement. To check the independence of the RVs, we estimate numerically the normalized covariance $\frac{\overline{\sU_{t_j} ((1/2 + \varphi_8)t_j) \sV_{t_j} ((1/2 + \varphi_8)t_j)}^c}{\overline{\sU_{t_j} ((1/2 + \varphi_8)t_j)}  \times \overline{\sV_{t_j} ((1/2 + \varphi_8)t_j)}}$ and study its behavior as a function of $t$. Although fluctuations are large, the normalized covariance clearly decays to $0$ with increasing $t$, a signature of the independence of the RVs. In Fig.\ref{fig:num12} we report similarly satisfying results for the set of parameters with $q' = 0.8$ and in the direction $\varphi = \varphi_2 = -0.3$ (there the appropriate boundary parameter is found to be $q_b^* \simeq 0.667665$).

\section{Conclusion}

In this paper we have obtained the stationary measure of the Inverse-Beta polymer, an exactly solvable, anisotropic finite temperature model of DP on the square lattice recently introduced in \cite{usIBeta}. As we discussed, the stationary model can be either studied on $\mathbb{Z}^2$ with a random initial condition for the polymer such that the free-energy of the DP performs a random walk with inverse-beta distributed increments, or also conveniently in a model on the upper-right quadrant $\mathbb{N}^2$ with special boundary conditions. This thus confers to the IB polymer a second exact solvability property that complements the coordinate Bethe ansatz solvability shown in \cite{usIBeta}.

\smallskip

In parallel we introduced a new model of zero temperature DP on the square lattice, the Bernoulli-Geometric polymer. It is obtained by appropriately discretizing the distributions of random energies of the zero temperature limit of the Inverse-Beta polymer from Bernoulli-Exponential distributions to Bernoulli-Geometric distributions. This model is thus canonically dual to the IB polymer. In two different limits the model becomes either a first passage percolation problem, or a last passage percolation problem. We showed that its stationary measure could be exactly obtained, thus conferring to this new model at least one exact solvability property.

\smallskip

We also showed that the two stationary measures are reversible and satisfy detailed balance. We obtained the mean quenched free-energy (resp. optimal energy) in the IB (resp. BG) polymer. For the IB polymer, the obtained result (\ref{FEresultBIS}) coincides with a previously obtained result of \cite{usIBeta}, therefore confirming the validity of the non-rigorous approach of \cite{usIBeta}, while in the BG polymer case, (\ref{T0decomp11BIS}) is genuinely new. In both cases, these results allowed us to discuss the convergence of each model to their stationary measure. Finally in Sec.~\ref{Sec:Num} we reported the results of numerical simulations of the BG polymer and compared them with a very good agreement to our results.

\medskip

Many possible research directions remain for the future. One interesting direction would be to understand if the models studied in this paper possess other exact solvability properties. Indeed for both models it is not clear whether or not combinatorial mappings similar to RSK and gRSK correspondences could be developed, although they both interpolate between models for which these correspondences can be applied (gRSK at finite temperature \cite{logsep2,StrictWeak2} and RSK at $0$ temperature \cite{Johansson2000,OConnell2005}). The question of the Bethe ansatz solvability of these models is also interesting. For the IB polymer it was shown in \cite{usIBeta} that the moment problem is exactly solvable by coordinate BA but another BA solvability could exist. Indeed in the Log-Gamma case it was shown in \cite{usLogGamma} that the moment problem was BA solvable, but is is also known that the partition sum of the Log-Gamma polymer can be interpreted \cite{PetrovMatveev2015} as an observable of a BA solvable interacting particles system on $\mathbb{Z}$, the q-Push TASEP \cite{PetrovBorodin2013,PetrovCorwin2015}. The same is true for the Strict-Weak polymer which can be mapped onto an observable of the q-TASEP \cite{StrictWeak1}. Exhibiting a similar mapping for the IB polymer case remains an open question. For the BG polymer introduced in this paper the question of BA solvability is also open. We note that in the isotropic limit of the model, i.e. last passage percolation with geometric weights, the optimal energy can be interpreted as waiting times of the TASEP with geometric waiting times and step initial condition, which can be solved by BA. Furthermore, we note that a version of the q-TASEP with Bernoulli and Geometric waiting times was already considered in \cite{BorodinCorwinDiscretTimeQ} where the authors notably proved BA solvability. Although it is not clear how to map the optimal energies in the BG polymer to the waiting times of an interacting particles system (since the random energies can be both positive and negative), this could be an interesting approach.

\smallskip

Another interesting direction of research would be to understand how to obtain more systematically models of directed polymers with exact solvability properties and how to classify them. For the case of Bethe ansatz solvability of the moment problem for models of DPs at finite temperature, this was already mostly accomplished in \cite{usIBeta}. The question remains open for BA solvability of models at zero temperature and for other type of exact solvability properties such as the possibility of writing down the stationary measure exactly. If this was accomplished it would be interesting to see whether or not the two classes coincide. We note that in the related context of zero-range-processes (ZRP) with simultaneous updates, it was recently shown that all BA solvable models have factorizable steady-states, but the converse is not true \cite{povolotsky2013integrability}. For the DP case, a step in this direction was already made since in \cite{Seppalainen2012} it was shown that the Log-Gamma was the unique model at finite temperature with {\it on site} disorder for which it is possible to write down exactly the SM, and the Log-Gamma also appeared as the unique finite temperature model with on site disorder exactly solvable by BA in the classification of \cite{usIBeta}. More generally it would be interesting to gain a better understanding of the links between different exact solvability properties.

\acknowledgments

This paper would have never existed without the numerous discussions I had with Timo Sepp\"{a}l\"{a}inen, discussions during which he kindly took the time to explain to me the techniques and results developed and obtained by him and his coworkers for the Log-Gamma polymer. These were a great source of inspiration for this work. He also took an active part during the first stages of research on the stationary measure of the Inverse-Beta polymer and shared with me related new results on the Beta polymer \cite{SeppalainenInprep}. I warmly thank him for that. I am also grateful to Guillaume Barraquand for many discussions and remarks on the existing mathematical literature, as well as to Francis Comets, Ivan Corwin, Thomas Gueudr\'e, Vivien Lecomte, Jeremy Quastel and Leonid Petrov for interesting discussions. Last but not least, I would like to warmly thank Pierre Le Doussal who introduced me and taught me most of the things I know on this topic through multiple discussions and collaborations on related subjects. I also thank him for useful comments on a first version of this manuscript. I acknowledge the KITP in Santa Barbara for hospitality during the first stages of redaction of this work. This research was supported in part by the National Science Foundation under Grant No. NSF PHY11-25915.

\appendix

\section{Proof of the properties of the finite temperature reversibility-stationarity map}\label{app:Stationarity}
In this appendix we prove Prop.~\ref{Prop:stationarityMap:Stationarity}, Prop.~\ref{Prop:stationarityMap:Reversibility} being trivial. We thus consider three independent random variables $(U,V,W)$ distributed as in (\ref{Statio-Inv-Beta2}) and (\ref{Statio-Inv-Beta3}) and consider the RVs $(U',V',W') = \phi(U,V,W)$ as given in (\ref{stationarityMap}). The Jacobian of the transformation $(U,V,W) \to (U',V',W')$ is easily computed as, schematically,
\bea \label{app1-1}
{\rm det}\left( \frac{\partial \phi(U,V,W)}{\partial(U,V,W) }  \right) = -\frac{U W+U+V W}{U V} <0 \ .
\eea
The PDF of the triplet $(U',V',W')$ is then directly evaluated as
\bea\label{app1-2}
P(U',V',W') = P_U( \phi^{(1)}(U',V',W') )  P_V( \phi^{(1)}(U',V',W') )  P_W( \phi^{(1)}(U',V',W') ) \times  \frac{U V}{U W+U+V W}    \    .
\eea
Where we introduced the PDF of the independent RVs $(U,V,W)$ as noted in (\ref{Statio-Inv-Beta2}) and (\ref{Statio-Inv-Beta3}) and used the fact that $\phi$ is an involution. It is then directly checked that 
\bea\label{app1-3}
P(U',V',W') = P_U( U')  P_V( V' )  P_W( W' )    \    ,
\eea
hence showing that $U'$, $V'$ and $W'$ are independent and distributed as $U' \sim U$, $V' \sim V$ and $W' \sim W$.

\section{Proof of the properties of the zero temperature stationarity map}  \label{app:T0Stationarity}

In this Appendix we prove Prop.~\ref{prop:T0statMap:Stat} and Prop.~\ref{prop:T0statMap:DB}. Let us first prove the detailed balance property Prop.~\ref{prop:T0statMap:DB}b. We thus consider $\sU \perp \sV \perp (\su, \sv)$ distributed as in (\ref{T0DistEnergyuv1}) and (\ref{T0DistEnergyUV1}).
%  Namely
% \bea 
% && \su \sim  (1- \zeta_{\su \sv}) (1+ G_{q'})  -\zeta_{\su \sv} G_{q} \in \JZ   \  ,    \\
% && \sv \sim -\zeta_{\su \sv} G_{q} \in \JZ_-   \  , \nn \\
% && \sU \sim  (1- \zeta_{\sU}) (1+ G_{q_b q'})  -\zeta_{\sU} G_{q/q_b} \in \JZ \ ,  \\
% && \sV \sim -\zeta_{\sV} G_{q_b}  \in \JZ_- \ ,
% \eea
% where $G_{q}$, $G_{q'}$, $G_{q_b q'}$, $G_{q/q_b}$ and $G_{q_b}$ are independent Geometric random variables distributed as in (\ref{T0probaGeo}) with parameters $q$, $q'$, $q_b q'$, $q/q_b$ and $q_b$ such that $0<q<q_b<1$ and $0<q'<1$, while $\zeta_{\sU}$, $\zeta_{\sV}$ and $\zeta_{\su \sv}$ are independent Bernoulli random variables distributed as in (\ref{T0probaBernou}) with parameters $p_{\sU} =\frac{1-q_b q'}{1-qq'}$, $p_{\sV}= \frac{1-q'}{1-q_b q'}$ and $p_{\su \sv}= \frac{1-q'}{1-qq'}$. 
Let us first compute the conditional probability
\bea
\Psi(k_{\sU'} , k_{\sV'} , k_{\sU} , k_{\sV}):= Proba\left(( (\sU', \sV' )= (k_{\sU'} , k_{\sV'} ) |  (\sU, \sV )= (k_{\sU} , k_{\sV} ) \right)
\eea

where $k_{\sU}  \in \mathbb{N}$, $k_{\sV} \in \JZ_-$, $\sU' = {\rm min}\left( \su , \sv + \sU - \sV \right)$ and $ \sV' = {\rm min}\left( \su+\sV-\sU , \sv \right)  = \sU' + \sV - \sU$. We have
\bea
\Psi(k_{\sU'} , k_{\sV'} , k_{\sU} , k_{\sV})=&&  p_{\su \sv}  \sum_{G_q =0}^{\infty} (1-q) (q)^{G_q}   \delta( k_{\sU'} =  {\rm min}\left( -G_q , -G_q +  k_{\sU}-  k_{\sV} \right) ) \delta( k_{\sV'} = k_{\sU'}+ k_{\sV}- k_{\sU}  )  \nn \\
&& + (1-p_{\su \sv})  \sum_{G_q' =0}^{\infty} (1-q') (q')^{G_q'}\delta( k_{\sU'} =  {\rm min}\left( 1+ G_q' , k_{\sU}-  k_{\sV} \right) ) \delta( k_{\sV'} = k_{\sU'}+ k_{\sV}- k_{\sU}  )  \nn \\
 && p_{\su \sv}  \sum_{G_q =0}^{\infty} (1-q) (q)^{G_q}   \delta( k_{\sU'} =   -G_q  )  \delta(k_{\sU}  > k_{\sV}  ) \delta( k_{\sV'} = k_{\sU'}+ k_{\sV}- k_{\sU}  )   \\
 && + p_{\su \sv}  \sum_{G_q =0}^{\infty} (1-q) (q)^{G_q}   \delta( k_{\sU'} =  -G_q +  k_{\sU}-  k_{\sV} )  \delta(k_{\sU}  \leq  k_{\sV}  ) \delta( k_{\sV'} = k_{\sU'}+ k_{\sV}- k_{\sU}  )  \nn \\
&& + (1-p_{\su \sv})  \sum_{G_q' =0}^{\infty} (1-q') (q')^{G_q'}\delta( k_{\sU'} =   1+ G_q'  ) \delta( k_{\sU}-  k_{\sV} > 1+ G_q' )  \delta( k_{\sV'} = k_{\sU'}+ k_{\sV}- k_{\sU}  )  \nn \\
&& + (1-p_{\su \sv})  \sum_{G_q' =0}^{\infty} (1-q') (q')^{G_q'}\delta( k_{\sU'} =  k_{\sU}-  k_{\sV}  ) \delta( k_{\sU}-  k_{\sV} \leq 1+ G_q' )  \delta( k_{\sV'} = k_{\sU'}+ k_{\sV}- k_{\sU}  )  \ , \nn 
\eea
i.e.
\bea
\Psi(k_{\sU'} , k_{\sV'} , k_{\sU} , k_{\sV})=&& p_{\su \sv} (1-q) (q)^{-k_{\sU'}}   \delta( k_{\sU'}  \leq 0 )  \delta(k_{\sU}  > k_{\sV}  ) \delta( k_{\sV'} = k_{\sU'}+ k_{\sV}- k_{\sU}  )  \nn\\
 && + p_{\su \sv}   (1-q) (q)^{-k_{\sU'} + k_{\sU}-k_{\sV}}   \delta( -k_{\sU'} + k_{\sU}-k_{\sV} \geq 0 )  \delta(k_{\sU}  \leq  k_{\sV}  ) \delta( k_{\sV'} = k_{\sU'}+ k_{\sV}- k_{\sU}  )   \nn \\
&& + (1-p_{\su \sv})  (1-q') (q')^{ k_{\sU'} -1}\delta( k_{\sU'} \geq   1  ) \delta( k_{\sU}-  k_{\sV} >k_{\sU'} )  \delta( k_{\sV'} = k_{\sU'}+ k_{\sV}- k_{\sU}  )  \nn \\
&& + (1-p_{\su \sv}) (q')^{k_{\sU}-  k_{\sV}-1 }\delta( k_{\sU'} =  k_{\sU}-  k_{\sV}  ) \delta( k_{\sU'}  \geq 1 )  \delta( k_{\sV'} = k_{\sU'}+ k_{\sV}- k_{\sU}  )   \nn \\
&& + (1-p_{\su \sv})  \delta( k_{\sU'} =  k_{\sU}-  k_{\sV}  ) \delta( k_{\sU'} \leq 0  )   \delta( k_{\sV'} = k_{\sU'}+ k_{\sV}- k_{\sU}  )     \ .
\eea
Using this last expression and the expression of $Proba((\sU, \sV )= (k_{\sU} , k_{\sV} ) )$ given in (\ref{T0DistEnergyUV2}), we obtain
\bea
\tilde{\Psi}(k_{\sU'} , k_{\sV'} , k_{\sU} , k_{\sV})&& := Proba\left(( (\sU', \sV' )= (k_{\sU'} , k_{\sV'} ) ,   (\sU, \sV )= (k_{\sU} , k_{\sV} ) \right)  \nn \\
&&  =  \Psi(k_{\sU'} , k_{\sV'} , k_{\sU} , k_{\sV}) Proba((\sU, \sV )= (k_{\sU} , k_{\sV} ) ) \nn \\
&&  = \Psi(k_{\sU'} , k_{\sV'} , k_{\sU} , k_{\sV}) \times \nn \\
&&  \left(p_{\sU} \delta(k_{\sU} \leq 0 ) (1-q/q_b)(q/q_b)^{-k_{\sU}}  +     (1 - p_{\sU} ) \delta(k_{\sU} \geq 1 ) (1-q_b q')(q_bq')^{k_{\sU} -1} \right) \times \nn \\
&&  \left( p_{\sV} \delta(k_{\sV} \leq 0 )(1-q_b)(q_b)^{-k_{\sV}} + (1 - p_{\sV} ) \delta(k_{\sV} = 0 )  \right) 
\eea
and it is then straightforward (although technically complicated due to the large number of terms) to check the detailed balance property Prop.~\ref{prop:T0statMap:DB}. Namely one shows that the equality
\bea \label{appT0DB}
\tilde{\Psi}(k_{\sU'} , k_{\sV'} , k_{\sU} , k_{\sV}) = \tilde{\Psi}(k_{\sU} , k_{\sV} , k_{\sU'} , k_{\sV'}) 
\eea
holds. Let us emphasize here that this property is rather special: the fact that (\ref{appT0DB}) works requires a large number of cancellation between terms that are made possible by the choice of only three parameters $p_{\sU} =\frac{1-q_b q'}{1-qq'}$, $p_{\sV}= \frac{1-q'}{1-q_b q'}$ and $p_{\su \sv}= \frac{1-q'}{1-qq'}$, a characteristic sign of the existence of exact solvability properties for the model. Finally, summing (\ref{appT0DB}) on $k_{\sU'}$ and $k_{\sV'}$ gives the stationarity property Prop.~\ref{prop:T0statMap:Stat}:
\bea
Proba( (\sU , \sV) = ( k_{\sU} , k_{\sV}) ) = Proba( (\sU' , \sV') = ( k_{\sU} , k_{\sV}) ) \ .
\eea

\end{document}